\documentclass[11pt]{article}
\usepackage[utf8]{inputenc}
\usepackage[T1]{fontenc}
\usepackage{fixltx2e}
\usepackage{graphicx}
\usepackage[alpine]{ifsym}
\usepackage{longtable}
\usepackage{float}
\usepackage{marvosym}
\usepackage{mathdots}
\usepackage{fullpage}
\usepackage{wrapfig}
\usepackage{enumitem}
\usepackage{epigraph}
\usepackage[normalem]{ulem}
\usepackage{amsmath, amsthm}
\usepackage{textcomp, stmaryrd}
\usepackage{marvosym}
\usepackage{wasysym}
\usepackage{amssymb}
\usepackage[colorlinks=true]{hyperref}
\usepackage[breakable]{tcolorbox}

\definecolor{block-gray}{gray}{0.85}
\newtcolorbox{mybox}{colback=block-gray,grow to right by=-10mm,grow to left by=-10mm,
boxrule=0pt,boxsep=0pt,breakable}

\newtcolorbox{mybox2}{colback=blue!5!white,grow to right by=-10mm,grow to left by=-10mm,
boxrule=0pt,boxsep=0pt,breakable}

\theoremstyle{definition}
\newtheorem{exercise}{Exercise}[section]
\newtheorem{statement}{Box}[section]

\hypersetup{
  pdfkeywords={},
  pdfsubject={},
  pdfcreator={Emacs 25.1.1 (Org mode 8.2.10)}}
\begin{document}

  \begin{center}
  \begin{Large}
    \textbf{Why is a soap bubble like a railway?} \\
  \end{Large}
  \vspace{10pt}
  \begin{large}
        David Wakeham\footnote{\href{mailto:david.a.wakeham@gmail.com}{david.a.wakeham@gmail.com}} \\
  \end{large}
\vspace{10pt}
\begin{small}
\emph{Department of Physics and Astronomy \\
  University of British Columbia \\
  Vancouver, BC V6T 1Z1, Canada \\}
\end{small}
\vspace{-5pt}
  \end{center}
\hrulefill

  \begin{abstract}
    \noindent
At a certain infamous tea party, the Mad Hatter posed the following 
riddle: why is a raven like a writing-desk?
We do not answer this question.
Instead, we solve a related nonsense query: why is a soap bubble
like a
railway?
The answer is that both minimize over graphs.
We give a 
self-contained
introduction to graphs and minimization, 
starting with minimal networks on the Euclidean plane and ending with 
close-packed structures for three-dimensional foams.
Along the way, we touch on algorithms and complexity, the physics of
computation, curvature, chemistry, space-filling polyhedra, and bees
from other dimensions.
The only prerequisites are high school geometry, some algebra,
and a spirit of adventure. 
These notes should therefore be suitable for high
school enrichment and bedside reading.
  \end{abstract}
\vspace{150pt}
  \begin{center}
\includegraphics[scale=0.11]{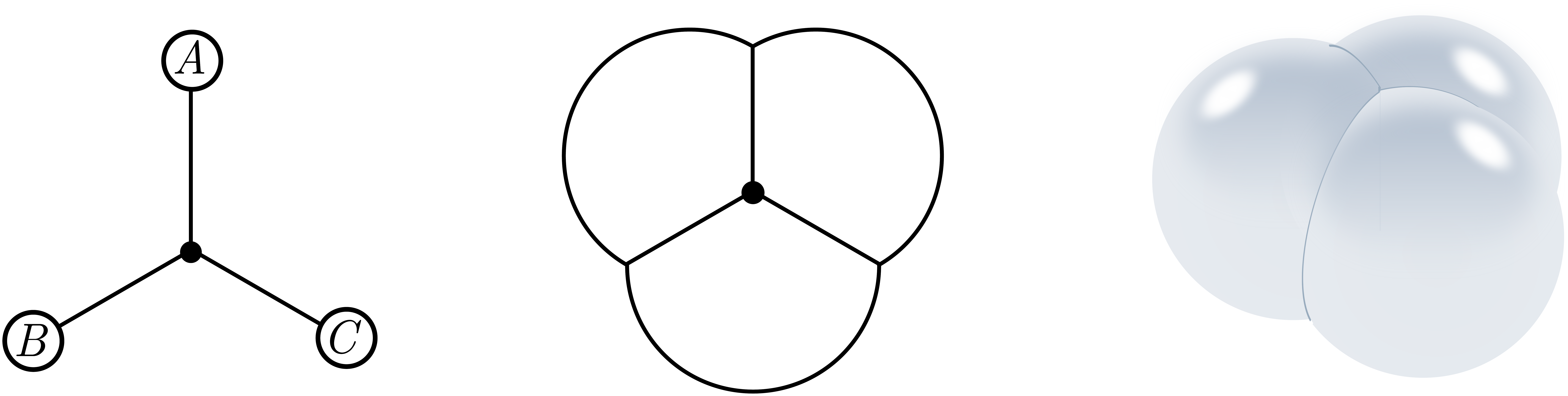} 

  \end{center}

\thispagestyle{empty}

\newpage

\tableofcontents

\vspace{15pt}
\begin{center}
  \textbf{Acknowledgments}
\end{center}

\noindent These notes were inspired by conversations with Rafael
Haenel, Pedro Lopes and Haris Amiri of
\href{https://www.digitalsupercluster.ca/programs/capacity-building-program/diversifying-talent-in-quantum-computing/}{DTQC},
and the hands-on Steiner tree activity they created for the
\href{https://outreach.phas.ubc.ca/events/metro-vancouver-physics-circle}{UBC
  Physics Circle}.
It got me curious about the computing power of bubbles!
I would particularly like to thank Rafael and Pedro for encouragement,
the Physics Circle students for letting me test material on them, and
Scott Aaronson for detailed comments on the draft.
I am supported by an International Doctoral Fellowship from UBC.

\thispagestyle{empty}

\newpage

\pagenumbering{arabic}

\section{Introduction}
\label{sec:intro}

\vspace{10pt}

\begin{quote}
    \begin{small}
The Hatter opened his eyes very wide on hearing this; but all he said was, ``Why is a raven like a writing-desk?''
``Come, we shall have some fun now!'' thought Alice. ``I’m glad they’ve begun asking riddles.—I believe I can guess that,'' she added aloud.
``Do you mean that you think you can find out the answer to it?'' said the March Hare.
``Exactly so,'' said Alice.
  \begin{flushright}
\emph{Lewis Carroll}
  \end{flushright}
    \end{small}
\end{quote}

\noindent Why is a soap bubble like a railway?
I believe we can guess that.
Suppose we are designing a rail network which joins three cities.
If stations are cheap, our biggest expense will be rail itself,
and to minimize cost we should make the network as short as
possible.
For three cities $A$, $B$ and $C$, the cheapest network
typically looks like the example below left.
In addition to stations at each city, we add a \emph{hub}
station in the middle to minimize length.
For a general triangle of cities, hub placement follows a simple
rule: outgoing rail lines are equally spaced, fanning out at angles
of $120^\circ$.

\begin{center}
  \includegraphics[scale=0.11]{pics/intro}
\end{center}

A two-dimensional bubble, with walls made of soapy water, solves the same
problem.
The molecules in the water are attracted to each other, creating surface tension.
Tension pulls the surface taut, and length is once again minimized,
due to the budgetary constraints of Nature itself.
Like rail lines, bubble walls converge at junctions three at
a time, separated by $120^\circ$.
The rule even works for the soapy walls of a \emph{three-dimensional}
bubble.

Of course, rail networks in the real world connect many cities, and
the problem is more complicated.
But it remains true that for the cheapest network, any time we
introduce a hub it must have three rail lines emerge at angles of
$120^\circ$, with the same going for multiple bubbles. 
This makes the connection between soap bubbles and railways \emph{useful}: by drilling screws through plexiglass, we can make a
soap bubble computer, and solve network planning problems with soapy
water!

\vspace{5pt}
\begin{center}
    \includegraphics[scale=0.22]{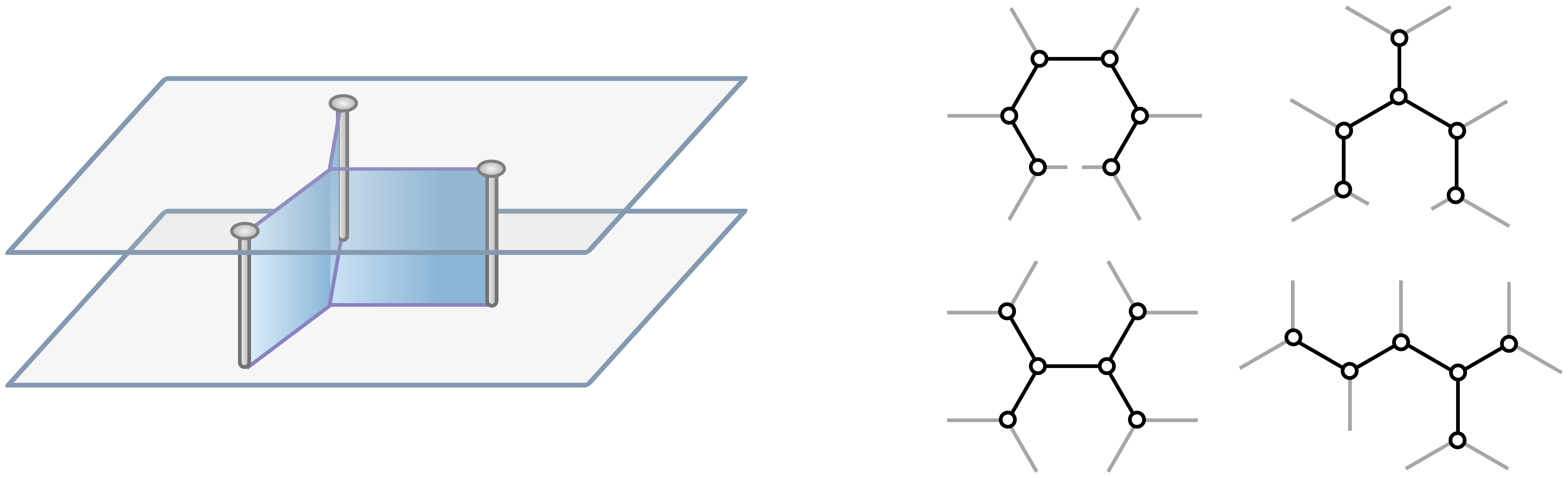}
\end{center}

While soap bubbles can find small railways almost instantaneously,
there is a deep but subtle reason they aren't useful for finding the
best way to connect every city in North America.
In principle, we just place a screw at the position of every city, dip
into soapy water, and withdraw.
In practice, it will probably take longer than the age of the universe for the
bubbles to settle down! 
The problem is just too hard.
Although we know what hub stations look like \emph{locally}---a
trident of three rail lines---there are many different ways to
arrange a given number of hubs.
We show a few examples 
above right.
As the number of hubs gets large, there are so many that
\emph{no physical mechanism}, soap bubbles or quantum computers or
positronic brains, can quickly search them all to find the shortest
candidate, unless there is a wildly clever algorithm we have overlooked.
This is called the \emph{\textsf{NP} Hardness Assumption}.
Ultimately, this is a physical hypothesis, because it makes predictions
about the behaviour of physical objects which compute, such as soap bubbles.

If it takes arbitrarily large amounts of computing power to find the
cheapest network, it is no longer cheap.
\emph{Approximate} answers are preferable if they can be found
quickly, and we will give two methods for rapid (but suboptimal) rail
planning below. 
But bubbles still hold surprises.
Once we remove the plexiglass and screws, genuine bubbles are free to
form, each cell enclosing some fixed volume.
The laws of physics will now try to minimize the total area of the cell 
surfaces, so poetically speaking, the forms flowing out of the bubble blower
are \emph{conjectures} made by Nature about the best
(i.e. smallest-area) way to enclose some air pockets.

For example, the humble spherical bubble harbours the following
conjecture: of all surfaces of fixed volume $V$, the sphere has the
smallest area.
This is the \emph{isoperimetric inequality}, a result we will prove later.
But surprisingly little is known about more bubbles.
While the symmetric \emph{double bubble} shown below is the most
economic way to enclose two equal volumes, 
no one knows if the symmetric triple bubble is optimal for three equal
volumes.

\vspace{10pt}
\begin{center}
\includegraphics[scale=0.12]{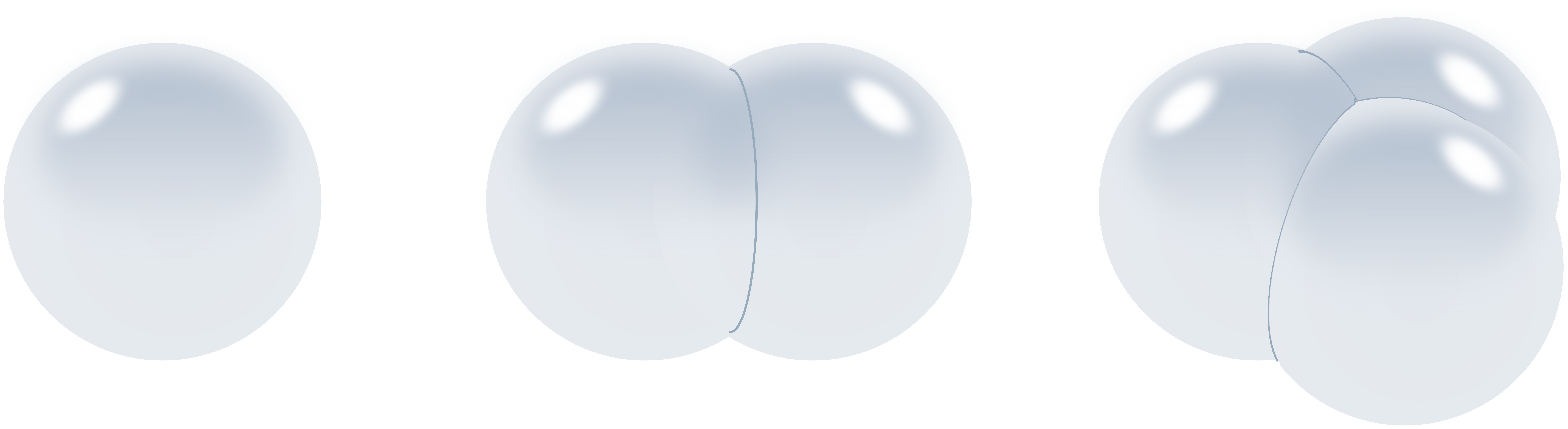}
\end{center}

Our comparative ignorance of bubbles will not stop us launching,
undaunted, into the problem of partitioning not two, not three, but an
\emph{infinite number} of equal volumes.
As a warm-up, we can consider the problem for two-dimensional
bubbles.
We will show that in a large foam of bubbles, the $120^\circ$
rule means that most cells are hexagonal.
This helps explain why bees prefer a hexagonal lattice for building
their hives.
They are trying not to waste wax!
In fact, the hexagonal tessellation, where each hexagon is identical,
provably requires the least amount of wax per equal volume cell.

\vspace{5pt}

\begin{center}
\includegraphics[scale=0.18]{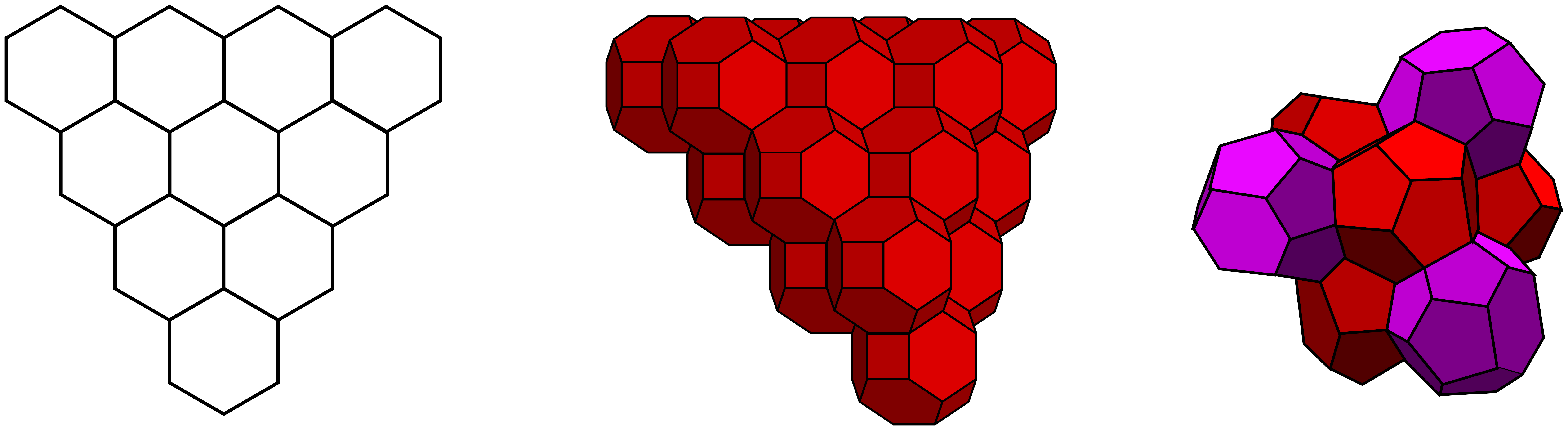}
\end{center}

In three dimensions, things are more interesting. 
In addition to the $120^\circ$ rule, we need a few other rules for
bubbles which together make up \emph{Plateau's laws}.
Unlike two dimensions, these laws don't tell us precisely how many
faces a bubble has, but they do give some constraints.
We can use these constraints to eliminate all but one space-filling
pattern, the \emph{Kelvin structure} (above middle),
made from pruned octahedra.
Surprisingly, this is \emph{not} the best way to separate an infinite
number of cells of equal volume. 
There is a mutant tessellation made from weaving together two
different equivoluminous shapes called the \emph{Weaire-Phelan structure}, shown above right.
Although there are no four-dimensional bees to store their honey in Weaire and
Phelan's cells, Nature uses this structure to make superconductors and
trap gas.
No one knows if there is a way to beat it.


\subsection{Outline}
\label{sec:overview-2}

Let's outline the contents a little more formally.
In \S \ref{sec:triangles}, 
we start our study of minimization with the suprisingly rich problem of minimal
networks on the triangle.
In \S
\ref{sec:equil-triangl}, we analyze the equilateral
triangle using symmetry, and argue that a hub should be placed in the
center.
We deform this solution in \S \ref{sec:deforming-triangle}, and give
some loose arguments that the hub collides with a vertex when an
internal angle opens to $120^\circ$.
This is generalized in \S \ref{sec:hubs-spokes} to give the
$120^\circ$ rule for general minimal networks.
Finally, in \S \ref{sec:historical-notes}, we give a brief history of
minimal networks and related problems.

In \S \ref{sec:graphs}, we use tools from graph theory to take
the $120^\circ$ rule, which is a local constraint, and turn it into a global
constraint on the structure of the network.
Trees and their basic properties are introduced in \S \ref{sec:trees},
and exploited in \S \ref{sec:max-hubs} to put a bound on the maximum number
of hubs. This allows us to solve some small but nontrivial networks.
In \S \ref{sec:tinkertoys}, the bound is turned into a rough argument
for the computational hardness of finding minimal networks,
while \S \ref{sec:minim-spann-trees} provides some easily computable
alternatives, namely the minimal spanning tree and Steiner insertion heuristic.

With \S \ref{sec:soap-bubbles}, we move laterally
into the realm of soap bubbles.
We build soap bubble computers in \S \ref{sec:comp-with-bubbl} to solve our
minimal network problems, where our computational hardness results
resurface as predictions about physics. 
In \S \ref{sec:foams}, we introduce Euler's formula and apply it to
bubble networks, while in \S \ref{sec:hexagon}, we make a simple scaling
argument that most bubbles in a large foam are hexagonal.
This is related to the fact that bees build hexagonal hives, and the
\emph{honeycomb theorem}
that bees know the best way to partition the plane into cells of equal
size.
The \emph{planar minimal bubble problem} 
make its appearance in \S
\ref{sec:circles-bubbletoys}, along with a heuristic proof of the isoperimetric
inequality.

The last section, \S \ref{sec:plateaus-laws}, considers
three-dimensional bubbles.
After defining mean curvature in \S \ref{sec:mean}, we state Plateau's laws
in \S \ref{sec:laws}, motivating them by analogy with bubble networks.
With \S \ref{sec:spheres-bubbletoys}, we describe the three-dimensional
bubble problem 
and Plateau's problem for wireframes and bubble blowers.
Finally, in \S \ref{sec:infinite-foam} we generalize Euler's
formula to study network constraints on three-dimensional foams, and
conclude with a whirlwind tour of regular tessellations
of space, the Kelvin problem, 
the Weiare-Phelan surprise, and the chemistry of tetrahedrally closed-packed
structures.

\textbf{Prerequisites.} 
The only prerequisites for these notes are high school algebra,
geometry and a little physics.
You can do a lot of minimization without calculus!
The material should therefore be suitable for high school enrichment
in math or physics, and parts of sections \ref{sec:triangles}--\ref{sec:graphs} have been successfully trialled in a physics outreach
program.
We often resort to heuristics, pictures, and
physical intuition, which may deter some readers.
But the price of admission is lower, and we hope the rides no less
fun!

\textbf{Exercises.} There are around 40 problems of varying difficulty.
Many of these are 
used subsequently in the text.
I hope this is not a weakness, but rather than an incentive to solve them!
Difficult exercises are labelled with a mountain (\Mountain), or an
icy mountain (\IceMountain) in the case of greater abstraction or
required background.
Mountain ranges (\VarMountain $\,$ and \VarIceMountain) inflect for length.
For solutions, please contact me
\href{mailto:david.a.wakeham@gmail.com}{by email}.
They will hopefully be included in a future iteration.

\newpage

\section{Trains and triangles}
\label{sec:triangles}

Suppose we want to join up three towns $A$, $B$ and $C$ by rail.
Building railways is expensive, since we not only need to design and
build the rail itself, but acquire the land beneath it.
In contrast, stations can be reasonably cheap: we just slap together some sidings, a
platform, and a bench or two.
To minimize cost we should make the total length of the
rail network as short as possible.

\begin{figure}[h]
  \centering
  \includegraphics[scale=0.45]{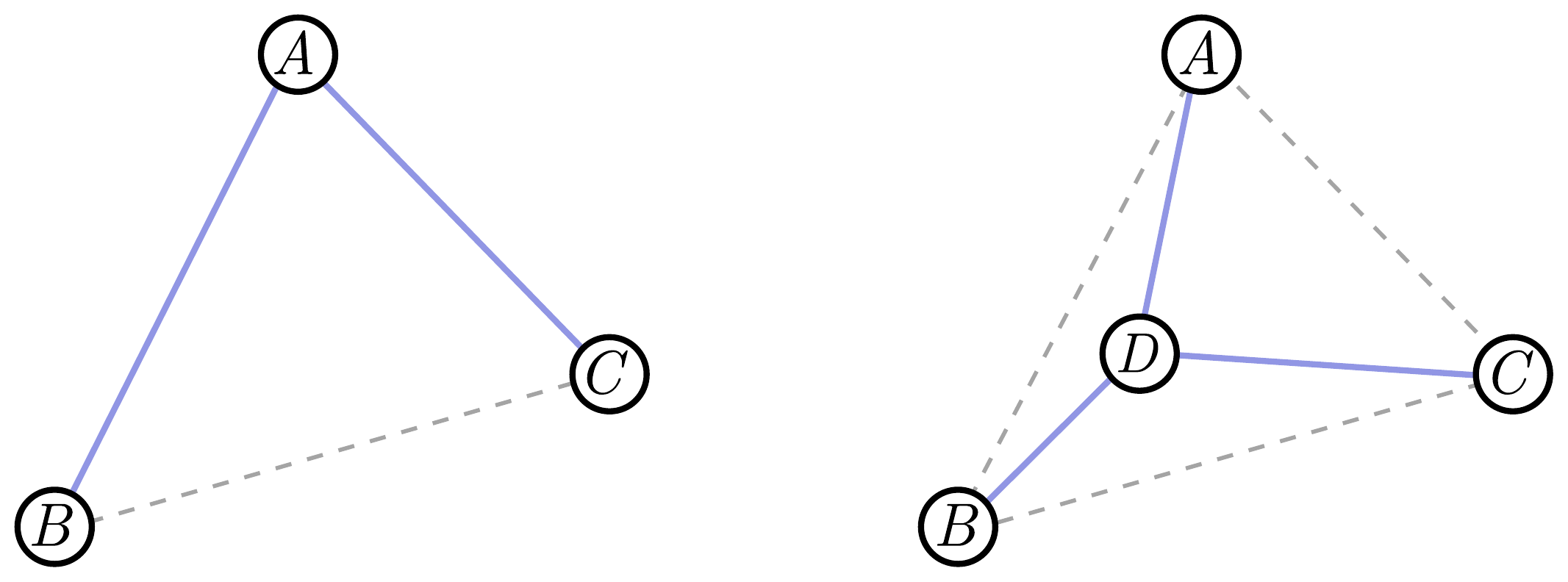}
  \caption{Rail networks (triangle and
    trident) connecting three towns.}
  \label{fig:tri}
\end{figure}

If the railway lets us travel from one town to any
other, we say that the rail network is \emph{connected}.
A \emph{hub} is a station built solely to connect rails.
A connected rail network of minimal length is called a \emph{minimal network}
or \emph{Steiner tree}.\footnote{We will see where the term ``tree''
  comes from in \S \ref{sec:graphs}.}
Two possible networks for the three towns are shown in
Fig. \ref{fig:tri}.
The ``triangle'' network is built from two sides of the triangle formed by
the three towns, while the ``trident'' network adds a hub (also called
a \emph{Steiner point}) in the middle.

\vspace{10pt}
\begin{mybox}
  \begin{exercise}
    \emph{Choosing sides.}
    \label{ex:sides}
  \end{exercise}
  Suppose $A$, $B$ and $C$ are separated by distances $AB$, $AC$ and
  $BC$. A triangular network consists of two sides of the triangle.
  Which ones should we choose?
\end{mybox}

\vspace{10pt}
\begin{mybox}
  \begin{exercise}
    \emph{Triangle or trident?}
    \label{ex:trides}
  \end{exercise}
  In Fig. \ref{fig:tri}, we have two networks connecting the same
  towns: two sides of the triangle, and the trident-shaped network
  with a hub $D$ in the middle.  Check the trident is shorter. \emph{Hint.}
  Measure lengths with a ruler. Simple but it works!
\end{mybox}
\vspace{5pt}

Already, there is a surprise.
Although the simplest network consists of two sides of the triangle,
this is not minimal, since (to spoil Exercise \ref{ex:trides}) the trident in Fig. \ref{fig:tri} is shorter.
We can go further and optimize the placement of the hub $D$.
The case for a general triangle is tricky, but we can build most of the intuition we need by focusing on the
special case of an \emph{equilateral} triangle.

\subsection{Equilateral triangles}
\label{sec:equil-triangl}

\begin{figure}[h]
  \centering
  \includegraphics[scale=0.45]{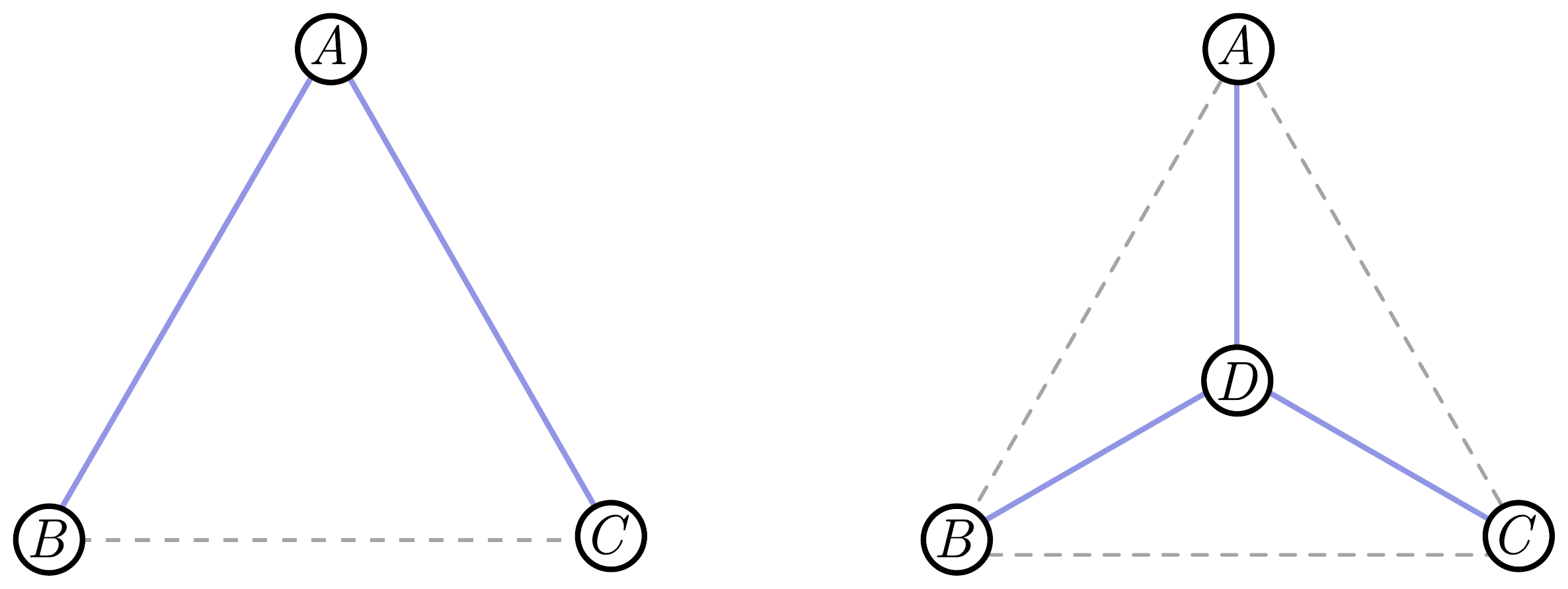}
  \caption{Rail networks on an equilateral triangle.}
  \label{fig:eq}
\end{figure}

Suppose $A$, $B$ and $C$ sit on the corners of an
equilateral triangle of side length $d$, as in Fig. \ref{fig:eq}.
The triangular network has total length $L_\Lambda = 2d$.
For the trident network on the right, we place the hub $D$ directly in
the middle.
Let's trade our engineering for math hats, and find the length of the
trident network using trigonometry.

\vspace{10pt}
\begin{mybox}
  \begin{exercise}
    \label{ex:eq-tri}
    \emph{Equilateral trident.}
  \end{exercise}
  Show that the length of the trident network is
$$
L_{\text{Y}} = \sqrt{3}d.
$$
Since $\sqrt{3} \approx 1.7 < 2$, the trident is shorter than the
triangle.
\end{mybox}
\vspace{5pt}

\begin{figure}[h]
  \centering
  \includegraphics[scale=0.45]{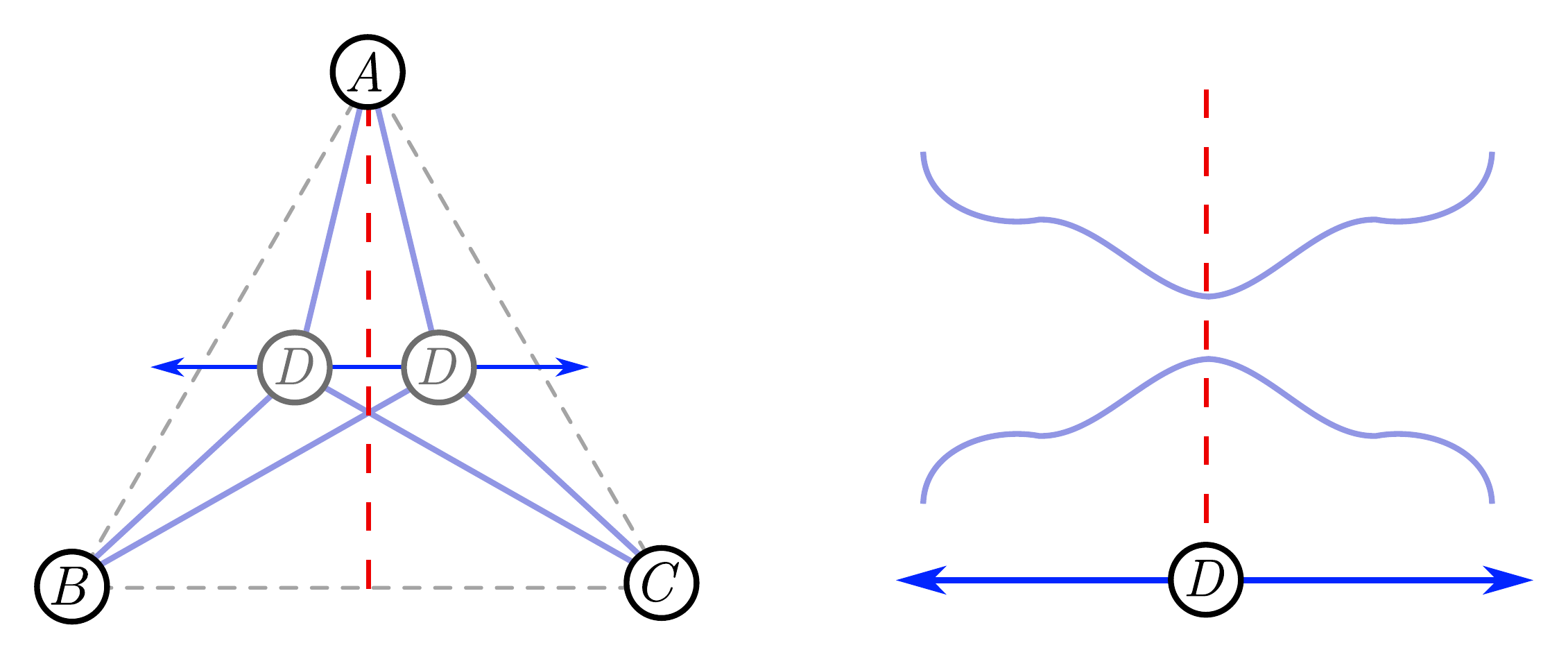}
  \caption{\emph{Left.} Wiggling the hub. \emph{Right.} Length is an even function of wiggle.}
  \label{fig:wiggler}
\end{figure}

Although this beats the triangle network, it's possible that placing $D$ somewhere other than the center could make the network even
shorter.
But as it turns out, the center is optimal, and we can argue this from
\emph{symmetry}.
We draw one of the triangle's axes of symmetry\footnote{This cuts the
  triangle into two mirror-image halves.} in red in
Fig. \ref{fig:wiggler}.
We can wiggle the hub $D$ left and right along the dark blue line in Fig. \ref{fig:wiggler}.

\begin{figure}[h]
  \centering
  \includegraphics[scale=0.45]{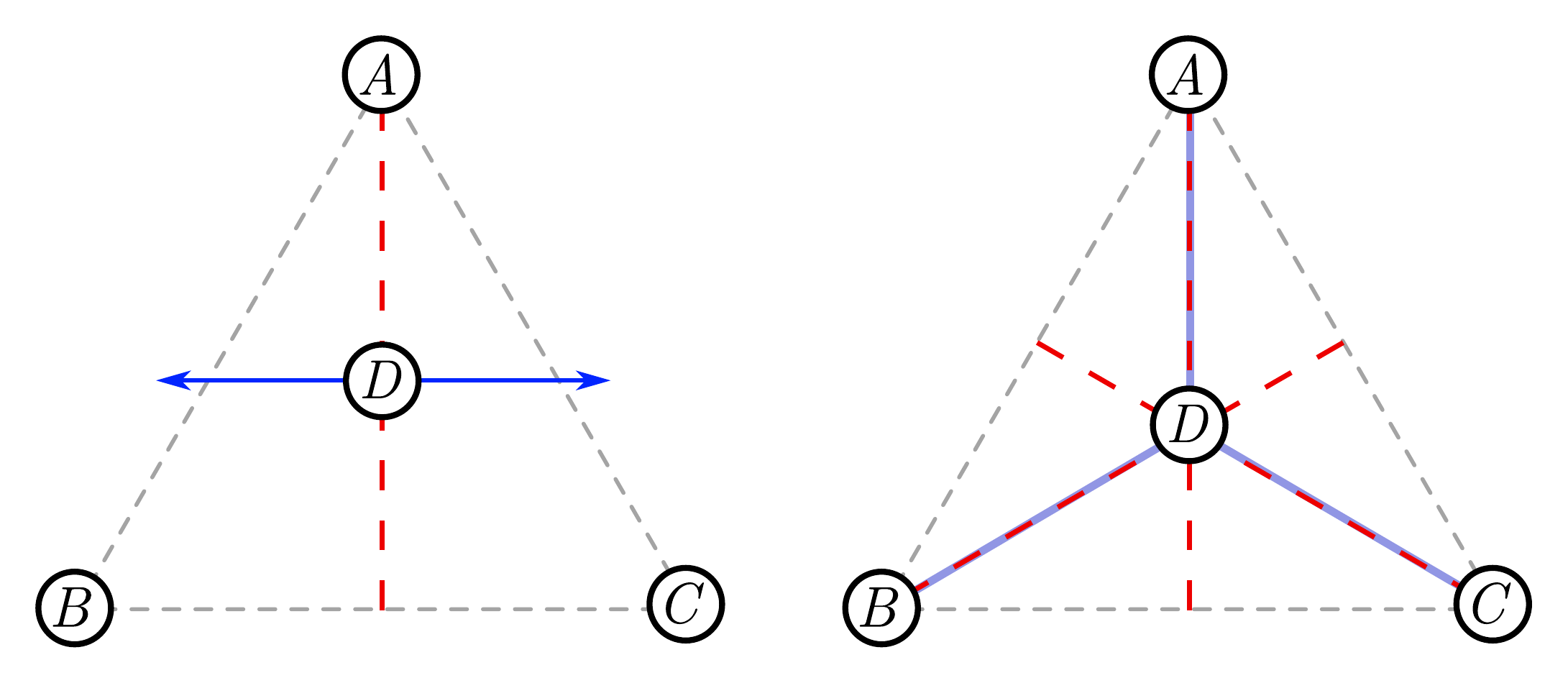}
  \caption{\emph{Left.} Length is minimized on the red
    line. \emph{Right.} Total length is minimized at the intersection
    of the red lines.}
  \label{fig:center}
\end{figure}

Because of symmetry, the total length of the network (light blue
lines) is an \emph{even function} of how far we have moved $D$ along
the dark blue line.
On the right in Fig. \ref{fig:wiggler}, we depict two possibilities
for an even function.
Length could either be a \emph{minimum} on the red line, like the
curve on top, or a \emph{maximum}, like the curve on the bottom.
Of course, if we move the hub along the blue line outside the
triangle, the network becomes very long.
This suggests it is a minimum\footnote{This does not \emph{prove} it
  is a minimum, since there may be other minima we have missed. 
See Exercise \ref{ex:vector} for a rigorous proof.
} on the red line, and for a minimal
network it should lie on that line, as in Fig. \ref{fig:center}
(left).
But there are two other axes of symmetry, associated with $B$ and $C$.
All three intersect at the center of the triangle, as shown in
Fig. \ref{fig:center} (right).
Since $D$ should lie on each of these lines, it must lie at the
center!

\subsection{Deforming the triangle}
\label{sec:deforming-triangle}

\begin{figure}[h]
  \centering
  \includegraphics[scale=0.44]{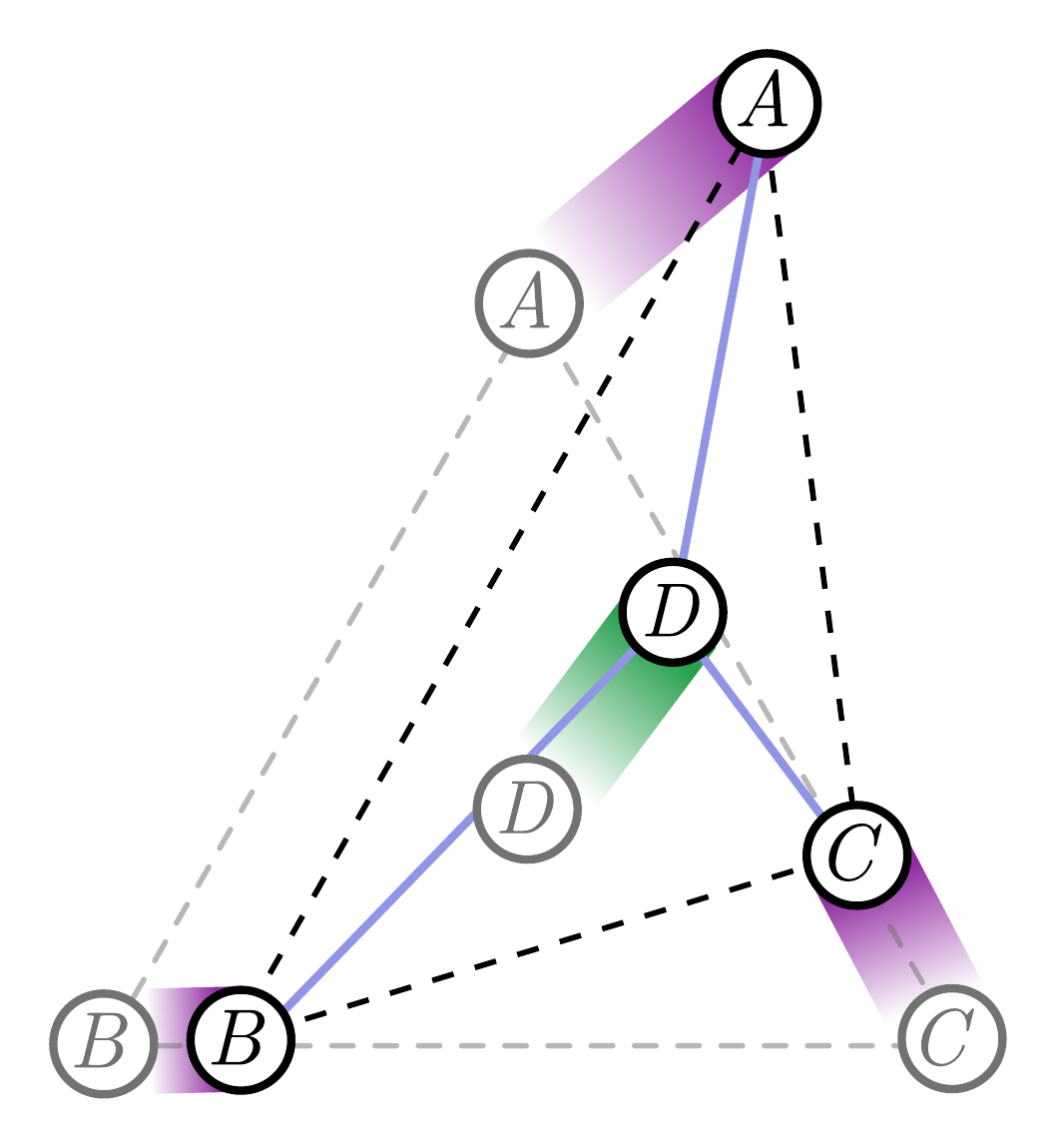}
\vspace{-12pt}
  \caption{Optimal hub position slides as we slide the corners of the triangle.}
  \label{fig:slide}
\end{figure}

We are now going to take our solution to the equilateral triangle and
slowly \emph{deform} it, sliding the corners so that the triangle
so it is no longer equilateral.
What will happen to the optimal position of the hub $D$?
Since everything is sliding continuously, the optimal hub
should slide continuously as well.
In Fig. \ref{fig:slide}, we give an example, with the paths of the
corners are depicted in purple, and the corresponding continuous change of hub in green.

\vspace{-5pt}
\vspace{-5pt}

\begin{figure}[h]
  \centering
  \includegraphics[scale=0.44]{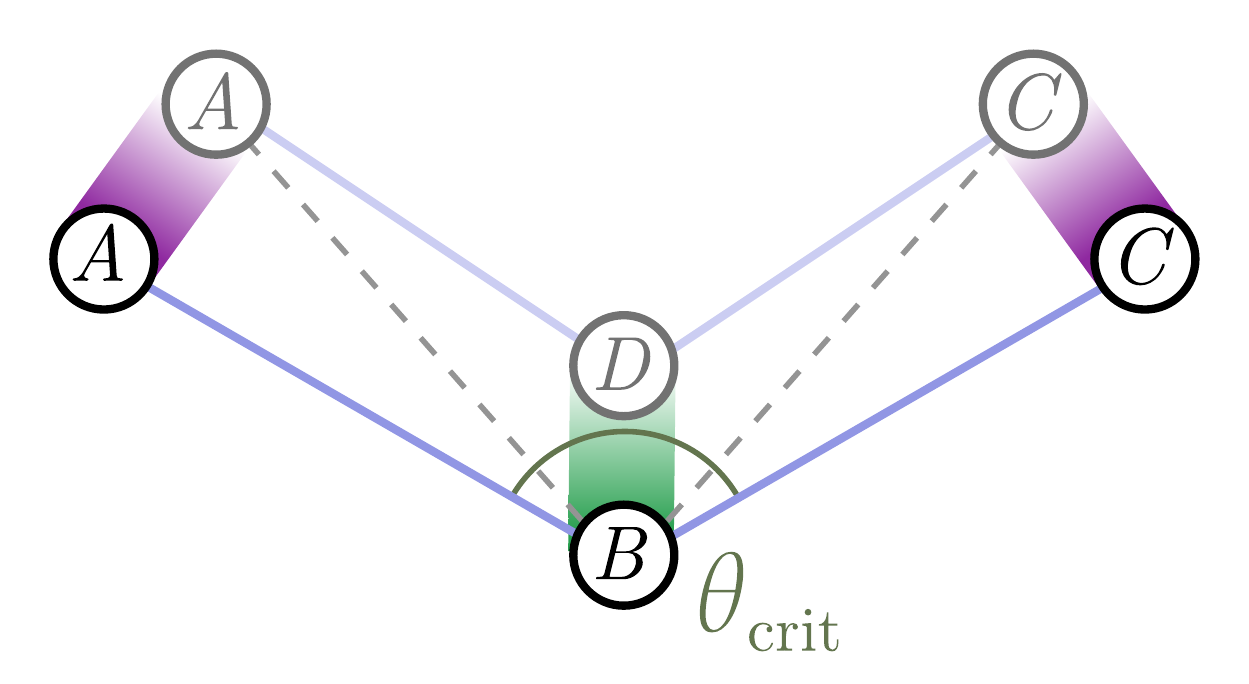}
  \caption{At some critical angle $\theta_\text{crit}$, $D$ collides
    with $B$.}
  \label{fig:crit}
\end{figure}

Since the hub position changes continuously, it should stay inside the
triangle for small deformations of the corners.
But for triangles which are far from equilateral, the sliding hub might
\emph{collide} with a corner!
In this case, the trident network collapses into a simpler triangular
network, formed from two sides of the triangle.
In Fig. \ref{fig:crit}, $B$ remains fixed in position, but $A$ and $C$
lower symmetrically and open out the angle of the triangle, with the optimal hub $D$
moving vertically down as they do so.
At some critical angle $\theta_\text{crit}$, it will coincide with $B$.

\begin{figure}[h]
  \centering
  \includegraphics[scale=0.44]{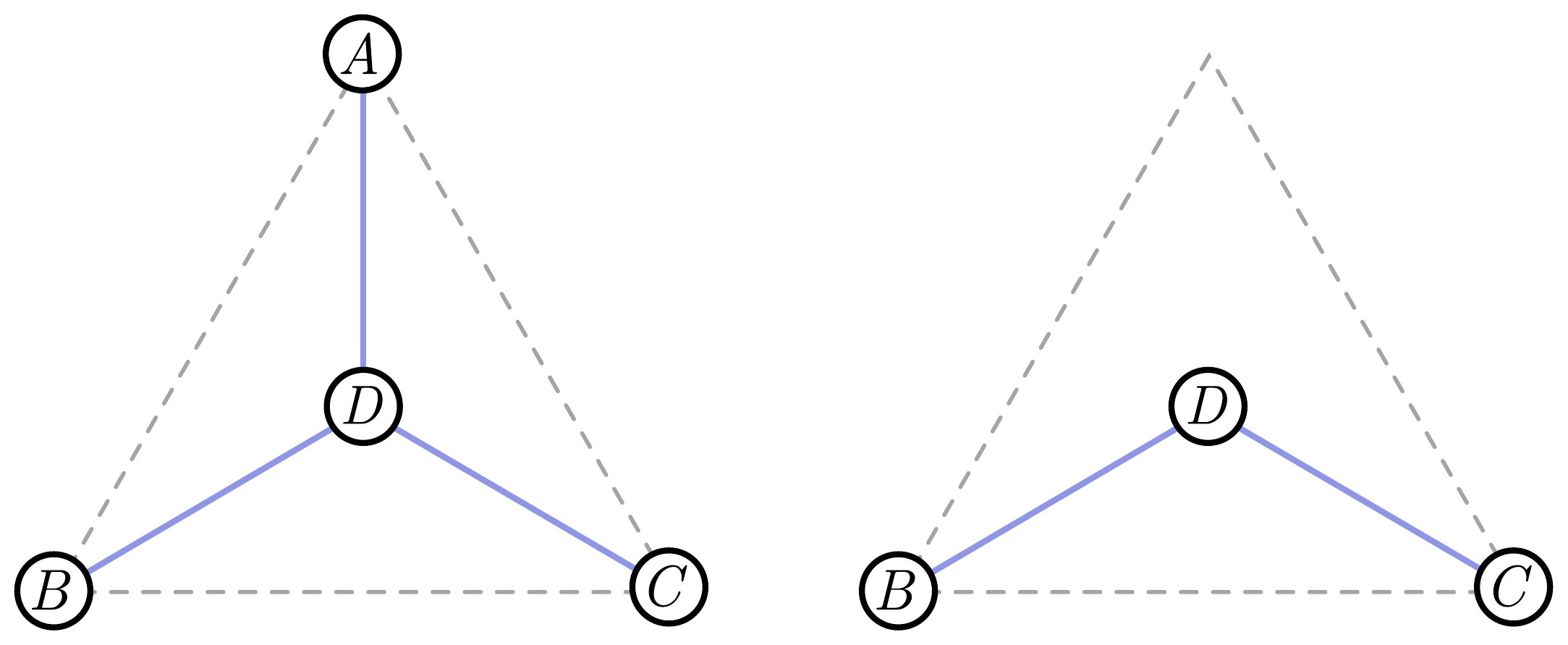}
  \caption{Removing a corner city removes a leg from the equilateral
    trident.}
\vspace{-5pt}
  \label{fig:critri}
\vspace{-10pt}
\end{figure}

It turns out this critical angle is $\theta_\text{crit} = 120^\circ$. 
Although we won't provide a watertight proof just yet, we can give a
plausibility argument.
Let's return to the equilateral triangle.
Instead of adding a hub in the middle, suppose that $D$ is in fact a
\emph{fourth city} fixed in place.
Clearly, the solution in Fig. \ref{fig:critri} (left) is still optimal, since if we could
add more hubs to reduce the total length, we could add more hubs to
improve the network for the equilateral triangle.
If we now remove a corner city, such as $A$, the optimal
network removes the corresponding leg of the trident, as in
Fig. \ref{fig:critri} (right).

\vspace{10pt}
\begin{mybox}
  \begin{exercise}\label{ex:corners}
    \emph{Cutting corners.}
  \end{exercise}
  Suppose that in Fig. \ref{fig:critri} (right), we can add a new hub
  $E$ which reduces the total length of the network.  Explain how
  adding $E$ could reduce the length of the network in
  Fig. \ref{fig:critri} (left), and thereby improve our solution for
  the equilateral triangle. 
\end{mybox}
\vspace{5pt}

\noindent Exercise \ref{ex:corners} is an example of a
\emph{proof by
   contradiction}, a favourite proof method among mathematicians. To
 show something is false, we assume it is true and use it to derive a
 contradiction with known facts.
 We then reason backwards to
 conclude that it cannot be true!
The next exercise gives a slightly stronger indication that the
critical angle is $120^\circ$.
This is the best we can do without more involved math
(Exercises \ref{ex:fermat} and \ref{ex:vector}).

\vspace{10pt}
\begin{mybox}
  \begin{exercise}
    \emph{Critical isosceles.} \Mountain
  \end{exercise}
  The argument above really only establishes that
  $\theta_\text{crit} \leq 120^\circ$.  In principle, the triangular
  network might become optimal at some angle
  $\theta_\text{crit} < 120^\circ$.  In this exercise, we will show
  for an isosceles triangle that this is not the case.
  We will need the \emph{law of cosines},
  \[
  c^2 = a^2 + b^2 - 2ab \cos \theta,  
  \]
  for the triangle depicted below left:

  \begin{center}
    \includegraphics[scale=0.34]{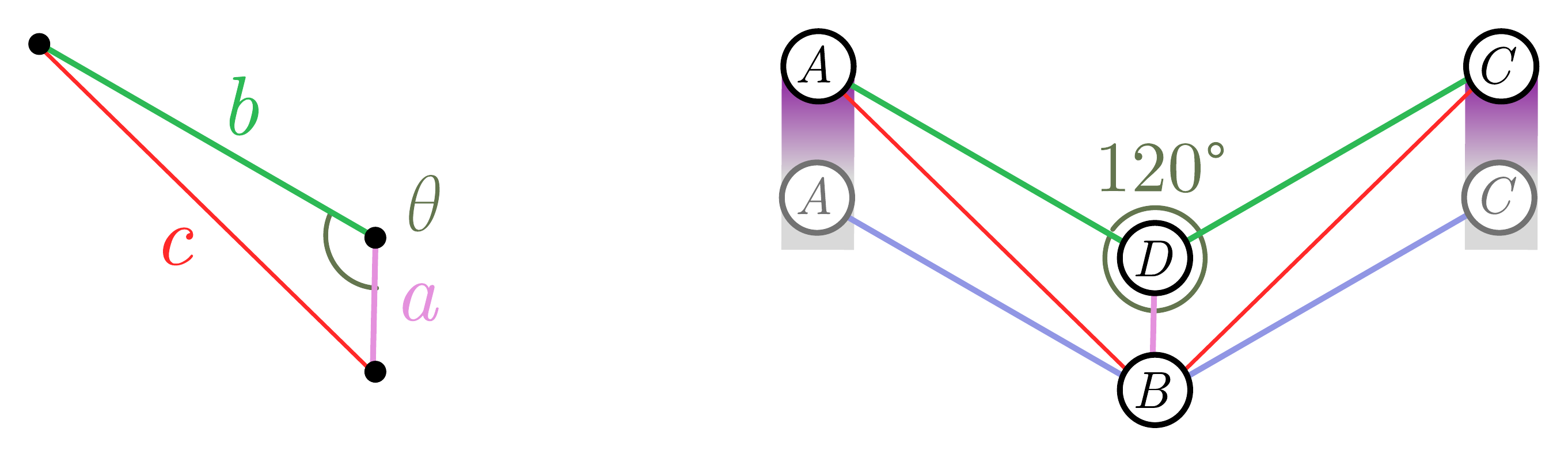}
    \label{fig:critiso}
  \end{center}
\vspace{-15pt}

  Above right, we have a triangular network (blue lines) $ABC$,
  forming an angle of $120^\circ$.  We now raise the two nodes $A$ and
  $B$ symmetrically so that the angle $ABC$ is less than $120^\circ$.
  You can prove that the green and purple lines are shorter than the
  red lines, so that an interior hub $D$, making an angle $120^\circ$
  with green and purple lines, yields a shorter network.
  \begin{enumerate}[label=(\alph*), itemsep=0pt]
  \item Show using the law of
    cosines 
    (or otherwise) that
$$
c^2 = a^2 + b^2 + ab.
$$
\item From part (a), argue that
$$
a + 2b < 2c.
$$
\item Conclude that for an isosceles triangle $ABC$, the critical
  angle is $\theta_\text{crit} = 120^\circ$.
\end{enumerate}
\vspace{0pt}
\end{mybox}

\subsection{The $120^\circ$ rule}
\label{sec:hubs-spokes}

Let's state the general, $n$-city version of the problem we've been studying:

\vspace{10pt}
\begin{mybox2}
  \begin{statement}
    \emph{Minimal networks.} \label{ex:min-net}
  \end{statement}
  Suppose we have $n$ cities on the plane.
  The \emph{minimal network} or \emph{Steiner tree} is a configuration
  of edges connecting these cities which has minimal total length.
  We can introduce additional hubs in order to minimize this total length.
\end{mybox2}
\vspace{5pt}

\noindent Our work with triangles pays off with a remarkable
conclusion about minimal networks connecting \emph{any} number of
cities called \emph{the $120^\circ$ rule}.
Readers who are not interested in the proof may simply internalize the
the contents of the following blue box 
and move on.

\vspace{10pt}
\begin{mybox2}
  \begin{statement}
    \emph{The $120^\circ$ rule.}
  \end{statement}
In a minimal network, every hub has
  three edges separated by angles of $120^\circ$.
\end{mybox2}
\vspace{5pt}

\begin{figure}[h]
  \centering
  \includegraphics[scale=0.5]{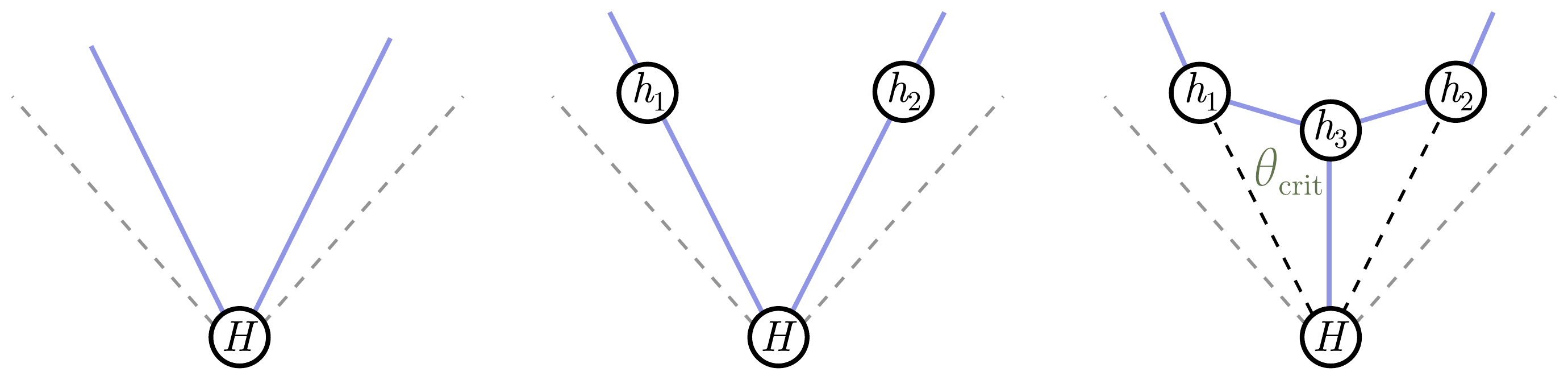}
  \caption{\emph{Left.} A hub with incoming angle less than
    $\theta_\text{crit}$. \emph{Middle.} Adding two extra stations. \emph{Right.} A shorter network.}
  \label{fig:rule1}
\end{figure}

The argument is 
ingenious.
Our first step is to show that it is impossible for a hub to have
edges separated by less than $120^\circ$.
Suppose we have cities or \emph{fixed nodes} $A_1, A_2, \ldots, A_n$ connected by a
minimal network, and a hub station $H$ with incoming rail lines
separated by less than
$\theta_\text{crit} = 120^\circ$, as on the left in Fig. \ref{fig:rule1}.
There may be other incoming lines, but these will play no role in our
proof and can be ignored.

We can build two new stations on these outgoing legs, $h_1$ and $h_2$,
without changing the length of track.
For simplicity, we take these new stations to be the same distance from $H$, as in
Fig. \ref{fig:rule1} (middle).
But from our work in the previous section, we know that the minimal
network connecting $h_1$, $h_2$ and $H$ is not the triangle network we have
drawn! Instead, it is a trident with another hub $h_3$ in the middle,
Fig. \ref{fig:rule1} (right).
This strictly decreases the length of the network, so our original
network could not be truly minimal.

\begin{figure}[h]
  \centering
  \includegraphics[scale=0.5]{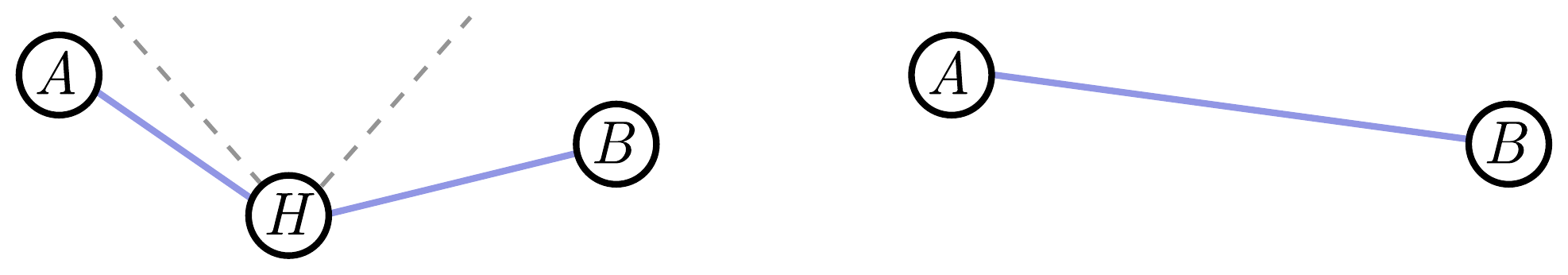}
  \caption{\emph{Left.} A hub with incoming angle greater than
    $\theta_\text{crit}$. \emph{Right.} A shorter network.}
  \label{fig:rule2}
\end{figure}

This means that any hub must have spokes separated by \emph{at least}
$120^\circ$.
How do we know that there are three, separated by exactly $120^\circ$?
Well, suppose two lines enter $H$, separated by \emph{more} than $120^\circ$.
Then there can only be two incoming edges, joining $H$ to some cities
$A$ and $B$, since any additional lines would have to be closer than
$120^\circ$ to one of these lines.
We have the situation depicted on the left of
Fig. \ref{fig:rule2}.

\vspace{5pt}

Hopefully you can see what goes wrong: if there is a
``kink'' in the blue line,
then we can obtain a shorter
network be deleting $H$ and directly connecting $A$ and $B$.
(Remember that $H$ is a hub, introduced only to shorten the network,
and not a city that needs to be connected.)
Once again, we have a contradiction!
Strictly speaking, we can have hubs with only two incoming edges,
separated by $180^\circ$.
But such a hub is always unnecessary, since all it does is sit on a
straight line.
If we delete these useless hubs, we have the general result advertised
above, namely that any hub in a minimal network has three equally
spaced spokes.

\vspace{10pt}
\begin{mybox}
  \begin{exercise}
    \emph{Outer rim.} 
\label{ex:rim}
  \end{exercise}
  Our proof applies to hubs only, but similar arguments apply to the
  cities $A_1, A_2, \ldots, A_n$.  Prove the following:
  \begin{enumerate}[label=(\alph*), itemsep=0pt]
  \item No incoming edges can be separated by less than $120^\circ$.
  \item The number of incoming edges is between one and three.
  \end{enumerate}
\end{mybox}
\vspace{5pt}

\subsection{A minimal history}
\label{sec:historical-notes}

French mathematician
\textsc{Pierre de Fermat} 
(1607--1665) was the first to ask about minimal networks on the
triangle, though he framed it as a geometric problem:

\vspace{10pt}
\begin{mybox2}
  \begin{statement}
    \emph{Fermat's problem.}
  \end{statement}
Given three points $A, B, C$ in the plane, find the point $D$ such
  that the sum of lengths $|DA| + |DB| + |DC|$ is minimal.
\end{mybox2}
\vspace{5pt}

\noindent He figured out the answer himself, but according to the
mathematical custom of the day, sent a letter to Galileo's student
\textsc{Evangelista Torricelli} 
(1608--1647), challenging him to solve it.
Torricelli found the same answer, but using a different method,
so the position of the hub is called the \emph{Fermat-Torricelli
  point} in joint honor of its discoverers.
\textsc{Jakob Steiner}
(1796--1863) generalized Fermat's question to $n$
points on the plane:

\vspace{10pt}
\begin{mybox2}
  \begin{statement}
    \emph{Steiner's problem.} \label{box:steiner}
  \end{statement}
Given $n$ points $A_1, \cdots, A_n$ in the plane, find the point $D$
  such that the sum of lengths $|DA_1| + \cdots + |DA_n|$ is minimal.
\end{mybox2}
\vspace{5pt}

\noindent Although minimal networks are also called \emph{Steiner} trees, Steiner's
problem is very different from the $n$-city problem we've been considering. 
Steiner wanted a \emph{single} point such that the sum of lengths
to that point is minimal, rather than a connected network of minimal
length.
Put differently, it is the minimal network when you are allowed to add
at most one hub.

\vspace{10pt}
\begin{figure}[h]
  \centering
  \includegraphics[scale=0.42]{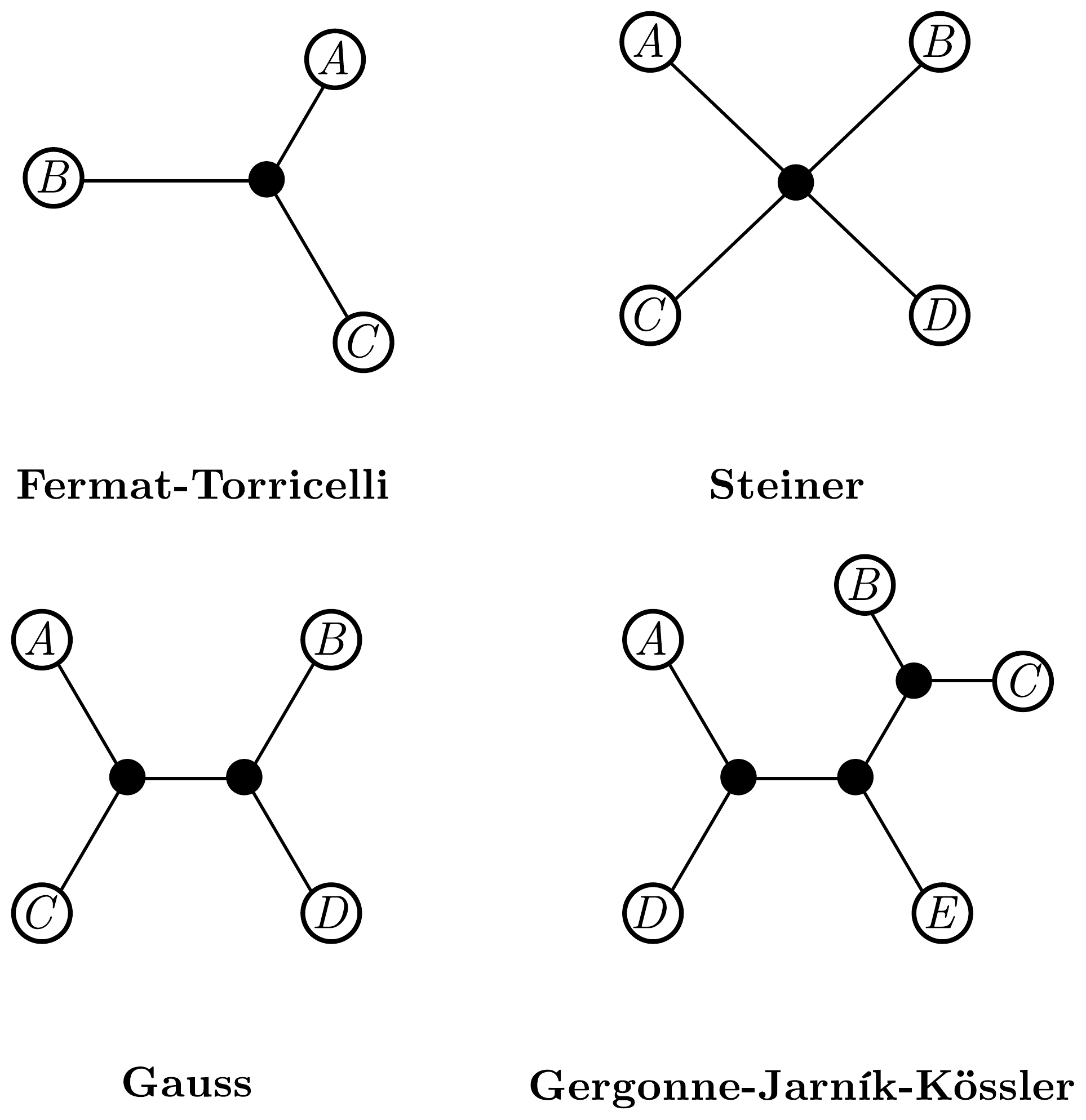}
\vspace{-5pt}
 \caption{A visual history of minimal networks.}
  \label{fig:history}
\end{figure}

In 1836, 200 years later, the great German mathematician
\textsc{Carl Friedrich Gauss}
(1777--1855) mulled on the design of a
minimal rail network between four German cities (Exercise \ref{ex:gauss}).
Around the same time, the French mathematician \textsc{Joseph Diez Gergonne} (1771--1859)
considered the general $n$ city problem (connecting them via
canals rather than railways) and discovered the $120^\circ$ rule.
The world
evidently paid
no attention until 1934, when Czech mathematicians
\textsc{Vojtěch Jarník} (1897--1970) and \textsc{Miloš Kössler}
(1884--1961) independently rediscovered Gergonne's results \cite{jk1934}.
The Gergonne-Jarník-Kössler version was popularized under the name
\emph{Steiner trees} by \textsc{Richard Courant} and \textsc{Herbert
  Robbins} in their classic 1941 text, \emph{What is Mathematics?}
\cite{courant1941mathematics}.
For a more in-depth history, see \cite{brazil2014}.

We finish this section by finding the Steiner point and Steiner tree
for regular polygons, a trigonometric construction of the Fermat-Torricelli point for the optimal hub placement (Exercise
\ref{ex:fermat}), and a proof that the $120^\circ$ rule does indeed
minimize total distance (Exercise \ref{ex:vector}).

\vspace{10pt}
\begin{mybox}
  \begin{exercise}
    \emph{Easy polygons.} 
  \end{exercise}
  Consider $n$ cities on the corners of a regular $n$-sided polygon. 
  \begin{enumerate}[label=(\alph*), itemsep=0pt] 
  \item Show that for $n \geq 6$, the network formed by removing a
    single edge from the perimeter satisfies the $120^\circ$ rule and
    requirement (a) from Exercise \ref{ex:rim}.
  It's harder to prove, but this is in fact the minimal
  network!\footnote{You might wonder why this doesn't follow
    immediately. As will explore in \S \ref{sec:graphs}, and
    particularly Exercise \ref{ex:rect}, it turns out that satisfying
    these rules does not guarantee a network is minimal.}
\item  Use the reasoning in \S \ref{sec:equil-triangl} to argue that
the center of the polygon solves
 Steiner's problem in Box \ref{box:steiner}.
\end{enumerate}
\end{mybox}

\vspace{10pt}
\begin{mybox}
  \begin{exercise}
    \emph{From straight lines to Steiner's problem.}  \VarIceMountain
    \label{ex:vector}
  \end{exercise}
Here, we give a rigorous proof of the $120^\circ$ rule, and
immediately extend it find the analogous rule for Steiner's problem.
The proof makes use of vectors and the dot product, hence the higher difficulty rating.
Recall that $|\mathbf{v}|$ is the length of the vector $\mathbf{v}$,
and $\hat{\mathbf{v}} = \mathbf{v}/|\mathbf{v}|$ is the unit vector
pointing in the same direction.
\begin{center}
  \includegraphics[scale=0.5]{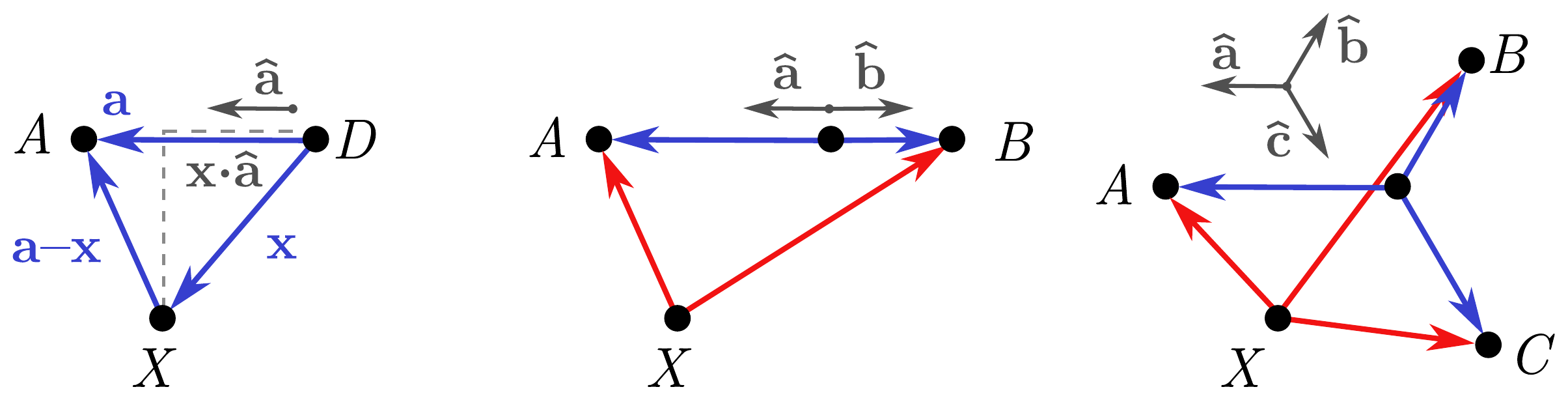}
\end{center}
Choose a point $D$ on the plane, which will act as the
  ``origin''.
Consider another point, $A$, making a vector $\mathbf{a} = DA$, with
unit vector $\hat{\mathbf{a}}$.
\begin{enumerate}[label=(\alph*), itemsep=0pt]
\item 
  Prove (visually or however you like) that for any other point $X$,
  with $\mathbf{x} = DX$,
  \[
    |\mathbf{a}| \leq |\mathbf{a}-\mathbf{x}| + \mathbf{x}\cdot \hat{\mathbf{a}},
  \]
  where as in the image above, $\mathbf{x}\cdot \hat{\mathbf{a}}$ is
  the length of $\mathbf{x}$ projected onto $\mathbf{a}$.
\item Consider two points $A$ and $B$ on the plane.
  Using the previous exercise, show that for any point $X$,
  \[
    |DA| + |DB| \leq |XA| + |XB| + \mathbf{x}\cdot (\hat{\mathbf{a}} + \hat{\mathbf{b}}).
  \]
\item Conclude that if we choose $D$ so that $\hat{\mathbf{a}} +
  \hat{\mathbf{b}} = \mathbf{0}$, the sum $|DA| + |DB|$ will be
  minimized.
  Geometrically, what does correspond to? Does this makes sense?
\item Let's now introduce \emph{three} points $A, B, C$ on the plane,
  with origin $D$.
  Generalize (c) to establish that $|DA|+|DB|+|DC|$ is
  minimized when
  \[
    \hat{\mathbf{a}} + \hat{\mathbf{b}} + \hat{\mathbf{c}} = \mathbf{0}.
  \]
  Show that this is precisely the $120^\circ$ rule.
    \item Finally, consider points $A_1,\ldots, A_n$ and corresponding
      vectors $\mathbf{a}_1, \ldots, \mathbf{a}_n$.
      Generalize (d) to conclude that if a point $D$ exists such that
      \[
        \hat{\mathbf{a}}_1 + \cdots + \hat{\mathbf{a}}_n = \mathbf{0},
      \]
      then it solves Steiner's problem (Box \ref{box:steiner}).
\item Exploit (e) to solve Steiner's problem for an arbitrary quadrilateral.
\end{enumerate}
\vspace{0pt}
\end{mybox}
\vspace{10pt}

\begin{mybox}
  \begin{exercise}
    \emph{Searching for Fermat-Torricelli.} \VarMountain
    \label{ex:fermat}
  \end{exercise}
Here, we give a geometric construction of the Fermat-Torricelli point
for any triangle.
Proceed if you like geometry!
So, we're going to find the interior point satisfying the $120^\circ$
rule for the blue triangle (below left).
Start by attaching equilateral triangles (green, red, yellow) on each
side, and drawing lines (dark blue) from the outer corners of the
equilateral triangles to the opposite corner of our original
triangle, as shown below right.
\vspace{-10pt}
\begin{center}
  \includegraphics[scale=0.4]{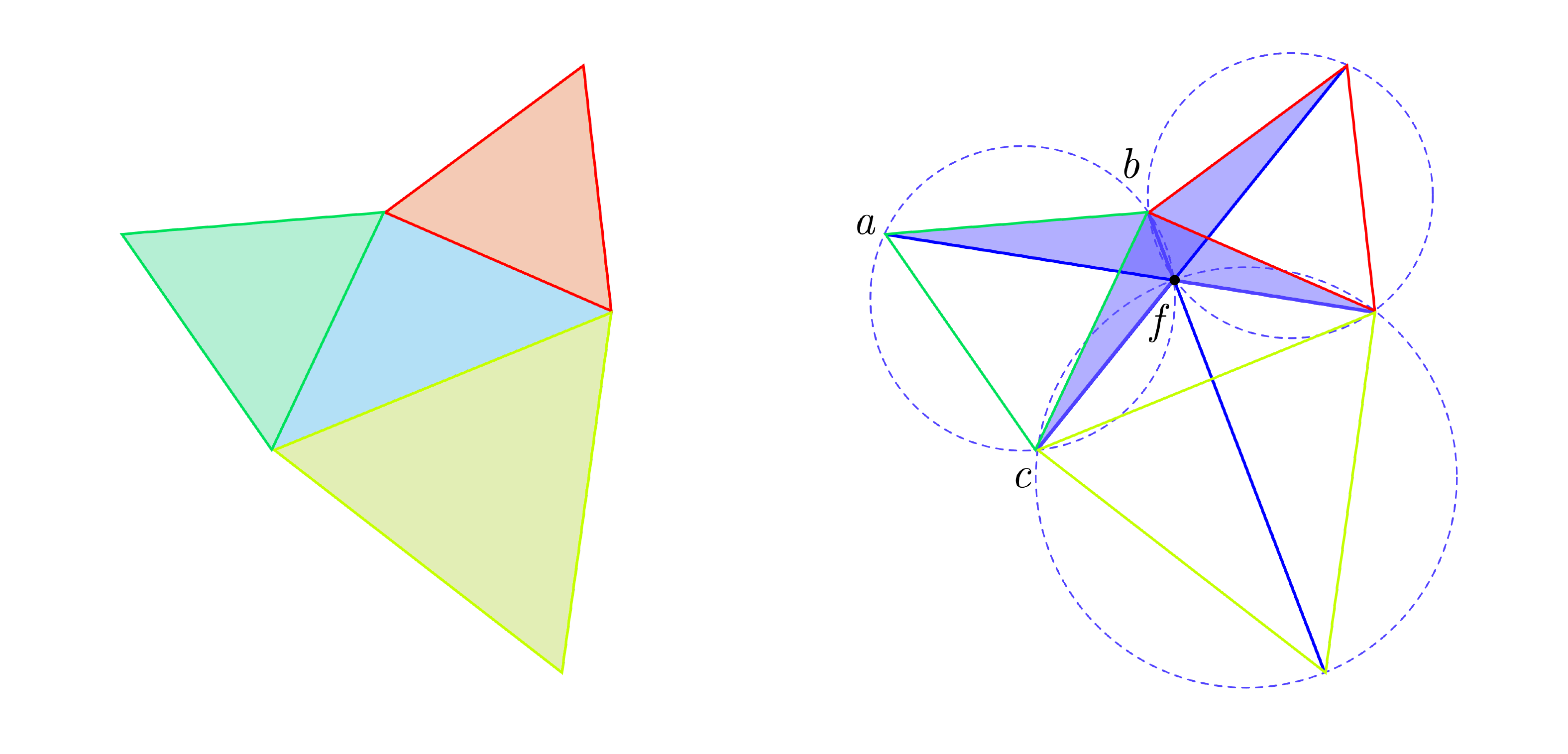}
\end{center}
\vspace{-10pt}

We claim these lines intersect at the point $f$, and moreover, are
separated by angles of $120^\circ$. 
To prove this, draw the dotted circles circumscribing each equilateral
triangle.  The exercises guide you through a demonstration that the
circles intersect at $120^\circ$ angles at $f$, using the
\href{https://en.wikipedia.org/wiki/Inscribed_angle}{inscribed angle
  theorem}. 

\begin{enumerate}[label=(\alph*), itemsep=0pt]
\item Show that the shaded triangles are congruent. Argue that, in
  consequence, the three blue lines do interesect at a single point.

\item From part (a), argue that $\angle baf = \angle bcf$.

\item From (b) and the inscribed angle theorem, argue that
  $a, b, c, f$ lie on a circle.

\item Since the triangle is equilateral, $\angle cab = 60^\circ$.
  Using the inscribed angle theorem once more, show that
  $\angle cfb = 120^\circ$.  Repeating this argument for the remaining
  two triangles gives our result!
\end{enumerate}
\vspace{-5pt}
This construction works provided all angles in the blue triangle are
$< 120^\circ$.  
\begin{enumerate}[label=(\alph*), itemsep=0pt] 
\item[(e)] What goes wrong if an angle is $\geq 120^\circ$?
\end{enumerate}
\vspace{0pt}
\end{mybox}

\newpage

\section{Graphs}
\label{sec:graphs}

In a sense, the $120^\circ$ rule solves the problem of minimal
networks, giving us a mathematical condition that hubs must obey.
But if I hand you a list of cities and tell you to start designing,
you will quickly see that the $120^\circ$ rule is not enough!
In this section, we will think more about the layout of
networks, including how many hubs we need to consider, the number of
network arrangements, the general difficulty of finding these networks
and methods for approximating them.

\begin{figure}[h]
  \centering
  \includegraphics[scale=0.42]{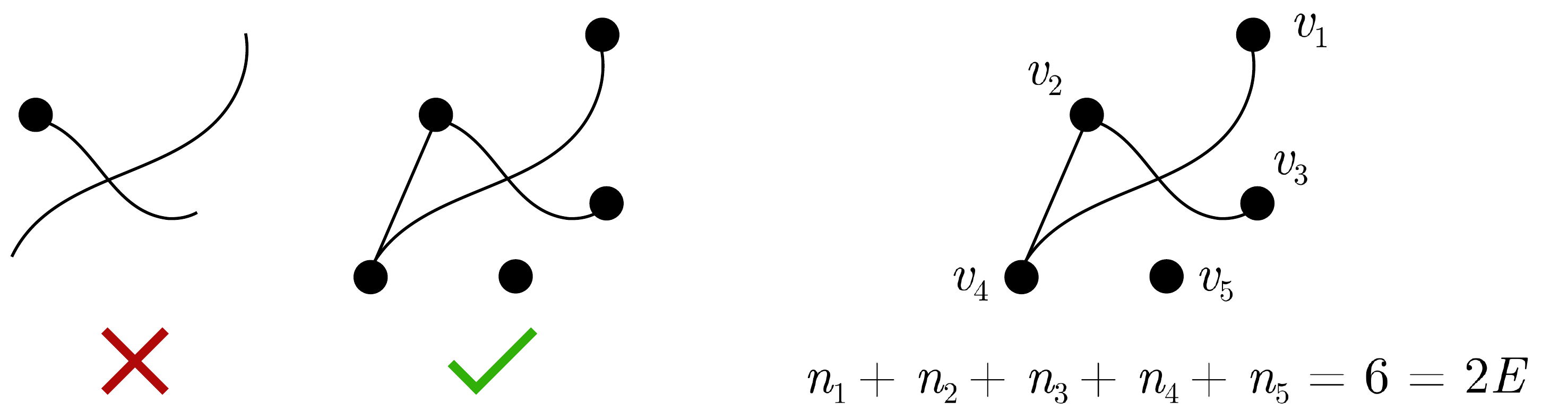}
\vspace{-10pt}
 \caption{\emph{Left.} ``Illegal'' and ``legal'' graphs. \emph{Right.}
   The handshake lemma in action.}
  \label{fig:graph}
\end{figure}

Studying network layouts is the domain of \emph{graph theory}.
A graph is a bunch of dots connected by lines, drawn on a page. 
The technical term for dots is \emph{vertices} or \emph{nodes},
and \emph{edges} for the lines.
If an edge joins two nodes, we say they are \emph{neighbours}.
Edges must start and end at different vertices, and are allowed to
overlap.
Vertices can be attached to any number of edges, including zero.
The rules are illustrated in Fig. \ref{fig:graph}.
We let $E$ denote the number of edges and $N$ the number of nodes.

\vspace{0pt}

We will need a simple, general result called the \emph{handshake
  lemma}.
There are two ways to count edges.
The first is simply to count the edges directly, yielding a number
$E$.
But our rules tell us that edges attach to a vertex at each end.
So instead, we can go through the vertices and count
the number of edges which attach to them.
This will hit each edge \emph{twice}, once for the vertex at either
end, so this way of counting gives $2E$.
That's the handshake lemma!
More precisely, suppose there are $N$ vertices $v_1, v_2, \ldots,
v_N$. If these have $n_1, n_2,
\ldots, n_N$ edges attached, the handshake lemma states that
\begin{equation}
  \label{eq:handshake}
  n_1 + n_2 + \cdots + n_N = 2E.
\end{equation}
The name, incidentally, comes from the fact that if vertices $v_1,
\ldots, v_N$ are people, and edges are handshakes, we add the number
of handshakes each person performs to get twice the total number of handshakes.

\subsection{Trees and leaves}
\label{sec:trees}

Some cities are not joined by rail, say Minsk and Darwin.
But in a \emph{connected} rail network, there is at least one route
between each pair of cities.
Anyone who has had the pleasure of exploring Tokyo's subway network will know
the dizzying extent to which more than one route from $A$ to
$B$ is possible.
But in a genuinely minimal train network, $A$ and $B$ will be joined
by a \emph{unique} route.

\begin{figure}[h]
  \centering
  \includegraphics[scale=0.52]{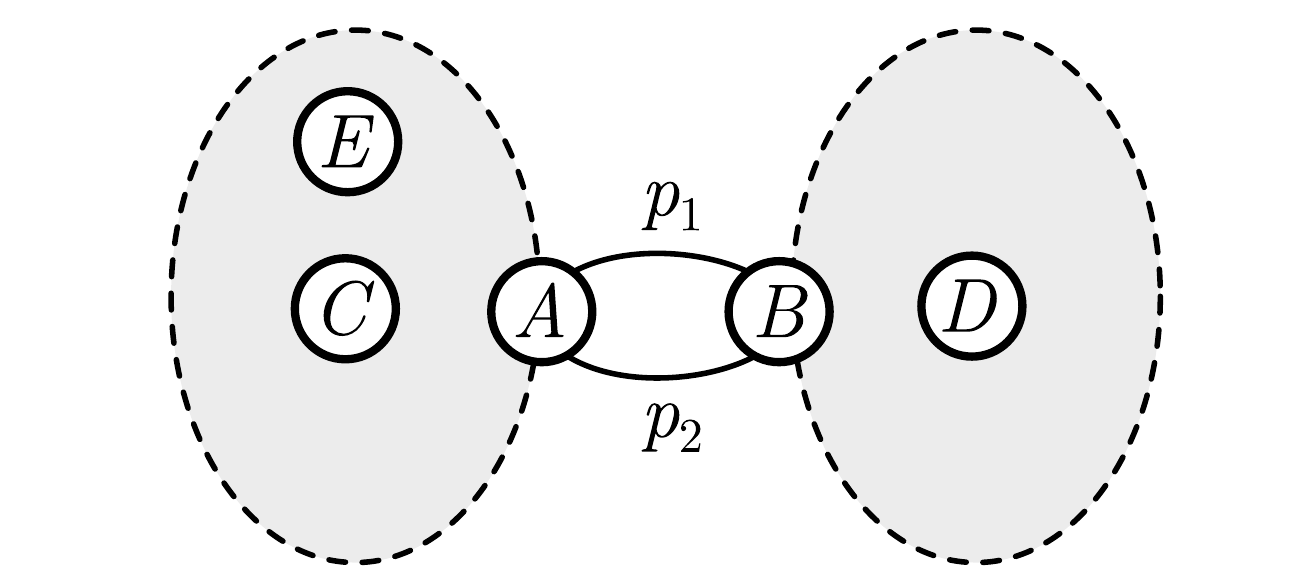}
\vspace{-10pt}
 \caption{Pruning unnecessary paths.}
  \label{fig:prune}
\end{figure}

The basic idea is to get rid of routes until one is left.
If there is more than one way to get from $A$ to $B$, the network has
unnecessary edges and can be ``pruned'' to get something shorter.
You might worry that pruning these unnecessary 
edges could accidentally disconnect other cities, but this is never the case!
Fig. \ref{fig:prune} shows why.
Suppose $A$ and $B$ are connected by two paths, labelled
$p_1$ and $p_2$, and potentially consisting of more than one edge.
The blob to the left is all the vertices whose paths to $B$ go
through $A$ first, and similarly, vertices on the right connect to $A$
through $B$.
If two nodes are in the same blob, such as $C$ and $E$, then pruning
path $p_2$ has no effect on whether they are connected.
If two nodes are in different blobs, like $C$ and $D$, they can still reach
each other using path $p_1$.
We can prune the redundant paths willy nilly!

\vspace{-0pt}

\vspace{10pt}
\begin{mybox}
  \begin{exercise}
    \emph{Pruning along the path.}
  \end{exercise}
Generalize the argument above to account
for nodes that lie between $A$ and $B$.
(These are nodes which can connect to either $A$ or $B$
without passing through the other, and schematically lie on paths $p_1$
or $p_2$.)
\end{mybox}
\vspace{5pt}

\begin{figure}[h]
  \centering
  \includegraphics[scale=0.5]{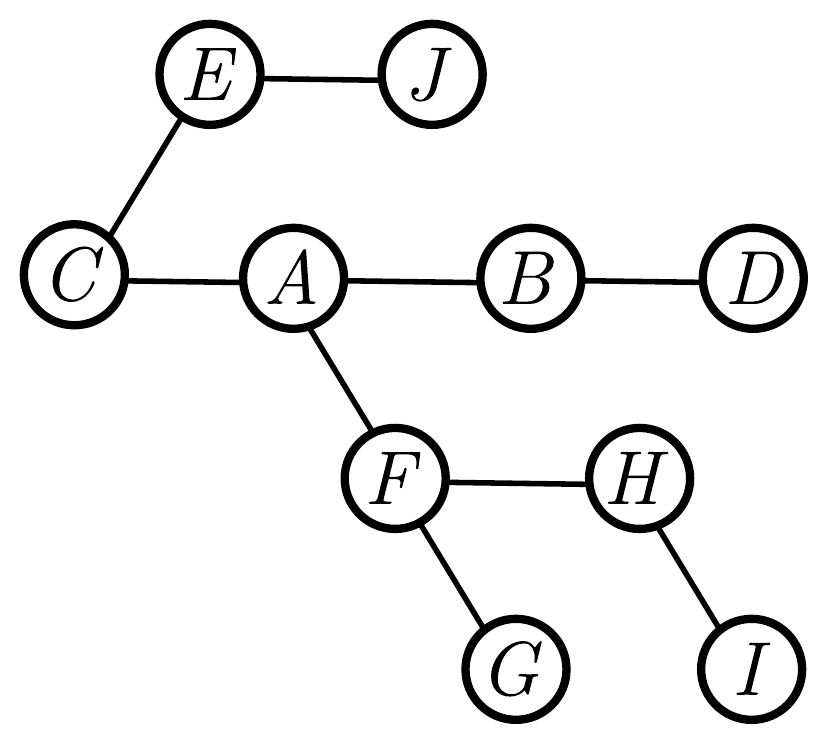}
 \caption{A tree network, with a unique path between each node.}
  \label{fig:tree1}
\end{figure}

Once we have completely pruned the network, there is only a single
path connecting any two nodes $A$ and $B$.
Such a network is called a \emph{tree} because it can be drawn so that
edges look like branches.
This finally explains why minimal networks are also called Steiner \emph{trees}!
An example of a tree is shown in Fig. \ref{fig:tree1}.
Every tree has a special node called a \emph{leaf}. 
As the name suggests, this is at the ``end'' of the tree's branches.
More formally, a leaf is a node with a single edge, like $J$, $D$,
$G$, and $I$ in Fig. \ref{fig:tree1}.
It may seem intuitive, but as an exercise in reasoning about trees,
let's \emph{prove} they must have leaves.

\vspace{10pt}

\vspace{10pt}
\begin{mybox}
  \begin{exercise}
    \emph{Finding leaves.} \VarMountain
  \end{exercise}
  To begin our proof that each tree has a leaf, we choose a node at
  random (red, below left) and count the number of steps to each other node.

  \begin{center}
    \includegraphics[scale=0.5]{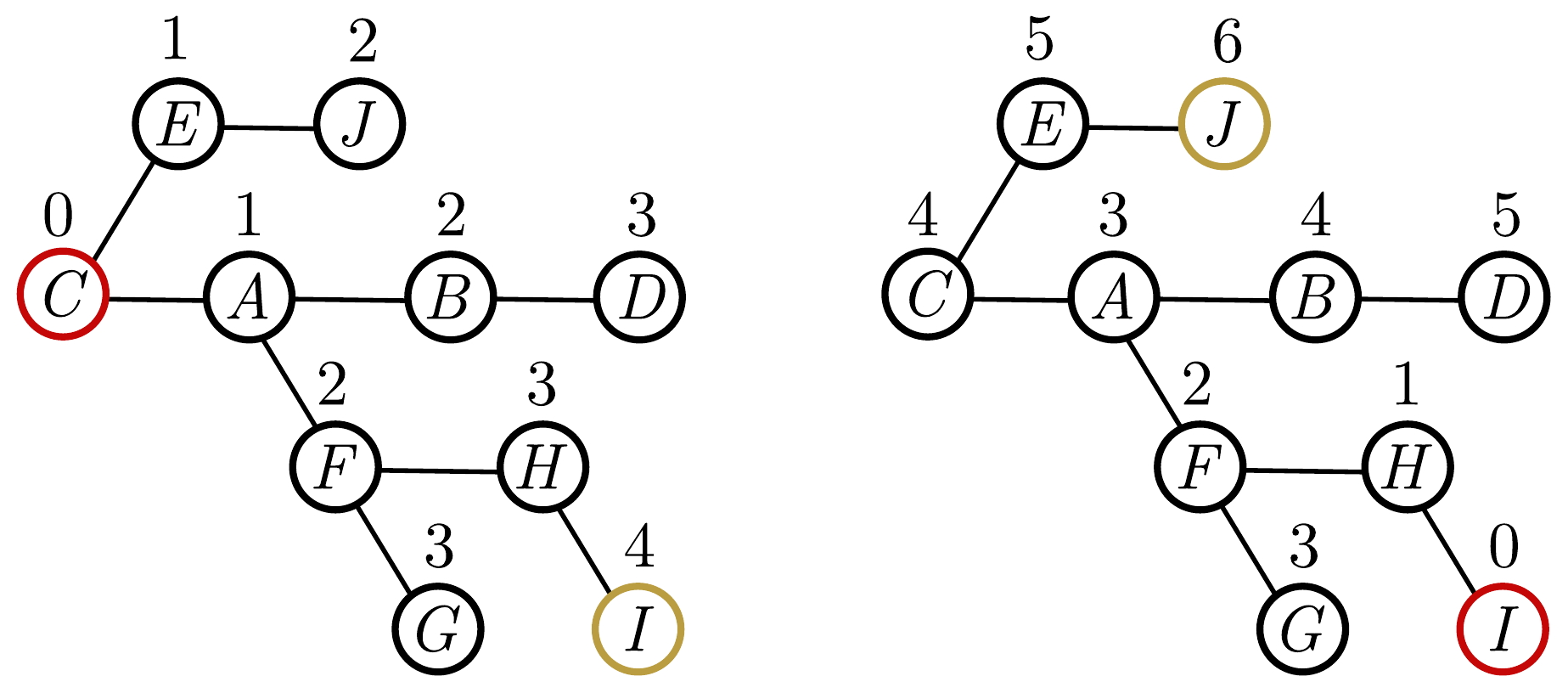}
  \end{center}
  \begin{enumerate}[label=(\alph*), itemsep=0pt]
  \item Explain why the number of steps from the red node to any other
    node is well-defined in a tree.
  \item Consider the node or nodes furthest from the red node
    (orange, above left). Argue that these must be leaves.
    \emph{Hint.} If they are not, what is the distance from red node to their neighbours?
  \item In fact, we can prove something stronger. 
    The previous question tells us how to find a leaf. Repeat the same
    procedure, but start with the leaf and find the furthest node.
    Conclude that every tree (with at least two nodes) has \emph{two} leaves.
  \item Show, using an example, that a tree need not have more than two leaves.
  \end{enumerate}
\vspace{-5pt}
\end{mybox}
\vspace{5pt}

\subsection{Hub caps}
\label{sec:max-hubs}

In Fig. \ref{fig:tree1}, you may have noticed the number of edges $E =
9$ is one less than the number of nodes, $N = 10$.
This is not a coincidence. 
For any tree, it turns out that $E =
N - 1$.
We can prove this fact using the existence of leaves.
The idea is simple: keep removing the leaf, and the single edge joining it
to the rest of the tree, until you have a single node left.
This requires the removal of $N-1$ nodes, and hence $N - 1$ edges.
Since there are \emph{no} edges now, and we removed one each time, we
must have started with $N-1$ edges.
Hence,
\begin{equation}
E = N - 1\label{eq:tree}
\end{equation}
for trees in general.
Equation (\ref{eq:tree}), along with the handshake lemma
(\ref{eq:handshake}), will allow us to place a cap on the maximum
number of hubs that can occur in the network.

Suppose we are trying to connect $n$ cities, and introduce $h$ hubs in
order to do so.
The total number of nodes is then $N = n + h$.
The $120^\circ$ rule tells that each hub attaches to exactly three edges.
Each of the $n$ cities attaches to at least one edge to ensure it is
connected to the rest of the network.
Thus, (\ref{eq:handshake}) gives
\begin{equation}
2E = n_1 + n_2 + \cdots + n_N \geq n + 3h.\label{eq:2E}
\end{equation}
From (\ref{eq:tree}), we know that $E = N -1  = n + h - 1$.
Combining this with (\ref{eq:2E}), we find 
\begin{align}
2(n + h - 1) = 2n + 2h - 2 & \geq n + 3h \quad      \Longrightarrow \quad n  - 2 \geq h.
\end{align}
In other words, the number of hubs $h$ is at most $n - 2$.

\vspace{10pt}
\begin{mybox}
  \begin{exercise}
    \emph{Hubs and nubs.}
  \end{exercise}
  While hubs always have three attached edges, Exercise \ref{ex:rim}
  tells us that cities (fixed nodes) have between one and three edges.
  \begin{enumerate}[label=(\alph*), itemsep=0pt]
 \item Show it is always possible to arrange $n$ cities so that $h =
   0$.
  \item At the other end of the spectrum, argue that the maximum number of hubs, $h =
    n - 2$, occurs when the fixed nodes are exactly the leaves of the network.
  \end{enumerate}
\vspace{0pt}
\end{mybox}
\vspace{5pt}

\begin{figure}[h]
  \centering
  \includegraphics[scale=0.2]{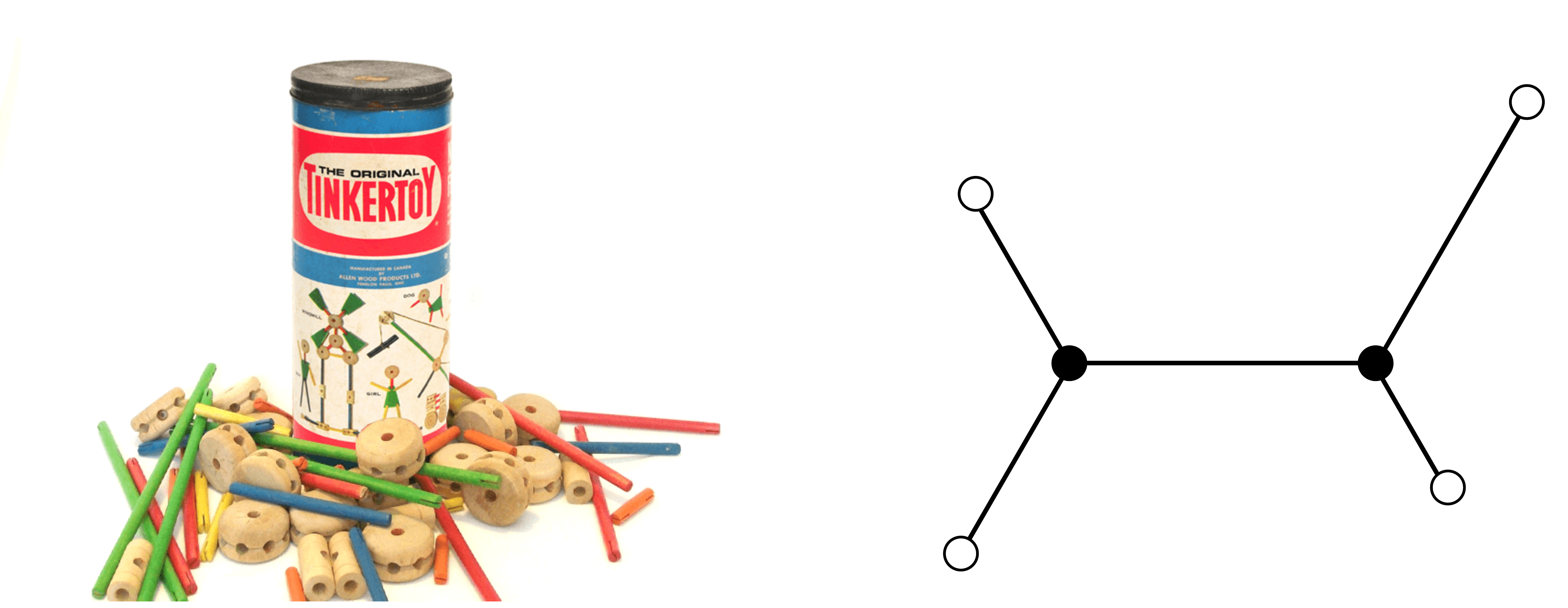}
 \caption{\emph{Left.} A real Tinkertoy\textsuperscript{TM}. \emph{Right.} A
   network tinkertoy.}
  \label{fig:tinkertoy}
\end{figure}

The $120^\circ$ rule and hub cap together give us a simple tool for
building minimal networks.
For $n$ fixed nodes, pick $h = n-2$ hubs, with spokes emerging at
angles of $120^\circ$, and connect them together to form a tree, with
the fixed nodes as leaves.
Although the angles are fixed, we can extend the spokes and legs, and
perform overall rotations of the network.
We call this extendable configuration of hubs and spokes a \emph{tinkertoy},
after the modular children's toy it vaguely resembles
(Fig. \ref{fig:tinkertoy}).\footnote{In the mathematics literature,
  a tinkertoy graph is related to what are called \emph{Steiner
    topologies}.
  They are slightly different, however, since the Steiner topologies
  are graphs which care about how they connect to fixed
  nodes.
}

\vspace{-5pt}

\noindent We can play with our network tinkertoys, or program a
computer to play with them, until they do what we want.
We give some examples in the following exercises.

\vspace{10pt}
\begin{mybox}
  \begin{exercise}
    \label{ex:rect}
    \emph{Minimal rectangular network.}
  \end{exercise}
Consider four cities on a rectangle of height $h$ and width $w \geq h$:
  \begin{center}
      \includegraphics[scale=0.22]{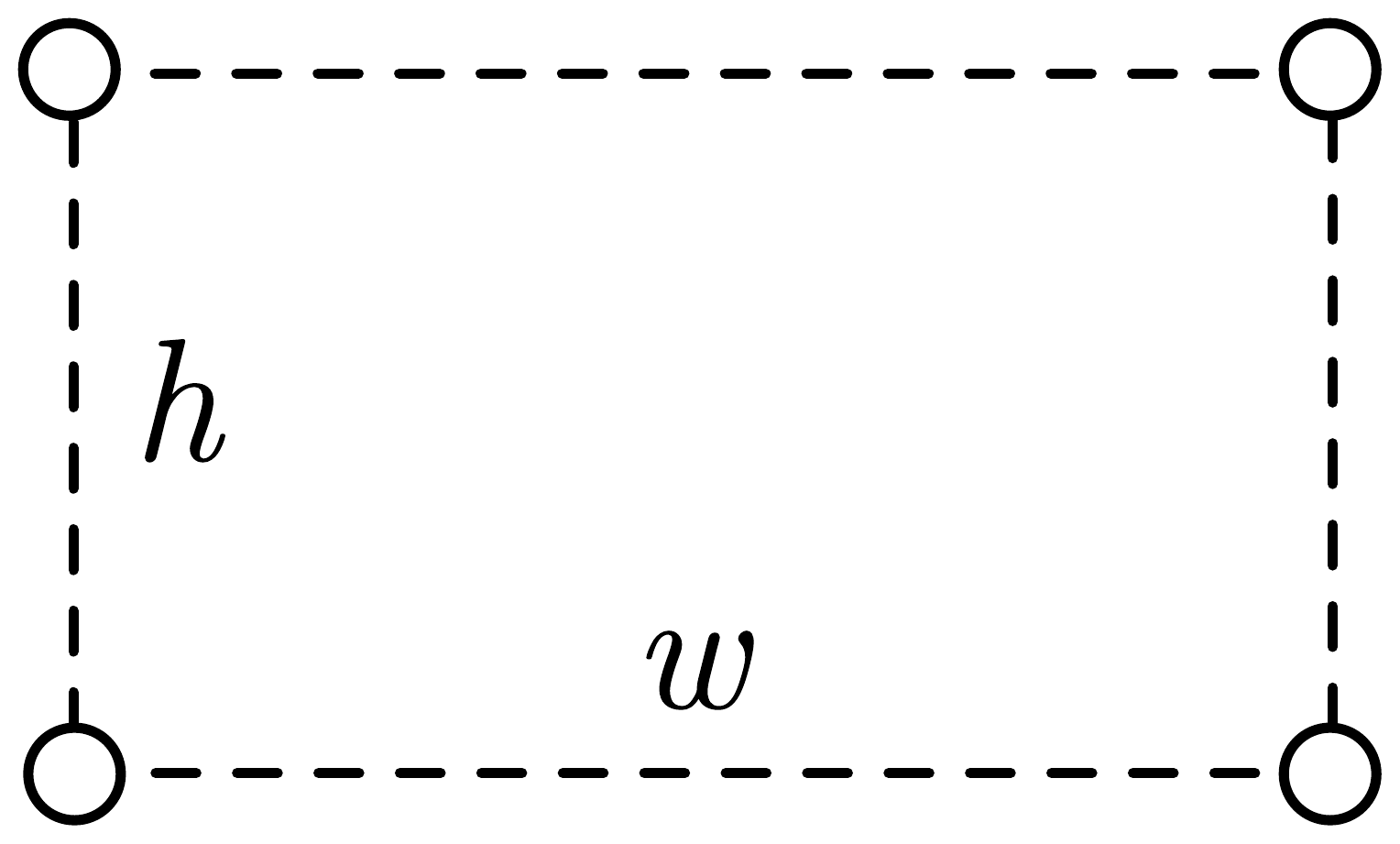}
  \end{center}
\vspace{-15pt}
  \begin{enumerate}[label=(\alph*), itemsep=0pt]
  \item Draw the single tinkertoy for $n = 4$, and argue from Exercise
    \ref{ex:rim} that this should describe the minimal network.
  \item Fit the tinkertoy to the city, and deduce that the minimal
    network has length
  $$ 
  L = w + \sqrt{3}h.
  $$
\item Show that the tinkertoy can be oriented in \emph{two} ways when
  $h < w < \sqrt{3}h$. Explain why the horizontal orientation is
  always minimal.
  \begin{center}
      \includegraphics[scale=0.21]{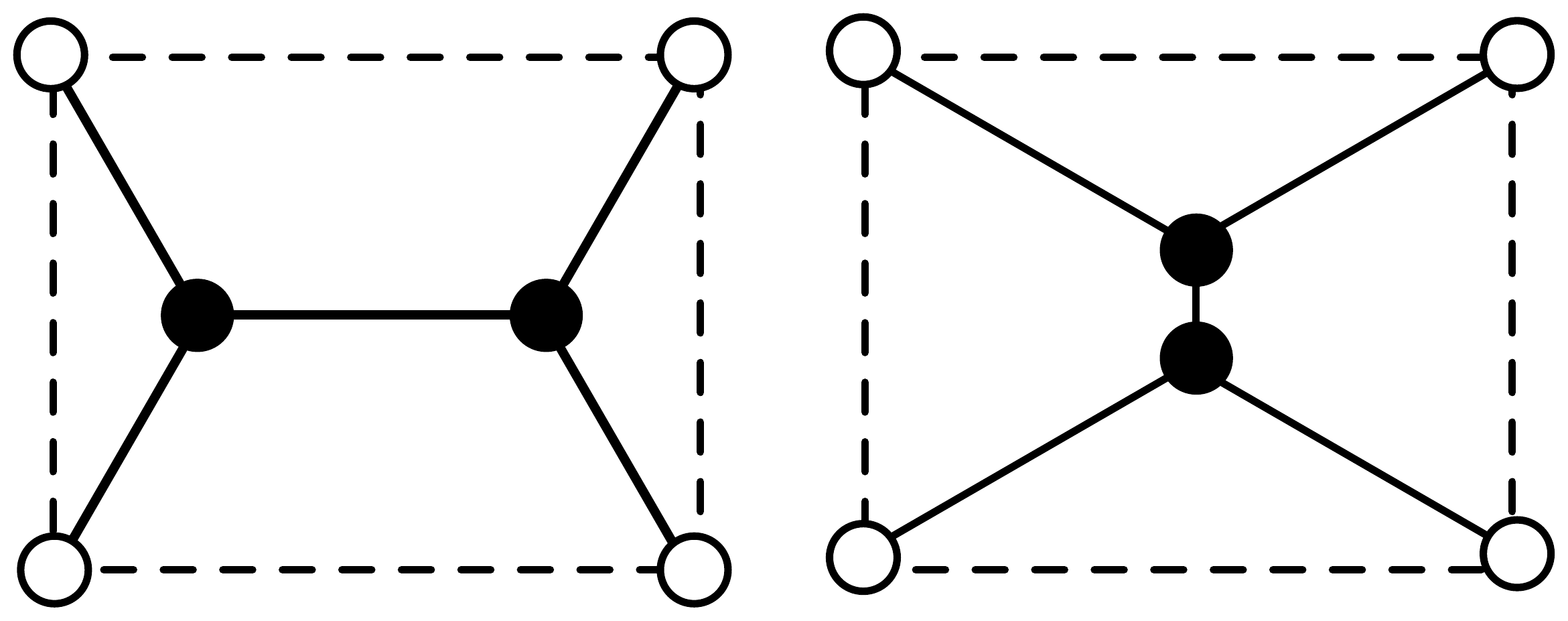}
\vspace{-10pt}
  \end{center}
  \end{enumerate}
Part (c) tells us something very
important.
Even if a tinkertoy fits, the configuration isn't necessarily the true
minimum!
Put different, the $120^\circ$ rule is not sufficient to guarantee
that a network is minimal.
\vspace{0pt}
\end{mybox}

\vspace{10pt}

\begin{mybox}
  \begin{exercise}
    \emph{Harder polygons.} \label{ex:hard}
  \end{exercise}
Fit a tinkertoy (or three) to the following shapes; no need for exact placement.
These networks are minimal, though it take a bit more work
to show.
  \vspace{-5pt}
  \begin{center}
      \includegraphics[scale=0.23]{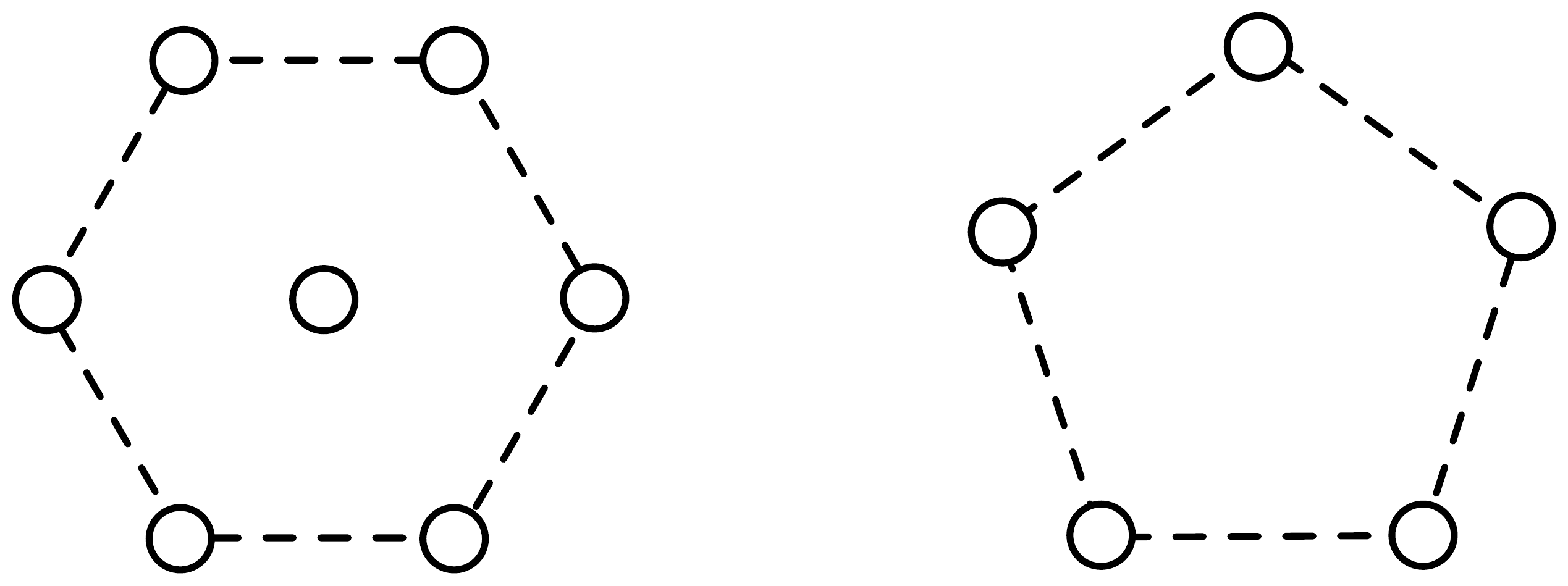}
  \end{center}
  \vspace{-5pt}
\vspace{0pt}
\end{mybox}

\subsection{Avoiding explosions}
\label{sec:tinkertoys}

\begin{figure}[h]
  \centering
  \includegraphics[scale=0.21]{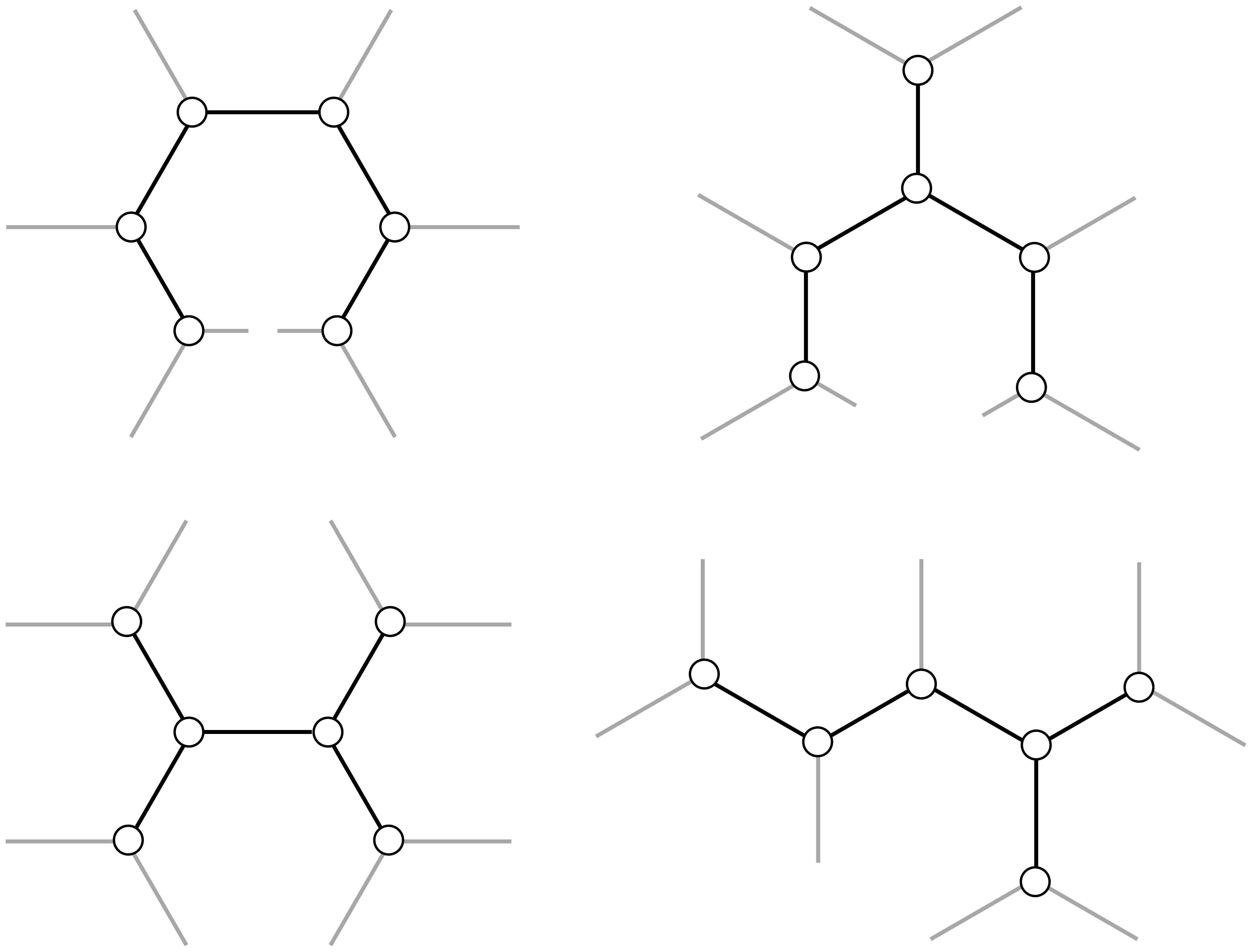}
\vspace{-5pt}
 \caption{A selection of tinkertoys for $h = 6$.}
  \label{fig:6-tinks}
\end{figure}

For a small number of hubs, tinkertoys are useful.
But are they useful for many hubs?
Suppose that fiddling with tinkertoys is a quick operation, and once a
tinkertoy is selected, a human or a computer can quickly check whether
the tinkertoy can be extruded to hit our fixed points.
If there are many tinkertoys, finding one that fits could still take a
while. 
In Fig. \ref{fig:6-tinks}, we show a few tinkertoys for $h = 6$,
suggesting that with more hubs, enumerating them all may turn out to
be hard.
In fact, as $h$ gets larger, the total number of tinkertoys $T_h$
suffers what is called a \emph{combinatorial explosion}, growing exponentially as a
function of $h$.
A brute force approach, which simply fiddles with each tinkertoy to
see if it can be made to fit the fixed points, will take an exponential
amount of time.
This is beginning to seem like a hard problem in
general!\footnote{There is a subtlety here. If most tinkertoys can be
  made to fit, then this brute force approach will run quickly! At
  least, it runs quickly if fiddling is a quick
  operation. In reality, the best fiddling algorithms are exponential
  in $n$, so the brute force approach remains exponential, irrespective of how
  many tinkertoys fit.
  While I'm not sure how many fit in general, for the purposes of our
  heuristic approach, we'll continue to assume the list is small.}

\begin{figure}[h]
  \centering
  \includegraphics[scale=0.21]{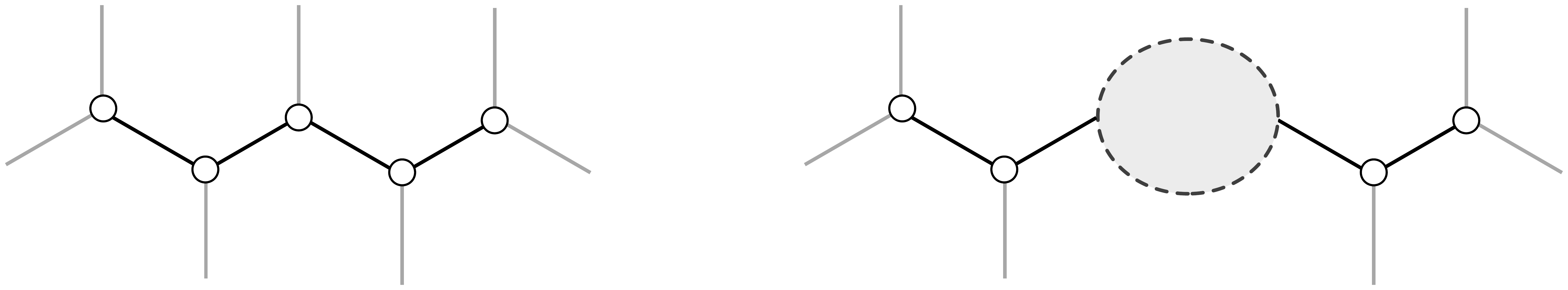}
 \caption{\emph{Left.} A linear tinkertoy. \emph{Right.} An oriented tinkertoy.}
  \label{fig:lin1}
\end{figure}

Counting the total number of tinkertoys is difficult.
To demonstrate this exponential growth, we are instead going to focus on a
subset of tinkertoys we can conveniently enumerate.
Trees in general have a complicated structure, so to simplify, we
consider only \emph{linear} tinkertoys.
These are tinkertoys where the hubs lie on a ``line'', so that no hub
has more than two neighbours, for instance  Fig. \ref{fig:lin1} (left). 
The next problem is that even these linear tinkertoys can be rotated
by $180^\circ$.
To avoid counting the same tinkertoy twice, we need some way of
knowing which end is which.
A simple method is to start and end with a $\vee$-shaped segment, as in
Fig. \ref{fig:lin1} (right).
If we rotate $180^\circ$, the tinkertoy is bookended by $\wedge$-shaped
segments, which is clearly
distinct.
Not every linear tinkertoy has this form, so we call these special
tinkertoys \emph{oriented}.

\begin{figure}[h]
  \centering
  \includegraphics[scale=0.21]{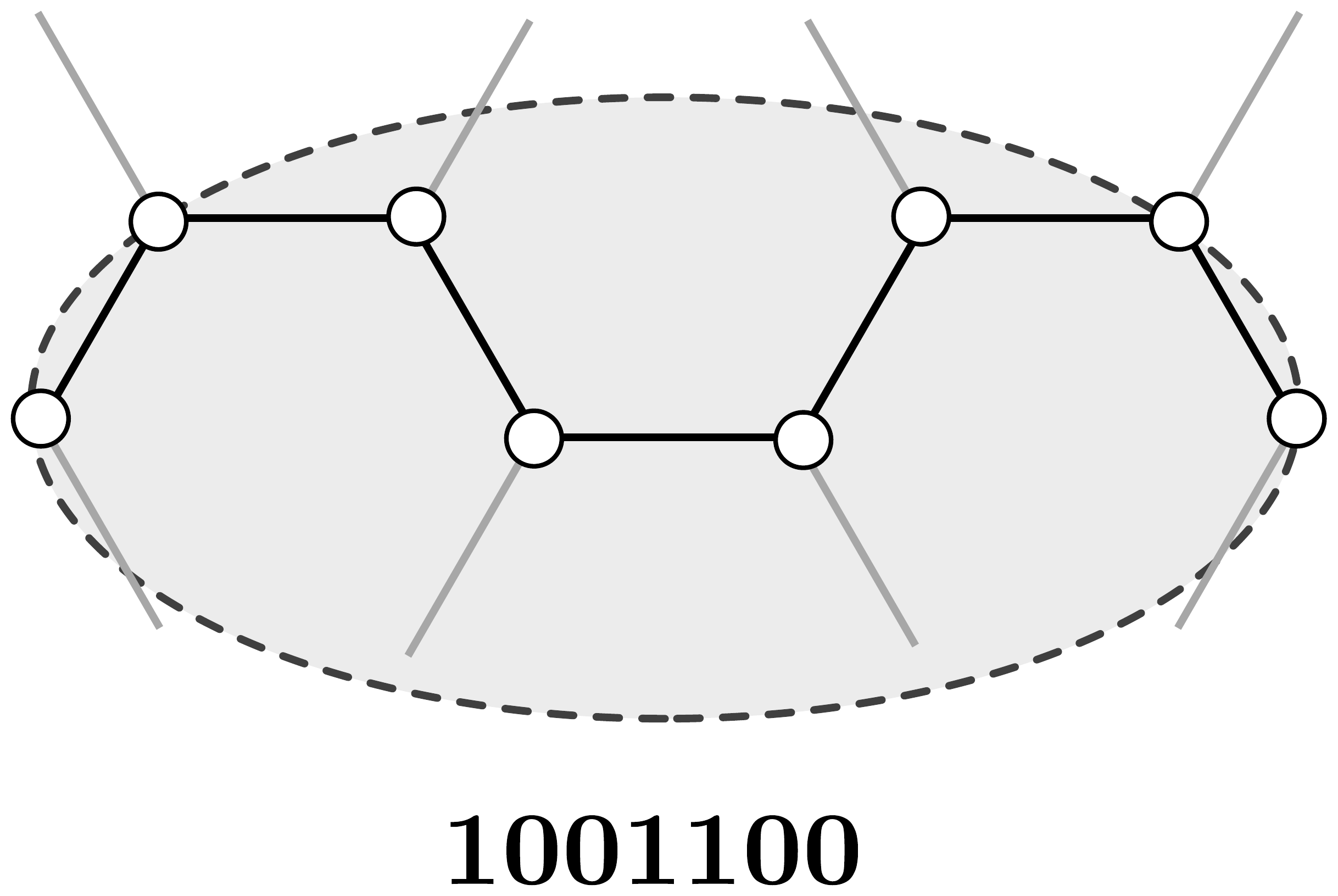}
 \caption{The binary sequence for an oriented tinkertoy.}
  \label{fig:lin2}
\end{figure}

With the notion of oriented tinkertoys, we can immediately find an
exponentially growing set!
There are $h - 1$ edges altogether since the hubs form a tree.
We fix four (two at each end) to ensure the tinkertoy is oriented.
That leaves $h - 5$ edges within the grey circle of Fig. \ref{fig:lin1} (right).
As we move along from the leftmost $\vee$, these edges constitute $h -
5$ turns left or right by $60^\circ$ before we exit again to hit the
final $\vee$.
At each point, either a left or a right turn is allowed, so there are
$2^{h-5}$ possible choices altogether.
To make this more transparent, we could label left and right turns
with $1$s and $0$s respectively, so that a tinkertoy is just a
sequence of binary digits, as in Fig. \ref{fig:lin2}.
Thus, there are an exponential number of oriented
tinkertoys.\footnote{You might worry
  that if we turn too many times, the tinkertoy will collide with
  itself and no longer be valid.
  For instance, after six right turns, edges of equal length will form
  a closed hexagon!
But we can always adjust the length of edges to prevent this from
happening, so the count remains valid.}
If you like drawing graphs, you can have a go at findiing the
\emph{total}
number $T_h$ in the next exercise.

\vspace{5pt}

\vspace{10pt}

\vspace{5pt}

\begin{mybox}
  \begin{exercise}
    \emph{Physicist's induction.} \IceMountain
\label{ex:phys-ind}
  \end{exercise}
Calculate the number of tinkertoys $T_h$ from $h = 0$ to $h = 6$. You
    should be able to find the general sequence $T_h$ by searching for
    these numbers in the \href{https://oeis.org/}{Online Encyclopedia
      of Integer Sequences}. At large $h$, the OEIS informs us that
    this sequence grows exponentially, with
$$
T_h \approx \frac{2^{2h-4}}{\sqrt{\pi}h^{5/2}}.
$$
If we count how the tinkertoys connect to the fixed nodes (``Steiner topologies''), there
are dramatically more arrangements: $\bar{T}_h = (2h)!/2^h h!$ to be
  precise!\footnote{This can be proved using \emph{mathematical
  induction}, rather than the physicist's induction we've used
here. We leave this as
a bonus exercise to the mathematically inducted (ahem).} 
\vspace{5pt}
\end{mybox}
\vspace{5pt}

By now, we should be confident that there are many tinkertoys.
If we have to consider even a fraction of them at large $h$, any
computer is doomed to failure.
For instance, using the counting in Exercise \ref{ex:phys-ind},
suppose a computer can check a billion tinkertoys per second, and
wants to design a railway network to connect the $\sim 800$ largest
cities in North America. 
If it has to check every tinkertoy, it will take an unimaginably long
\[
\frac{2^{2\cdot 800-4}}{\sqrt{\pi}800^{5/2}\cdot 10^9} \text{ s}
\approx 10^{456} \text{ years}.
\]
Would a faster computer help?
Not likely.
If you do more operations per second than there are atoms in the
universe, it still takes $\sim 10^{388}$ years!
No realistic improvements in processing speed will make this
problem solvable, 
unless we find a
\emph{much much} better algorithm.
As we'll discuss below, most computer scientists think no such
algorithm exists, but can't prove it!

Notice that there are two slightly distinct problems here.
The first is searching for tinkertoys that fit; and the second is
singling out the truly minimal network from the shortlist of fitting
tinkertoys. 
The two are not the same because, as we saw in Exercise \ref{ex:rect},
just because a tinkertoy fits doesn't mean it is minimal.
The first problem is easier because if somebody hands you a tinkertoy
and claims it fits, you can easily check.
In fact, you yourself could make a lucky guess and find a tinkertoy
which fits immediately.
There is an area of computer science called \emph{complexity theory}
which classifies problems according to how hard they are.
In the language of complexity theory, finding tinkertoys that fit is
called \textsf{NP}, for ``\textsf{N}ondeterministic \textsf{P}olynomial time''.
This is a fancy way of saying you can make a lucky guess and confirm
it immediately.

In fact, fitting tinkertoys is as hard as any problem in the
set \textsf{NP}.
``As hard as'' is a technical term in complexity theory,
meaning that you can transform any algorithm for finding good
tinkertoys into an algorithm for solving \emph{any other problem} in \textsf{NP}!
It is a key that unlocks the rest of the set.
We call such a task \textsf{NP-complete}, since it gives us ``complete''
access to every \textsf{NP} problem.
Now, if someone hands you a tinkertoy configuration and claims that it's
the minimal network, you must first check that it fits.
So finding a minimal network is at least as hard as fitting a tinkertoy.
But you can't stop there!
You have to keep searching to find \emph{all} the tinkertoys that fit,
checking the lengths, and verifying that the first configuration
really is the shortest.
The second problem is therefore \emph{at least} as hard as fitting
tinkertoys. This places it in a
class called \textsf{NP-hard} \cite{SteinerNP}, which is
complexity-ese for ``as \textsf{hard} as any problem in
\textsf{NP''}.\footnote{Note that if we can quickly fit tinkertoys, we
  can quickly find the minimal network. 
  So while finding minimal networks is hard, it's only marginally
  harder than \textsf{NP-complete}, and if \textsf{P} $=$ \textsf{NP},
  finding minimal networks is also in \textsf{P}!
  I thank Scott Aaronson for pointing this out.
}

\vspace{10pt}
\begin{mybox}
  \begin{exercise}
    \emph{Tiny tinkertoys.} \label{ex:tiny}
  \end{exercise}
We've been talking about fitting a \emph{single} tinkertoy, but as we
saw in Exercise \ref{ex:hard}, the minimal network is sometimes
obtained by cobbling together multiple ``tiny'' tinkertoys. 
Argue that including tiny tinkertoys makes finding minimal networks
harder, but the problem of fitting potentially easier.
\vspace{0pt}
\end{mybox}

\vspace{10pt}
\begin{mybox2}
  \begin{statement}
    \emph{Complexity I.}
  \end{statement}
Fitting tinkertoys is \textsf{NP-complete}.
Finding minimal
  networks is \textsf{NP-hard}.
\end{mybox2}
\vspace{-5pt}

\subsection{Minimum spanning trees}
\label{sec:minim-spann-trees}

All these heavy-sounding results about complexity theory make life
sound impossible for network planners.
But while finding the exact minimal network is
difficult, approximating is easy!
Life, and near-optimal rail travel, go on.
We'll discuss two simple approximation schemes, starting with a
generalization of the very first Exercise \ref{ex:sides}.
Recall that, for three cities, the triangle network consists of the
two shortest sides of the triangle.
Put differently, we draw an edge between each city, and select the two
shortest ones, which happen to form a tree which connects everything.

\begin{figure}[h]
  \centering
  \includegraphics[scale=0.42]{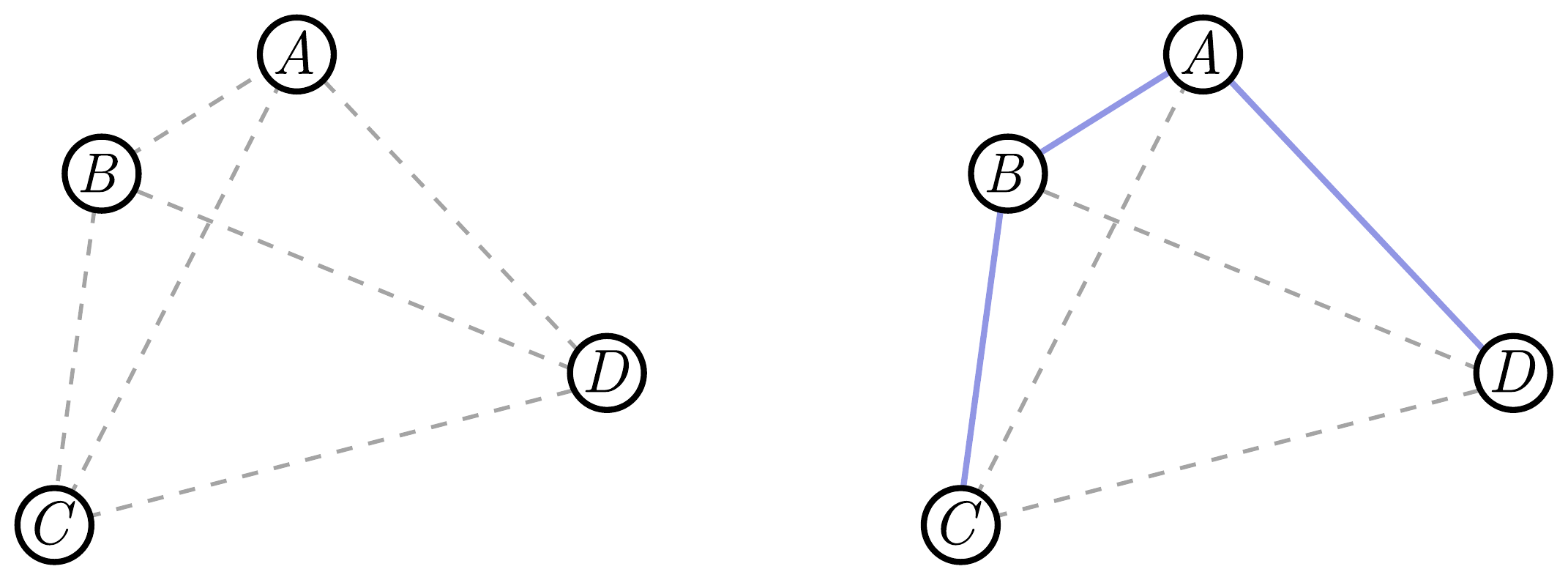}
\vspace{-5pt}
 \caption{\emph{Left.} The complete graph for four
   cities. \emph{Right.} The minimum spanning tree.}
  \label{fig:mst1}
\end{figure}

For $n$ cities, we do the same thing.
Draw an edge between each city, forming what is
called the \emph{complete graph} on $n$ nodes.
From these edges, we select a subset which form a tree, connecting
each city and of minimum total length. This is called a \emph{minimum
spanning tree (MST)}, since it ``spans'' the cities.
We illustrate the construction for $n = 4$ in Fig. \ref{fig:mst1}.

\vspace{5pt}

\begin{figure}[h]
  \centering
  \includegraphics[scale=0.42]{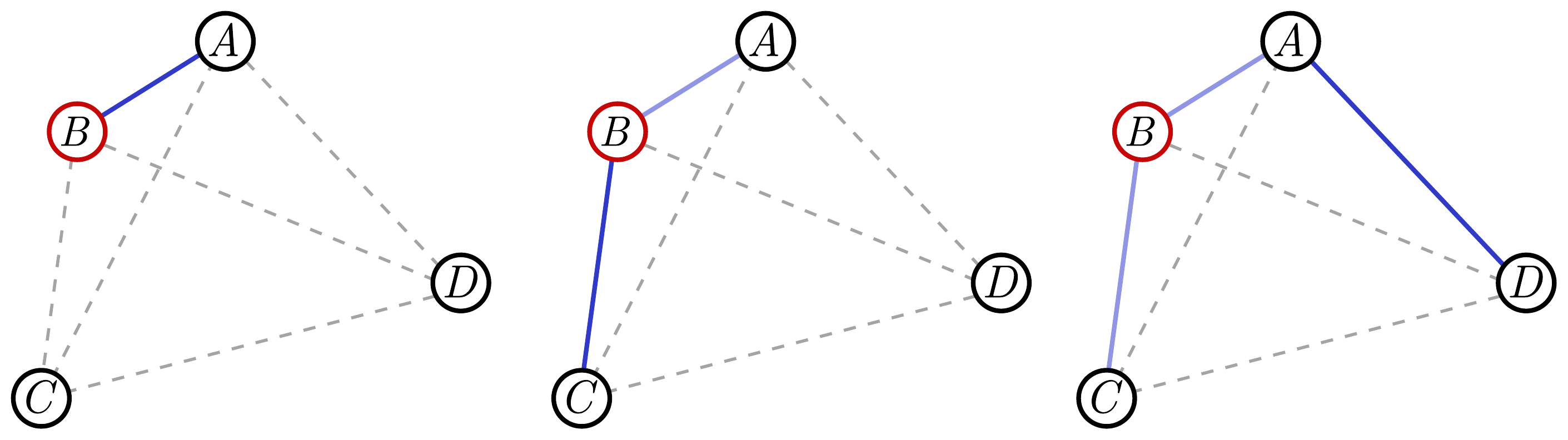}
\vspace{-5pt}
 \caption{The Prim-Jarník algorithm for $n=4$. Dark blue edges are
   added sequentially.}
  \label{fig:mst2}
\end{figure}

The usefulness of MSTs depends on whether they are fast to compute and
close to optimal.
We start with the first question.
Unlike tinkertoys, there is a procedure to construct the MST edge by edge.
This procedure is very simple: 
\begin{enumerate}[itemsep=-1pt]
\item[0.] Pick a random vertex $v_0$.
\item[1.] Add the shortest edge adjacent to\footnote{By ``adjacent to'', we just mean an edge which touches the tree but is
not already in it.} $v_0$ to form a tree $T_1$.
\item[$k\geq 2$.] Add the shortest edge adjacent to $T_k$ to form a tree $T_{k+1}$.
\end{enumerate}
Repeat the last step until we have a tree $T_{n-1}$ which spans all the nodes.
This algorithm was discovered in 1930 by Jarník \cite{jarnikmst}, but
subsequently rediscovered by \textsc{Robert Prim} in 1957 \cite{Prim1957}, so
it is called the \emph{Prim-Jarník algorithm}.
We implement it for $n = 4$ in Fig. \ref{fig:mst2}.

\begin{mybox}
  \begin{exercise}
    \emph{MST is easy.}
  \end{exercise}
  Here, we will give a very lazy bound on the number of steps
  required to perform the Prim-Jarník algorithm.
  \begin{enumerate}[label=(\alph*), itemsep=0pt]
  \item Using the handshake lemma (\ref{eq:handshake}), show the total
  number of edges in the complete graph on $n$ cities is $E_\text{complete} = n(n + 1)/2$.
  \item The algorithm has $n - 1$ steps where it adds an edge.
    For each step, it must consider the available edges.
    Call this a \emph{substep}.
    Give a very lazy argument that the total number of substeps for
    the algorithm is $\leq n^3$.
  \end{enumerate}
This is a polynomial function of $n$, rather than an exponential
function of $n$. 
\vspace{5pt}
\end{mybox}

\vspace{10pt}
\begin{mybox}
  \begin{exercise}
    \emph{Correctness of Prim-Jarník.} \IceMountain
  \end{exercise}
  Suppose that the Prim-Jarník algorithm produces a tree $T$ which is not
  minimal, with $T' \neq T$ the genuine MST.
  Then there must be a step in the construction where we first add an
  edge $e$ which is not in $T'$.
  We will show that the algorithm is \emph{correct} in the sense that
  this situation cannot occur!
  There will always be a shorter edge $e'$ it should add instead of
  $e$.
  The setup is shown below.
  \begin{center}
  \includegraphics[scale=0.55]{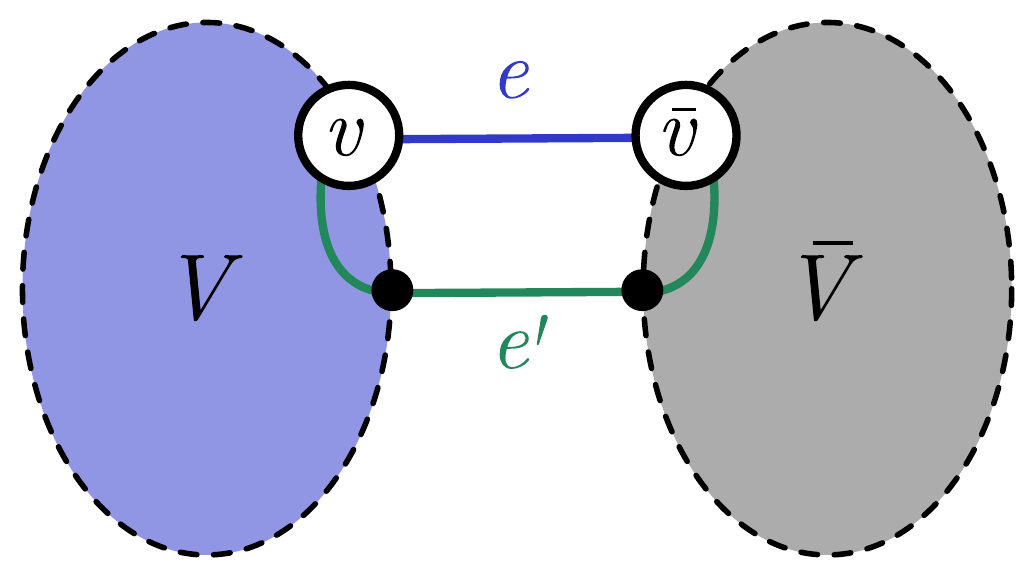}
  \end{center}
  \begin{enumerate}[label=(\alph*), itemsep=0pt]
  \item Suppose that, before the algorithm adds the ``bad edge'' $e$, it spans a
    set of cities $V$.
    The complementary set of cities is $\bar{V}$.
    Show that $e$ connects a vertex $v \in V$ to a vertex $\bar{v} \in
    \bar{V}$.
  \item Argue that the MST $T'$ has an edge $e'$ connecting $V$ to
    $\bar{V}$. \emph{Hint.} Use the fact that there is a path from $v$
    to $\bar{v}$ in $T'$.
  \item Explain why removing $e'$ from $T'$, and replacing it with $e$,
    results in a \emph{tree}.
    \emph{Hint.} Show there is still exactly one route between any
    two nodes.
  \item From part (c), conclude that the Prim-Jarník algorithm is correct.
  \end{enumerate}
\vspace{-5pt}
\end{mybox}
\vspace{5pt}

Finding MSTs is quick.
But are they any good, or can they be much longer than the minimal network?
Once again, our simple results on triangles provide some insight.
Let's start with an equilateral triangle of side length $d$.
In Exercise \ref{ex:eq-tri}, you found that the minimal network has
length $L_Y = \sqrt{3}d$.
The MST for the equilateral triangle just consists of any two sides,
and therefore has length $L_\Lambda = 2d$.
The \emph{ratio} of these two lengths is $\rho = L_\Lambda/L_Y
= 2/\sqrt{3} \approx 1.15$, so
the MST is about $15\%$ longer than the Steiner tree.
This is close enough for many practical purposes.

\begin{figure}[h]
  \centering
  \includegraphics[scale=0.42]{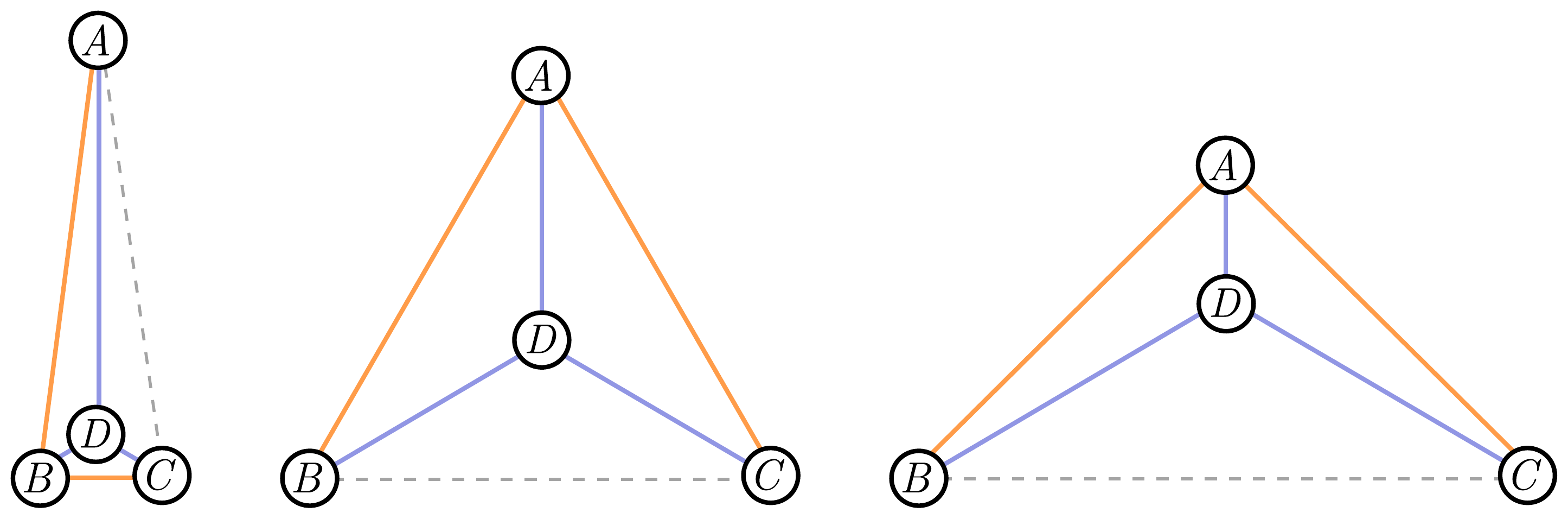}
\vspace{-10pt}
 \caption{Stretching and squeezing the equilateral triangle.}
  \label{fig:pollak1}
\end{figure}

You might wonder, in general, how bad this ratio can get.
To start with, let's see what happens when we squeeze or stretch the
triangle symmetrically.
If we squeeze it, like Fig. \ref{fig:pollak1} (left), the MST consists
of a long side of length $d$ and the short side which shrinks to zero.
Similarly, the minimal network consists of two short sides which
approach zero, and a long side which approaches $d$.
So the ratio of lengths approaches $1$.
Similarly, as we stretch the triangle out like Fig. \ref{fig:pollak1}
(right), the MST is the shorter two sides at the top, of total length $2d$,
while the hub eventually hits the top vertex, so it coincides with the
MST.
Once again, the ratio approaches $1$.

This hints that the equilateral triangle is the worst-case
scenario.
In fact, you can show in Exercise \ref{ex:gp-tri} that this ratio
is at most $2/\sqrt{3}$ for any triangle.
In \textsc{Gilbert} and \textsc{Pollak}'s magisterial study \cite{pollak}, they
conjecture that this holds for \emph{any number} of cities!
In other words, if $\rho$ is the ratio of the length of the MST to the
minimal network for any given set of cities, 
the \emph{Gilbert-Pollak
conjecture} 
states that
\begin{equation}
\rho \leq \frac{2}{\sqrt{3}}.\label{eq:1}
\end{equation}
The conjecture remains unproven. The best we can do right now is $\rho
\leq 1.21$ \cite{chung2006}.

What if we want to do better than $15\%$?
We can tweak the MST a little to get closer to the optimal network
length.
One particularly simple method is the \emph{Steiner insertion
  heuristic} \cite{dreyer}, which elegantly combines the MST and our
work with triangles.
The basic observation is that no edges in a minimal network
are separated by less than $120^\circ$, since hubs always have edges
separated by \emph{exactly} $120^\circ$, and edges at fixed nodes must
be separated by at least $120^\circ$ according to Exercise
\ref{ex:rim}(a).
The idea is to find edges with ``bad'' angles
($< 120^\circ$) and replace them with hubs.

In more detail, the insertion heuristic works as follows.
We first find the MST (using Prim-Jarník or another quick procedure), and then search for
the pair of edges with the smallest angle $< 120^\circ$.
If no such angle exists, we are done!
It such an angle does exist, the two edges connect a vertex, say $A$, to
vertices $B$ and $C$, as below in Fig. \ref{fig:mst3}.
We introduce a hub for these three vertices, which satisfies the
$120^\circ$ rule.
And then we do the whole thing again, looking for bad angles to
replace, until no more are left.
That's it!

\vspace{5pt}
\begin{figure}[h]
  \centering
  \includegraphics[scale=0.42]{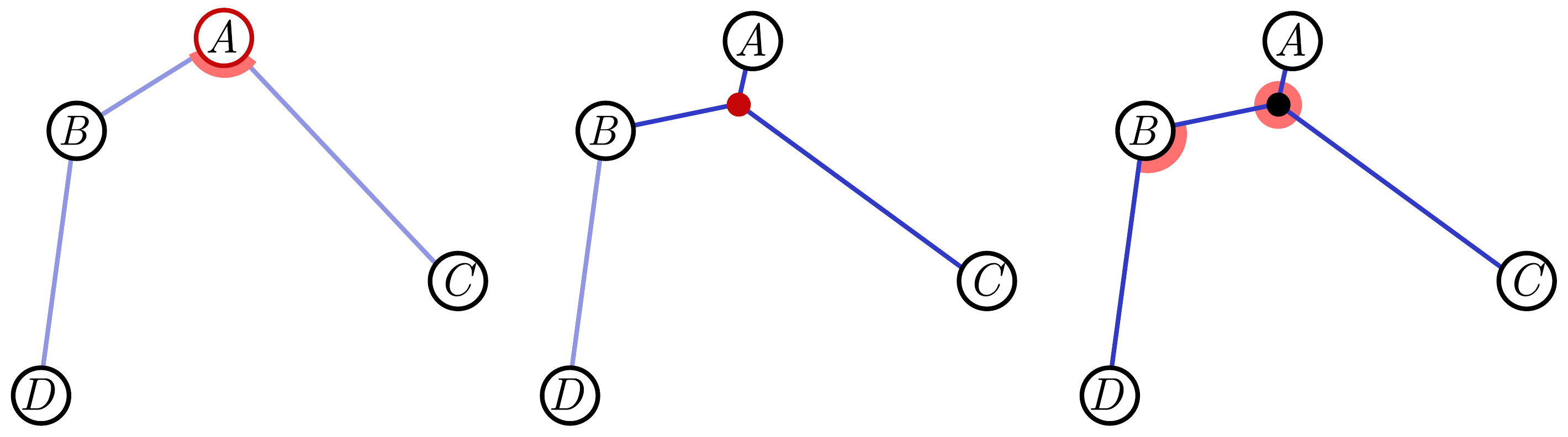}
\vspace{-5pt}
 \caption{Applying the Steiner insertion heuristic to our
   MST. First, we find the smallest angle $<
   120^\circ$. Then, add a hub. No more bad angles, so we're done!
}
  \label{fig:mst3}
\end{figure}

\begin{mybox}
  \begin{exercise}
    \emph{Steiner insertion heuristic.} \label{ex:insertion}
  \end{exercise}
Let's explore some general properties of the insertion algorithm.
  \vspace{-5pt}
  \begin{enumerate}[label=(\alph*), itemsep=0pt]
  \item Argue that the insertion of a hub can only \emph{decrease}
    length.
  \item Give an example showing that the insertion heuristic need not
    converge to the globally minimal network. \emph{Hint.} Exercise \ref{ex:rect}(c).
  \item Remember our earlier statement that fitting a tinkertoy to a
    set of fixed nodes is \textsf{NP-complete}.
    Explain why the Steiner heuristic can run quickly without
    contradicting this result.
    \emph{Hint.} Exercise \ref{ex:tiny}.
\end{enumerate}
\vspace{0pt}
\end{mybox}
\vspace{5pt}

Although Steiner insertion is quick, the optimality varies.
A different and less practical
 method \cite{arora} shows that, in
principle, you can approximate the minimal network on $n$
cities as \emph{closely as you like}, in some number of steps at most
polynomial in $n$.
For this reason, minimal networks belong to a complexity class called
\textsf{PTAS} (``\textsf{P}olynomial \textsf{T}ime \textsf{A}pproximation \textsf{S}cheme''), the problems
which can be easily approximated.
We can update our statement about complexity:

\vspace{10pt}
\begin{mybox2}
  \begin{statement}
    \emph{Complexity II.}
  \end{statement}
Finding minimal networks is \textsf{NP-hard}
  but also \textsf{PTAS}.
\end{mybox2}
\vspace{5pt}

\begin{mybox}
  \begin{exercise}
    \emph{Gilbert-Pollak for triangles.} \Mountain
\label{ex:gp-tri}
  \end{exercise}

Below, we give a visual proof of the Gilbert-Pollak conjecture for
triangles.
The basic idea is that, in an arbitrary triangle with angles $\leq
120^\circ$, we can attach a small equilateral triangle to the largest
angle (city $A$ below).
\vspace{5pt}
  \begin{center}
      \includegraphics[scale=0.45]{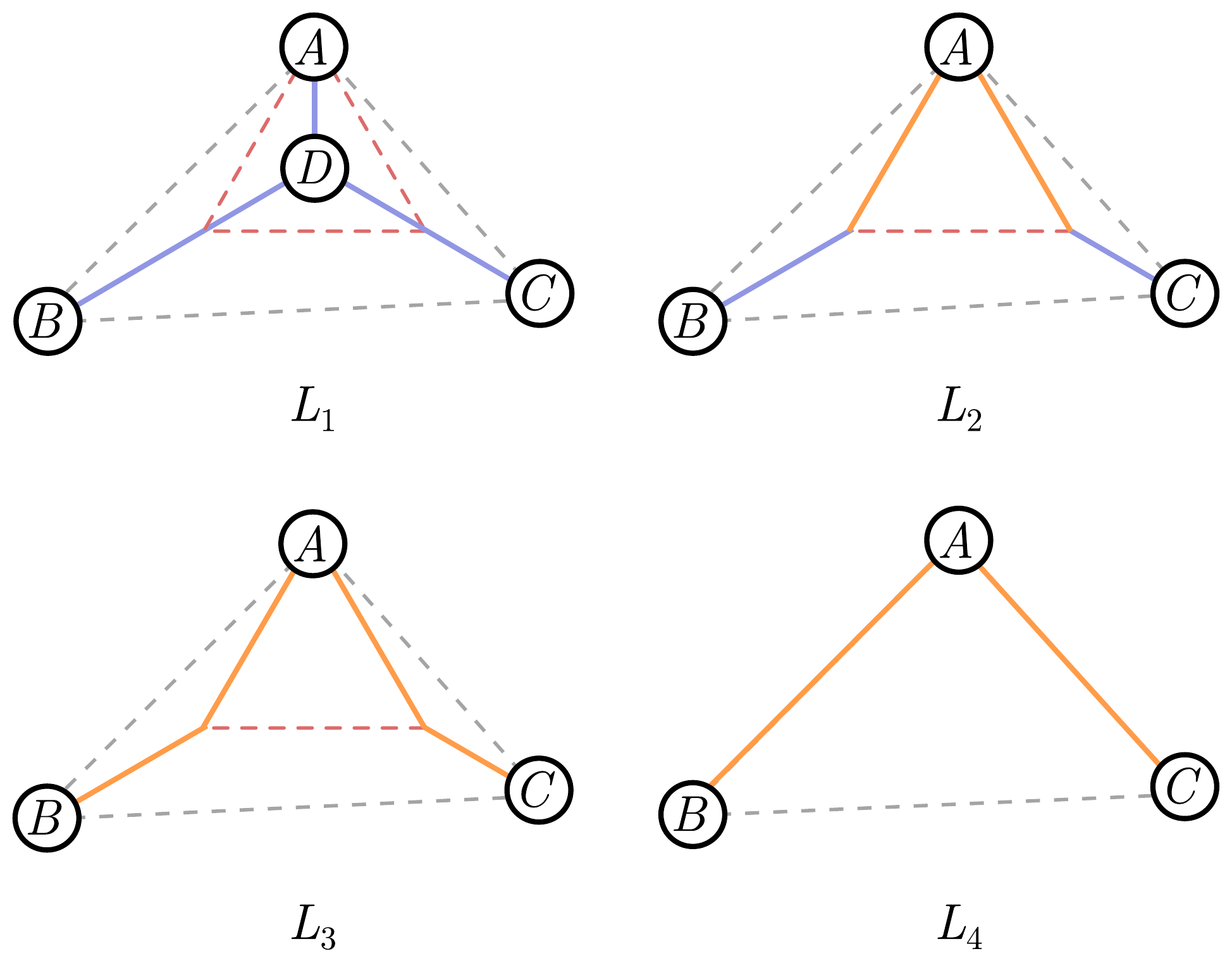}
  \end{center}
\vspace{-15pt}
The lengths $L_1, L_2, L_3, L_4$ are made up of lengths of coloured
lines, but blue lines have weight $1$, while orange lines have a weight $\sqrt{3}/2$.
For instance,
\[
L_1 = |DA|+|DB|+|DC|, \quad L_4 = \frac{\sqrt{3}}{2}(|AC| + |BC|).
\]
In other words, $L_1$ is the length of the minimal network, and $L_4$
is $\sqrt{3} /2$ times the length of the MST.
\begin{enumerate}[label=(\alph*), itemsep=0pt]
\item Argue that $L_1 \leq L_2 \leq L_3 \leq L_4$.
\item Use this (along with the case where some internal angle is $\geq
  120^\circ$) to establish the Gilbert-Pollak conjecture for triangles.
\end{enumerate}
\vspace{0pt}
\end{mybox}

\newpage

\section{Bubble networks}
\label{sec:soap-bubbles}

Humans are not the only players in the minimization game.
Nature is also cheap, or rather \emph{lazy}:
it does as little as possible, formally known as the Principle of
Least Action.
If we play our cards right, perhaps we can hack the laws of
physics to do our minimization for us.
In our case, it turns out we can do network planning with \emph{bubbles}.
Bubbles are formed when a film of liquid separates two volumes of air.
Surface tension tries to pull the bubble surface taut in all
directions, which results in the \emph{minimization of area}.
But if there are no constraints, then the surface will shrink
until nothing is left!
Really, we mean that bubbles minimize the area of the wall
\emph{subject to constraints}.

\begin{figure}[h]
  \centering
  \includegraphics[scale=0.33]{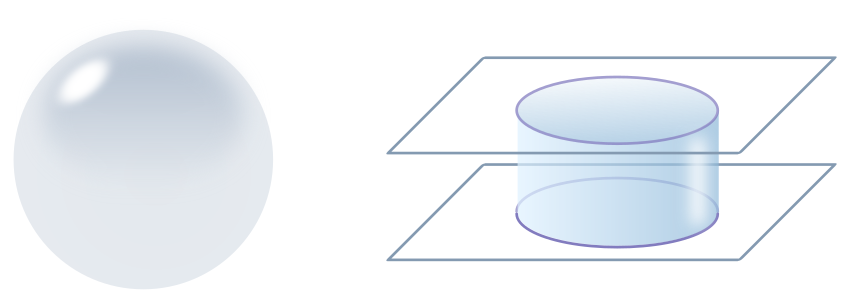}
 \caption{Soap bubbles in three and two dimensions.}
  \label{fig:2bubbles}
\end{figure}

For building railway networks, we want walls to be one-dimensional,
and the contraints to be fixed external nodes.
We'll talk about how to do this in a moment, but there is a more
natural constraint associated with blowing bubbles: they enclose a
pocket of air.
This explains why soap bubbles are spheres!
As we will show in \S \ref{sec:spheres-bubbletoys}, a sphere (Fig. \ref{fig:2bubbles} (left)) is the
smallest surface containing a fixed volume of air.
A lone bubble is direct proof of Nature's laziness.

\vspace{5pt}

If we sandwich the bubbles between two plexiglass plates, we will get
a \emph{two-dimensional} network of bubbles.
The vertical walls look like a graph from above, and a single bubble
will be a circle (Fig. \ref{fig:2bubbles} (right)).
There are two questions about these networks that immediately present
themselves.
First, what happens at a junction of bubble walls?
And second, what do walls look like \emph{away} from a junction?
The first question is easy to answer.
Imagine zooming in on a junction until the walls \emph{look straight}.
Since the bubbles try to minimize wall area, or viewed from above,
\emph{wall length}, they will obey the $120^\circ$ rule, since this is
the local rule any length-minimizing network obeys!\footnote{Zooming in
  enough means that edges can be reconfigured without having any
  practical effect on air enclosed.}

The situation away from junctions is a little trickier, but as we will
see, for both physical (Exercise \ref{ex:walls}) and
mathematical reasons (\S \ref{sec:circles-bubbletoys}), a bubble wall
can either be straight, or it can curve along the arc of a circle.
Viewing a straight line as the arc of an \emph{infinitely} large
circle, we can just say that walls are arcs of circles.

\subsection{Computing with bubbles}
\label{sec:comp-with-bubbl}

\begin{figure}[h]
  \centering
  \includegraphics[scale=0.25]{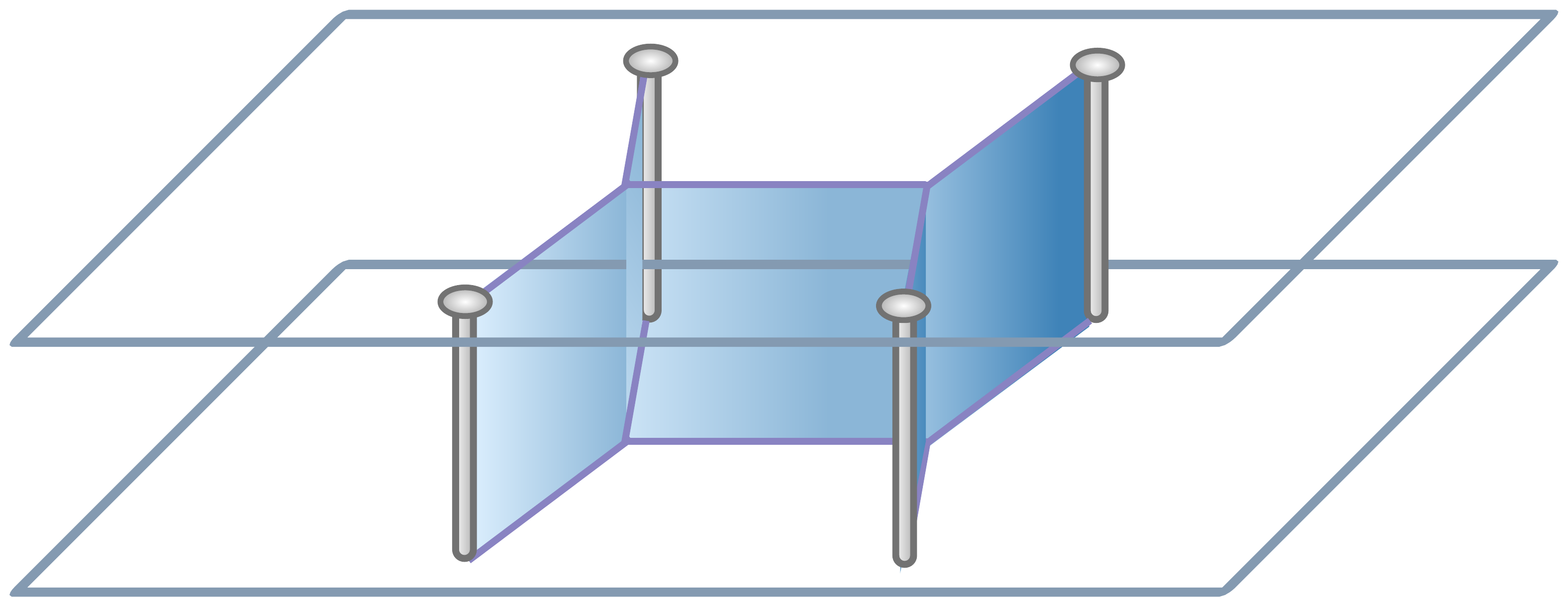}
 \caption{A soap bubble computer for finding minimal networks.}
  \label{fig:comp}
\end{figure}

Plexiglass gives us two-dimensional bubbles, and length rather than
surface area will be minimized.
But the constraint will generally be to enclose a fixed \emph{area} of air
per cell.
Can we hack this setup to make a \emph{soap bubble computer} for
finding minimal networks?
Yes!
The key is to give the bubble walls something to hold onto.
If we drill some screws between the plexiglass plates, these will act like
the cities, and a network of walls can form between them.\footnote{For a programmable soap bubble computer, you can use
  suction cups with rods between them. Thanks to Pedro Lopes for
  pointing this out.}
Fig. \ref{fig:comp} shows an example with four screws, and the
junctions that can form between bubble walls.

\vspace{5pt}

You can use a soap bubble computer to solve the original railway planning problem.

\vspace{10pt}
\begin{mybox}
  \begin{exercise}
    \emph{Railways and soap bubbles.} \label{ex:gauss}
  \end{exercise}
  As advertised in \S \ref{sec:historical-notes}, the mathematician
  Gauss wanted to connect four cities with a
  minimal rail network.
  In an 1836 letter to his friend, the astronomer \textsc{Heinrich
    Schuhmacher} (1780--1850), Gauss asked:
  \begin{quote}
\emph{How does a railway network of minimal length connect the
    four German cities of Bremen, Harburg, Hannover, and Braunschweig?}
  \end{quote}
  The cities are drawn, along with their GPS coordinates, below:
  \vspace{0pt}
  \begin{center}
    \includegraphics[scale=0.2]{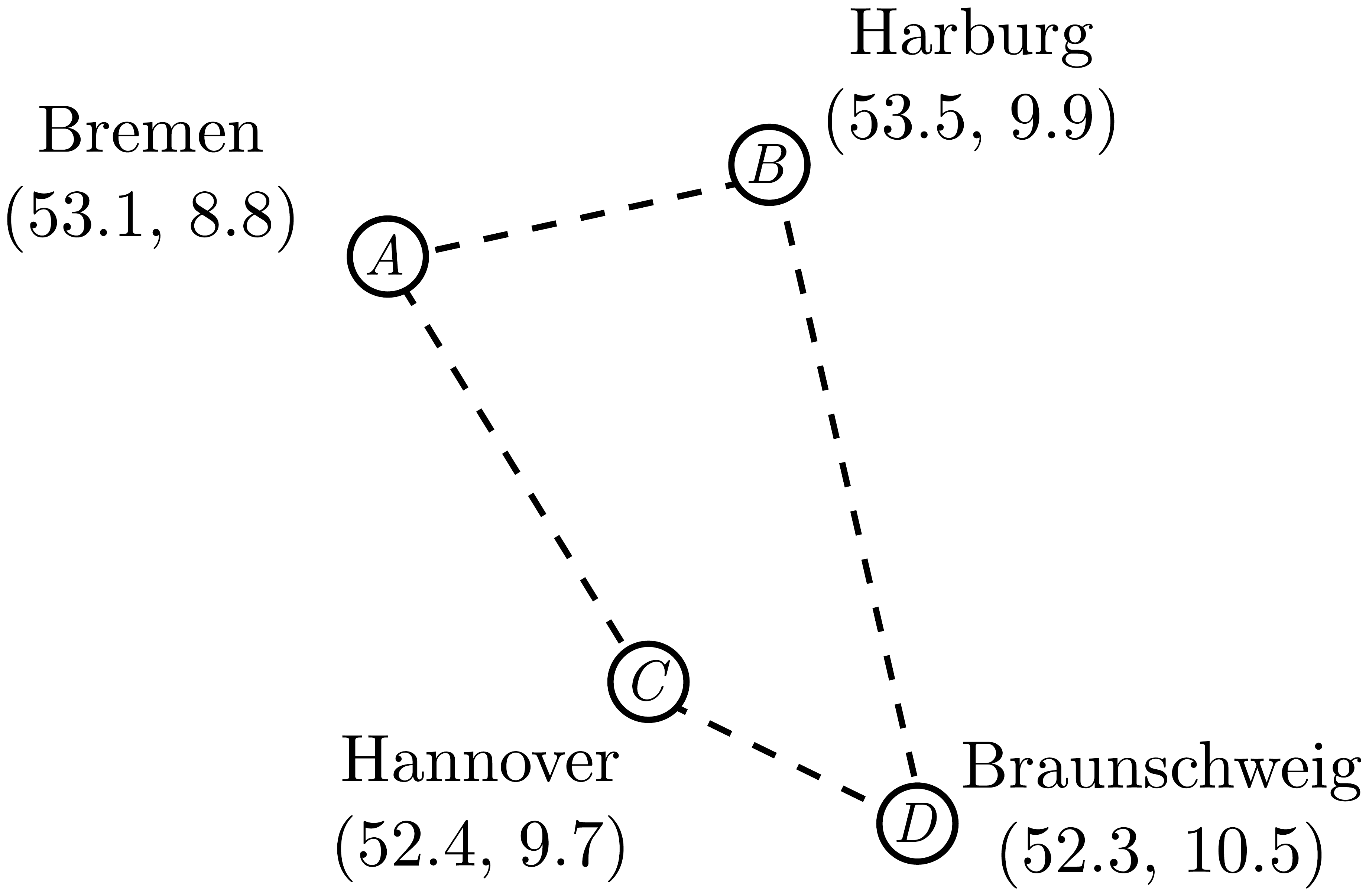}
  \end{center}
  \vspace{-15pt}
  \begin{enumerate}[label=(\alph*), itemsep=0pt]
  \item Find the minimum spanning tree using the Prim-Jarník
    algorithm.
  \item Assuming Gilbert-Pollak, lower bound the
    length of the minimal network.
  \item Improve the MST using the Steiner insertion heuristic.
  \item Build a soap bubble computer 
and solve the Gauss' railway problem.
How does this compare to the results of the Steiner insertion heuristic?
\end{enumerate}
\vspace{0pt}
\end{mybox}
\vspace{5pt}

For small networks, the soap bubble works almost instantaneously,
making it easy to believe it will quickly give the right answer for
large networks as well.
Sadly, this is very unlikely to be true!
To see why, recall that in \S \ref{sec:tinkertoys}, we argued that
finding a tinkertoy is \textsf{NP-complete}, and finding the genuine minimal network is
\textsf{NP-hard}.
Both problems are at least as hard as everything in \textsf{NP}, the
class of problems where lucky guesses can be checked quickly.
But just because a lucky guess can be checked quickly does not mean
your chances of making a lucky guess are good.
In fact, computer scientists are almost certain that most problems in
\textsf{NP} \emph{cannot} be solved quickly on a regular digital
computer.
The set of problems which can be solved quickly is called \textsf{P},
for ``\textsf{P}olynomial time''.
To summarize, computer scientists believe that \textsf{P}$\neq$\textsf{NP}
through proving it is the most important open problem in computer
science.\footnote{So important that there is a
  \href{https://www.claymath.org/millennium-problems/p-vs-np-problem}{\$1
    million bounty} on its head!}

But, you might object, a soap bubble is not a regular digital computer;
it is built out of the laws of physics rather than 1s and 0s.
Could it do things quickly that would take a digital computer longer
than the age of the universe?
The answer is probably no. 
Computer scientist \textsc{Scott Aaronson} hypothesized \cite{Aaronson2005}
that the problems in \textsf{NP-complete} (and hence \textsf{NP-hard})
cannot be solved quickly by \emph{any} computer,
digital or analogue.
This is called the \emph{\textsf{NP} Hardness Assumption}.

One piece of evidence is that every time we think we have a loophole
for quickly solving \textsf{NP-complete} problems, the loophole
disappears on closer examination.
The devil is in the details!
But there is broader philosophical reason for believing \textsf{NP}
Hardness: roughly, \emph{\textsf{NP} is OP}.\footnote{Gamer speak for ``overpowered''.}
Many of the hardest problems we know are
\textsf{NP-complete}, and if we could solve them, then as Aaronson says,

\begin{quote}
\emph{$\ldots$we would be almost like gods. The \textsf{NP} Hardness Assumption is
the belief that such power will be forever beyond our reach.}
\end{quote}

\noindent This means we cannot quickly find the minimal rail network for 800 cities using
soap bubbles, a black hole, human DNA, a quantum computer, or any
other conceivable mechanism. No one will ever know what the network looks
like.

That raises the question: what do soap bubbles actually do?
They cannot quickly find minimal networks, since this problem is potentially even harder than \textsf{NP}.
But there are several ways for this to fail.
First, they can take a long time to settle down, which Aaronson saw
happening in his own soap bubble experiments, even for a few screws \cite{Aaronson2005}.
Secondly, they could relax into \emph{local minima} rather than the
true minima.
Since fitting tinkertoys is \textsf{NP-complete},
even this can take a long time, unless (like the Steiner insertion
heuristic) the tinkertoys are small.\footnote{See Exercise
  \ref{ex:tiny} for more on this subtlety.}
A final possibility is that we simply solve the wrong problem,
e.g. by introducing small bubbles which change the network configuration.
Based on my own experiments with soap bubble computers, it appears
that \emph{all} of these failure modes can be realized!

I am not trying to 
skewer soap bubbles.
Indeed, the rest of these notes are really just a love letter to 
their physico-mathematical beauty.
Rather, the moral is that physics and computation \emph{interact in interesting ways},
with results about computation leading to physical predictions
(see Exercise \ref{ex:predict} for some non-bubbly examples).
Going in the other direction, physics can lead to new insights into
computer science, the most spectacular example being 
\emph{quantum computers}.
These are machines based on the laws of quantum mechanics
rather than the classical logic of 1s and 0s.
Although in their infancy, thinking about quantum computers has
already taught us some remarkable things about computer science,
complexity classes, and the power of Nature's laziness.
Sadly, we must leave that story for another time!

\vspace{10pt}
\begin{mybox}
  \begin{exercise}
    \emph{\textsf{NP} Hardness and the laws of physics.} \label{ex:predict}
  \end{exercise}
Here are a few fun ways to solve \textsf{NP-complete} problems:
\begin{enumerate}[label=(\alph*), itemsep=0pt]
\item Create a time machine, and by sending a computer through it
  again and again, perform an arbitrary number of computations in
  finite time.
\item Build a ``Zeno hypercomputer'', performing one step in $1/2$ s, the second step in $1/4$ s, the third step in $1/8$
  s, etc., so an infinite steps take $1$ second.
\item Store information in infinite precision real numbers,
  e.g. points on a line, and manipulate them using basic
  arithmetic \cite{ram}.
\end{enumerate}
\vspace{-4pt}
If the \textsf{NP} Hardness Assumption is correct, none of these methods
works!
In each case, what do you think this is telling us about the nature of
the universe?
\vspace{4pt}
\end{mybox}

\subsection{The many faces of networks}
\label{sec:foams}

\begin{figure}[h]
  \centering
  \includegraphics[scale=0.35]{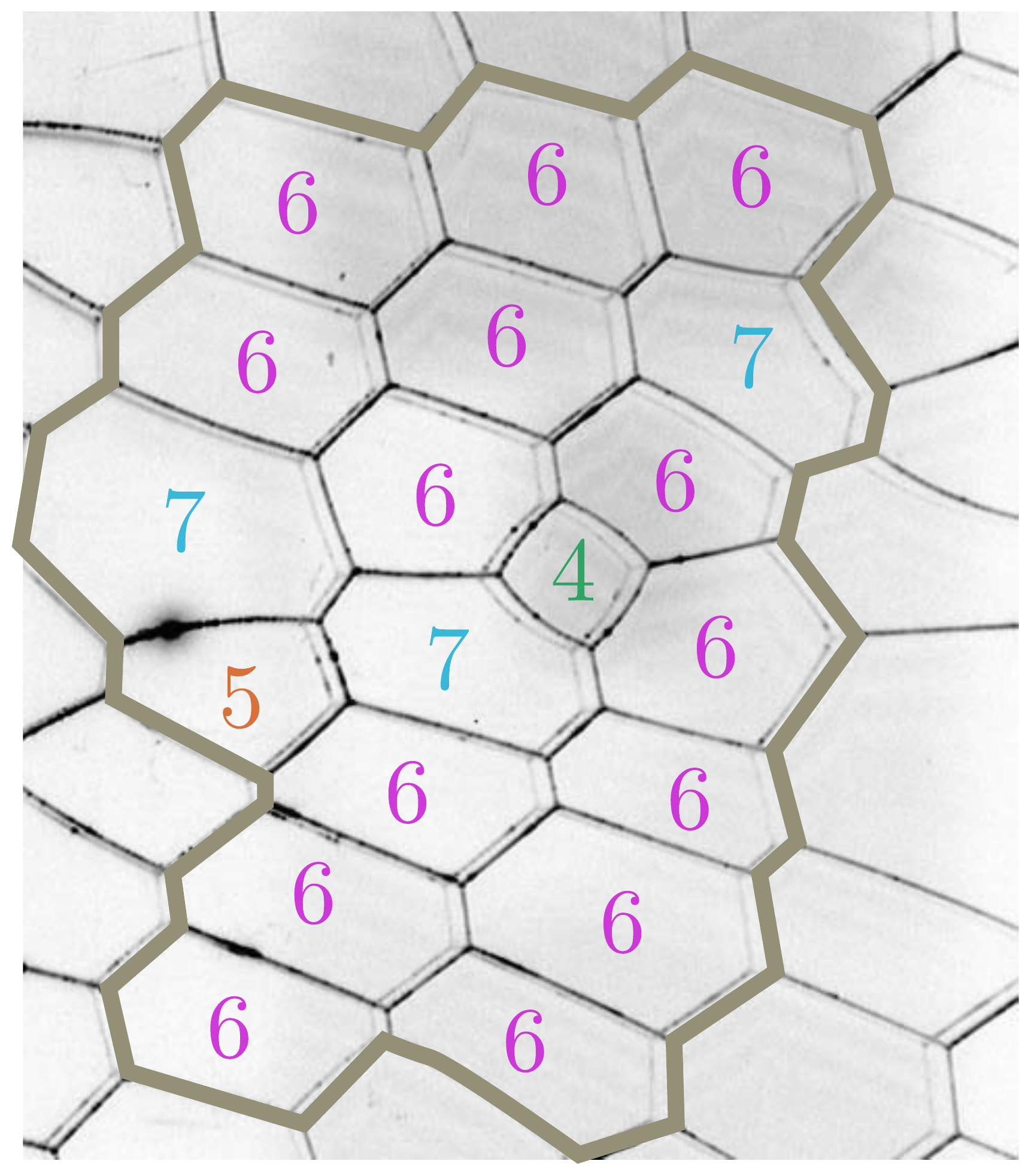}
 \caption{Most cells in a bubble network are hexagonal.}
  \label{fig:2foam}
\end{figure}

While we can use soap bubbles to learn about minimal networks, we can
arguably obtain more insight by going in the other direction.
What do minimal networks teach us about soap bubbles?
In this section, we consider the
two-dimensional bubble networks
with \emph{no screws}.
We'll just let the bubbles do their own
thing!
Fig. \ref{fig:2foam} shows a real two-dimensional soap
foam.\footnote{Based on a
  \href{https://commons.wikimedia.org/wiki/File:2-dimensional_foam_(colors_inverted).jpg}{photograph} by Klaus-Dieter Keller, Wikimedia Commons.}
We've counted the number of sides per cell, and surprisingly, most
seem to be hexagonal.
Is this is a coincidence, or is something deep going on?

\begin{figure}[h]
  \centering
  \includegraphics[scale=0.42]{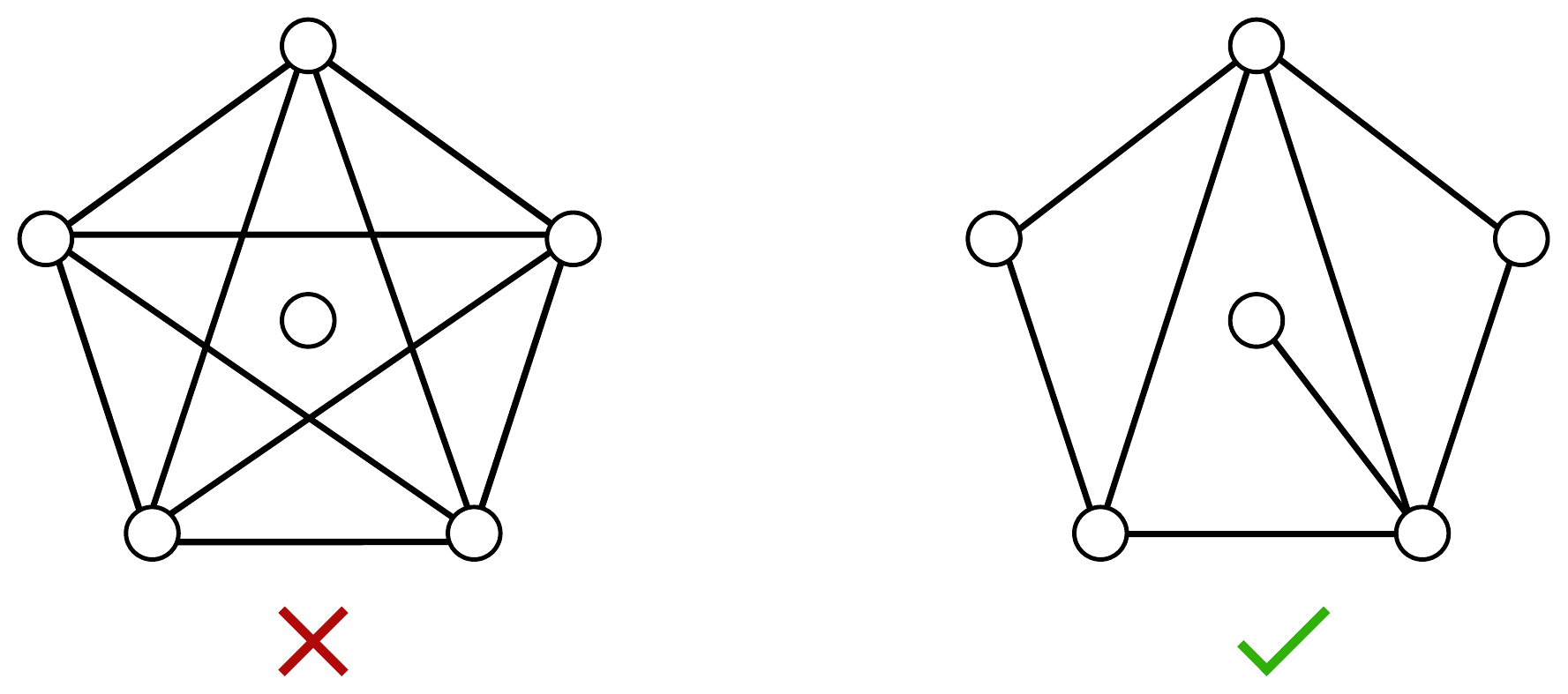}
 \caption{\emph{Left.} A disconnected graph which cannot be drawn without edge
   crossings. \emph{Right}. A planar graph. Euler's
   formula holds if we count the region outside the graph as a
   face.}
  \label{fig:planar}
\end{figure}

The answer is something deep.
We can actually \emph{prove} most bubbles are hexagonal using the
$120^\circ$ rule, some more graph theory, and a little physics.
The main result we will need from graph theory is \emph{Euler's
  formula}, discovered by the prolific Swiss mathematician \textsc{Leonhard
  Euler} (1707--1783) in 1735.
It states a relationship between the number of nodes $N$, edges $E$,
and faces $F$ in a graph, proved below in Exercise \ref{ex:euler}:
\begin{equation}
N - E + F = 2.\label{eq:euler}
\end{equation}
Importantly, this only holds for connected graphs which can be drawn
without any edges crossing, also called \emph{planar graphs} (Fig. \ref{fig:planar}).
A face is defined as any region enclosed by a loop of edges, including
(counterintuitively at first) the \emph{exterior} of the graph.

\begin{figure}[h]
  \centering
  \includegraphics[scale=0.2]{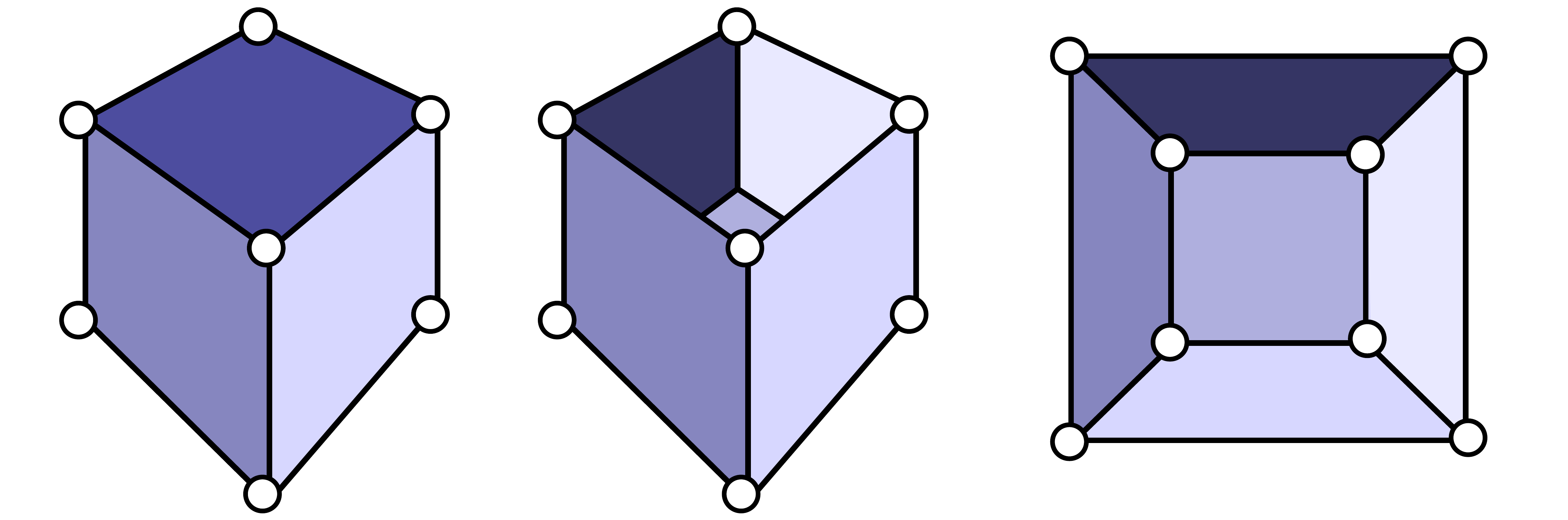}
 \caption{Remove the top of the cube and flatten.}
  \label{fig:cube}
\end{figure}

One way to obtain a planar graph is to take a three-dimensional
polyhedron, remove a single face, and flatten what remains onto the
plane.
This flattening process is shown for the cube in Fig. \ref{fig:cube}.
The removed face becomes the exterior region of the graph,
which is why we count it as a face.

\noindent Let's check that Euler's formula works. For the cube, we have $F = 6$ faces, $E = 12$ edges, and
$N = 8$ corners, so $F - E + N = 2$ just as Euler predicts.
We can count either using the cube itself, or the flattened graph,
provided we count the exterior as a face.

\vspace{10pt}
\begin{mybox}
  \begin{exercise}
    \emph{Euler's formula.} 
\label{ex:euler}
  \end{exercise}
Define the \emph{Euler characteristic}
\[
\chi = N - E + F.
\]
Our goal will be to show $\chi = 2$ for a graph without crossings.
First, we will establish Euler's formula for networks made out of
triangles.
We can then extend this to any graph without crossings.
Below we depict stages (a), (b), (c) and (e).
  \begin{center}
    \includegraphics[scale=0.27]{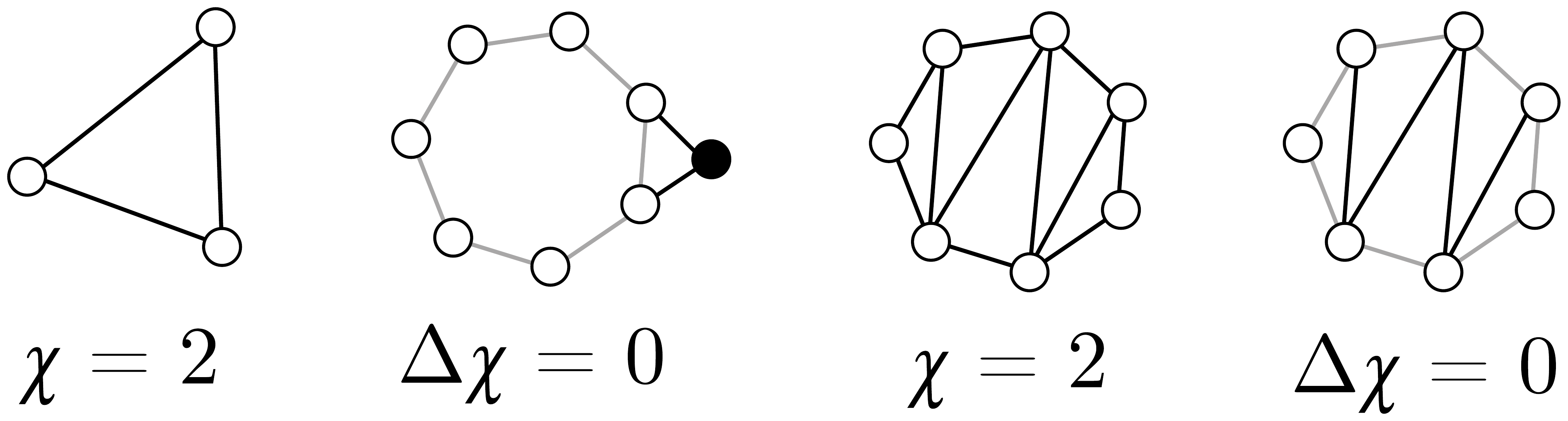}
  \end{center}
\vspace{-15pt}
\begin{enumerate}[itemsep=0pt]
\item[(a)] Show that a lone triangle in the plane obeys Euler's
  formula.  
\item[(b)] Suppose a network obeys Euler's formula.  Add a
  triangle (two edges and a node) to an external edge, and explain why
  the Euler characteristic doesn't change, $\Delta \chi = 0$.
  Conclude that the new network obeys Euler's formula.
\item[(c)] Explain why a network composed of triangles obeys
  Euler's formula.
\end{enumerate}
Now we can generalize to any network without crossings.
\begin{enumerate}[itemsep=0pt]\item[(d)] Consider a face, i.e. loop of
  edges, in such a network.
  Describe a procedure to add edges so that the face is split into
  triangles.
\item[(e)] Show that, after your procedure in part (d),
  $\Delta \chi = 0$.
\item[(f)] Conclude that any network without
  crossings obeys $\chi = 2$.
\end{enumerate}
\vspace{0pt}
\end{mybox}
\vspace{5pt}

When there are no screws, every node in the bubble network is a hub,
and therefore obeys the $120^\circ$ rule, with three bubble walls
meeting.
By the handshake lemma (\ref{eq:handshake}), we have $2E = 3N$.
Putting this into Euler's formula, we can eliminate $N$ and find a
relation between the number of faces and number of edges:
\begin{equation}
N - E + F = \tfrac{2}{3} E - E + F = 2 \quad \Longrightarrow \quad 3F -
E = 6.\label{eq:euler-hub}
\end{equation}
It will be useful to treat the external face a little differently.
Let $F'$ be the number of \emph{internal} faces, so that $F = F' + 1$.
Then (\ref{eq:euler-hub}) becomes $3F' - E = 3$.

Before proceeding, we need two additional properties
of our bubble networks.
First of all, an edge cannot dangle into the middle of a face.
If it did, the vertex at the end of the dangling edge would not have
three attached edges, only one, which is impossible by the $120^\circ$
rule.
It follows that every edge straddles two distinct faces.\footnote{Counterexamples like an edge
  cutting across two concentric circles are also ruled out
  by the $120^\circ$ rule.}

\begin{figure}[h]
  \centering
  \includegraphics[scale=0.2]{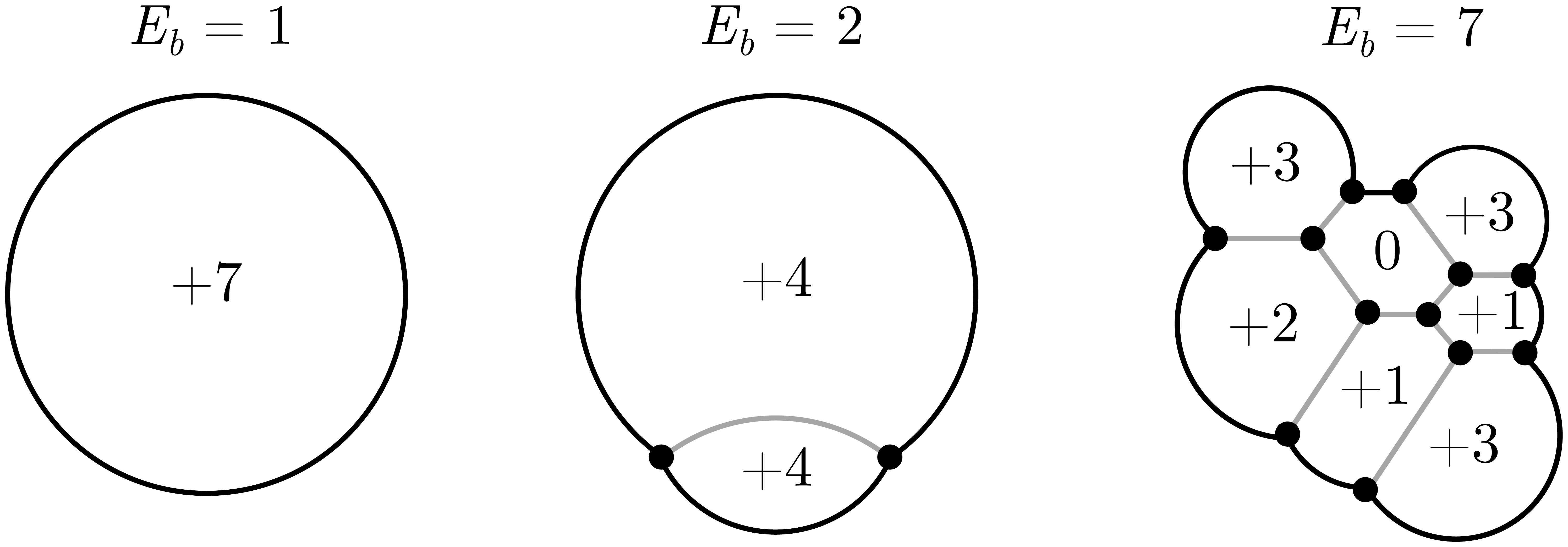}
 \caption{
$D_\text{hex}$, the sum of numbers in cells, is always $6$ more than $E_b$.}
  \label{fig:hex}
\end{figure}

Let $F_s$ denote the number of internal faces with
$s$ sides, and let $E_b$ stand for the number of edges of the outer
face of the collection of bubbles.
The total number of internal faces is
\begin{equation}
F' = F_1 + F_2 + F_3 + \cdots \,. \label{eq:F'}
\end{equation}
But since each edge is associated with \emph{two} faces, we can also
express edges as
\begin{equation}
2E = E_b + 1 \cdot F_1 + 2 \cdot F_2 + \cdots + s\cdot F_s + \cdots
\,.\label{eq:Eb}
\end{equation}
If we plug (\ref{eq:F'}) and (\ref{eq:Eb}) into $3F' - E = 3$, we finally get
\[
6 + E_b = 6F' - 2E + E_b = (6-1) \cdot F_1 + (6-2) \cdot F_2 +
\cdots + (6-s)\cdot F_s + \cdots \,.
\]
We will call the RHS the \emph{hexagonal difference} $D_\text{hex}$,
since it counts the number of edges which do not belong to a hexagonal
face, with a sign depending on whether the face is smaller ($+$) or
larger ($-$)
than a hexagon.
So, more simply, we have
\begin{equation}
D_\text{hex} = 6 + E_b.\label{eq:hexdiff}
\end{equation}
The hexagonal difference is $6$ more than the number of boundary
edges.
We give a few simple examples in Fig. \ref{fig:hex}, with the
contribution to $D_\text{hex}$ indicated in each cell.

\vspace{10pt}

\vspace{10pt}
\begin{mybox}
  \begin{exercise}
    \emph{Large and small faces.} \label{ex:faces}
  \end{exercise}
Equation (\ref{eq:hexdiff}) already tells us some interesting things
about bubble networks.
  \begin{enumerate}[label=(\alph*), itemsep=0pt]
  \item Explain why $D_\text{hex} \geq 6$.
  \item Deduce that
    \[
6 + 5 \cdot F_1 + 4 \cdot F_2 + \cdots + 1\cdot F_5 \geq 1 \cdot F_7 +
2\cdot F_8 + \cdots \,.
    \]
  \item Suppose a bubble network has two bubbles with four sides and no
    other small faces.
    What is the maximum number of $10$-sided bubbles?
\end{enumerate}
In general, once we count the ``small'' faces $F_1, \ldots, F_5$,
we can constrain the number of ``large'' faces $F_7, F_8, \ldots$.
\vspace{5pt}
\end{mybox}

\subsection{Hexagons and honeycomb}
\label{sec:hexagon}

\begin{figure}[h]
  \centering
  \includegraphics[scale=0.23]{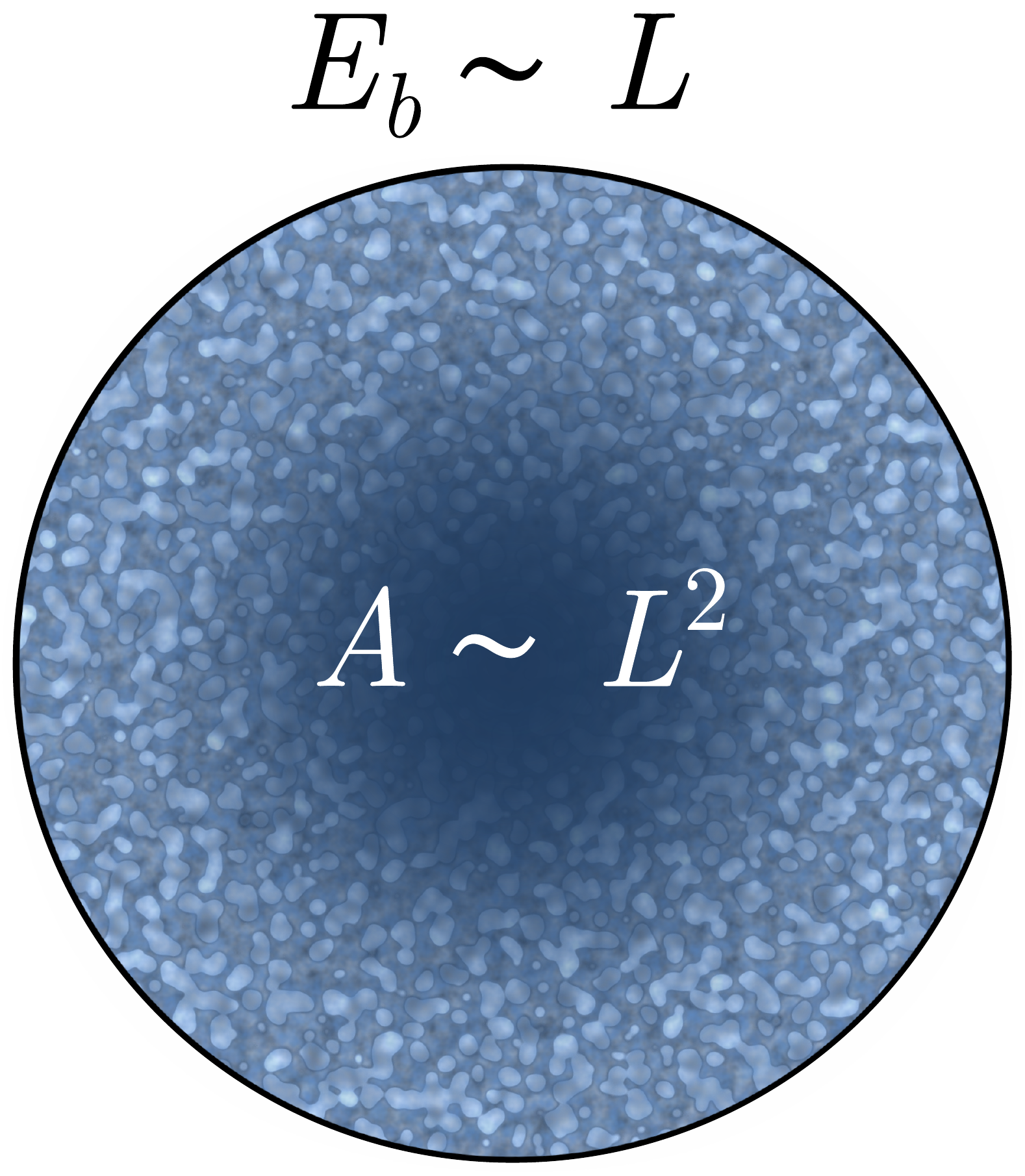}
\vspace{-5pt}
 \caption{As the foam gets large, the number of outer edges scales as
   $L$, and the area as $L^2$.}
  \label{fig:scaling}
\end{figure}

It's still not clear why most bubbles are hexagonal.
At this point, we need to introduce some basic physical intuition. 
Suppose the foam has overall size $\sim L$.
Assuming bubbles have a typical size independent of $L$, the
number of external edges $E_b \sim L$.
The total area of the bubble network should scale as $A \sim L^2$.
For instance, consider a roughly circular foam of radius $L$
(Fig. \ref{fig:scaling}).
If bubbles have average edge length $\ell$, independent
of $L$, then $E_b \approx (\pi/\ell)L$, while $A = \pi L^2$.

\vspace{5pt}

It follows that, for large $L$, the hexagonal difference $D_\text{hex} = 6 + E_b \sim L$.
The ``density'' of
non-hexagonal edges $d_\text{hex}$ is just
the total hexagonal difference divided by the area of the foam. Since
area scales as $L^2$, the density of non-hexagonal edges scales as
\begin{equation}
d_\text{hex} \sim \frac{D_\text{hex}}{L^2} \sim
\frac{1}{L}.\label{eq:7}
\end{equation}
As $L$ becomes larger, edges belonging to non-hexagons become increasingly rare.
This explains why a typical cell in a bubble network has six sides, just like
Fig. \ref{fig:2foam}.

\vspace{10pt}
\begin{mybox}
  \begin{exercise}
    \emph{Bubble blobs.}\label{ex:blobs}
  \end{exercise}
  A \emph{bubble blob} is a set of contiguous bubbles in a bubble network.  Let $E_o$ denote the number of edges
  extending outward from the boundary, and $E_i$ the number extending
  inward.
  \begin{enumerate}[itemsep=0pt]
\item[(a)] Explain why the difference from hexagonality is now given by
  \begin{equation}
      D_\text{hex} = 6 + E_i - E_o.\label{eq:blob}
    \end{equation}
  \item[(b)] Verify that the blob of cells in Fig. \ref{fig:2foam}
    satisfies (\ref{eq:blob}).
  \item[(c)] Repeat the scaling argument above, and conclude that in a
    large blob, departures from hexagonality become rare.
  \end{enumerate}
\end{mybox}
\vspace{5pt}

You may have wondered if the hexagonality of bubbles is related
to the fact that bees build honeycombs in a hexagonal lattice.
It is!
Bees have a clear evolutionary reason to minimize the amount of wax
used.
\textsc{Charles Darwin} (1809--1882) discusses the hive-making
instinct and its relation to fitness in his \emph{Origin of Species}
\cite{darwin1859}:
\begin{quote}
\emph{That motive power of the process of natural selection having been
  economy of wax; that individual swarm that wasted least honey in the
  secretion of wax, having succeeded best.}
\end{quote}
For honeycomb walls to be minimal, they must obey the $120^\circ$ rule.
If honeycomb cells are equal in size (which bees
might prefer for simplicity of construction), then a
natural guess at the optimal arrangement is the \emph{hexagonal
  lattice}. This is the only
regular tessellation 
satisfying the $120^\circ$ rule.

\vspace{0pt}
\begin{figure}[h]
  \centering
  \includegraphics[scale=0.2]{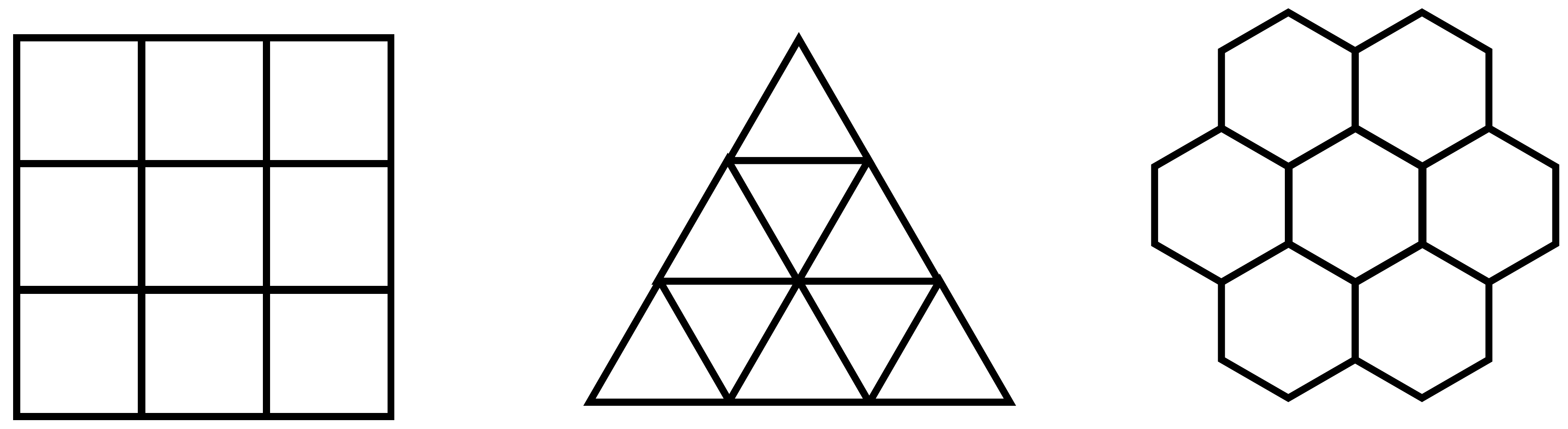}
\vspace{-5pt}
 \caption{The three regular tessellations of the plane: square,
   triangle, hexagon. 
}
  \label{fig:tess}
\end{figure}

The \emph{honeycomb conjecture}
states that the hexagonal lattice is the \emph{globally} minimal solution
among tessellations of the plane with equal cell size.
It is hard to verify this guess, since you need to check all possible
\emph{irregular} tilings as well as the regular ones. 
But in 1999, it turned from conjecture into theorem after \textsc{Thomas Hales}
gave a formal proof \cite{hales1999}.
In Exercise \ref{ex:hyp-hon}, we explore the analogous problem for the
saddle-shaped \emph{hyperbolic} plane.

\vspace{10pt}
\begin{mybox}
  \begin{exercise}
    \emph{Hyperbolic honeycomb.} \VarIceMountain \label{ex:hyp-hon}
  \end{exercise}
Our scaling argument assumed we were on a regular, Euclidean plane.
But we can see what happens if, instead of working on the Euclidean
plane, we work on the strangely curved \emph{hyperbolic plane}.
  \begin{center}
    \includegraphics[scale=0.73]{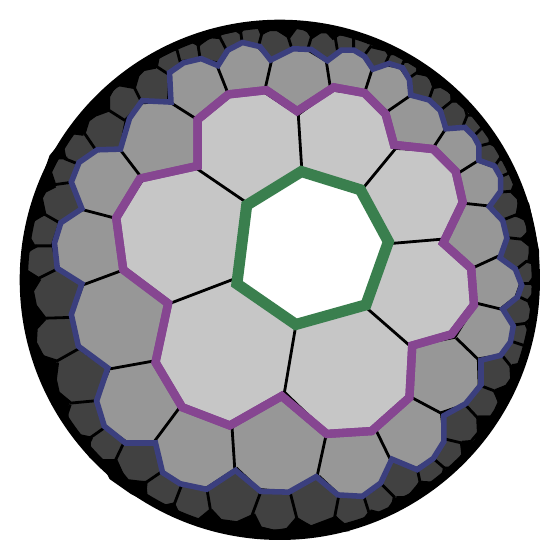}
  \end{center}
Above, we have tiled the hyperbolic plane with heptagons.
Each heptagon has the same area, and sides of equal length, but the
curvature means they must be drawn with different lengths on our flat page!
  \begin{enumerate}[label=(\alph*), itemsep=0pt]
  \item Find $A$ (in heptagon units) and $E_b$ for the regions enclosed in
    (i) green, (ii) purple, (iii) blue. Does the ratio $E_b/A$ appear
    to be decreasing?
  \item Argue that, in general, for $n > 1$ ``rings'' of heptagons,
    \[
      E_b = 4 \cdot 7^n, \quad A = \frac{1}{6}(7^{n+1} - 1) \approx
      \frac{7}{6} \cdot 7^n.
    \]
   \emph{Hint.} Use a geometric sum for $A$.
\end{enumerate}
This shows that on the hyperbolic plane, the scaling $E_b \sim L$, $A
\sim L^2$ no longer holds.
Instead, the boundary and area \emph{scale the same way}.
  \begin{enumerate}[itemsep=0pt]
  \item[(c)] Why does the $120^\circ$ rule still hold for minimal
    networks on the hyperbolic plane?
  \emph{Hint.} What happens when you zoom in on a node?
  \item[(d)] Show that for ring $n$, $E_o - E_i = 2 \cdot 7^n$.
    Using part (c) and similar reasoning to the plane, conclude that a
    large number of heptagonal rings,
    \[
      D_\text{hex} = 6 - 2 \cdot 7^n \approx - 2 \cdot 7^n.
    \]
    \vspace{-20pt}
  \item[(e)] Finally, show that our heptagonal tiling has
    \[
      d_\text{hex} = \frac{D_\text{hex}}{A} \approx - \frac{12}{7}.
    \]
  \item[(f)] Given Exercise \ref{ex:faces}(a), how can $D_\text{hex}$ be negative?
\end{enumerate}
\vspace{5pt}
\end{mybox}
\vspace{5pt}

The weird properties of hyperbolic space mean that the optimal
tessellation depends on the size of the cells.
The heptagonal tiling is optimal for the cell size pictured above, at
least among
\emph{regular} hyperbolic tilings \cite{hept}.
The ``hyperbolic honeycomb conjecture''---that this is optimal among
\emph{all} hyperbolic tessellations with this cell size, including the
irregular ones---remains open.\footnote{As far as I know, the problem
  is open for all cell sizes.}
Perhaps we should breed some hyperbolic bees, and inspect their 
honeycomb in a few million years!

\subsection{The isoperimetric inequality and bubbletoys}
\label{sec:circles-bubbletoys}

The preceding two sections studied two-dimensional
bubble foams, assuming there were no fixed nodes.
The total length is being minimized, but subject to what constraints?
The answer is suggested by our earlier discussion of air pockets, and
by the honeycomb conjecture.
The bees have no fixed nodes, since they are not trying to connect
anything.
Instead, they are trying to build cells to store honey.
To simplify the problem, we have considered an infinite number of
cells of the same size, but what if the bees only want six?
Or want to vary their serving sizes with cells of different area?
In general, we can ask for the \emph{minimal length bubble network}
enclosing cells of size $A_1, A_2, \ldots, A_n$.

\vspace{0pt}
\begin{figure}[h]
  \centering
  \includegraphics[scale=0.7]{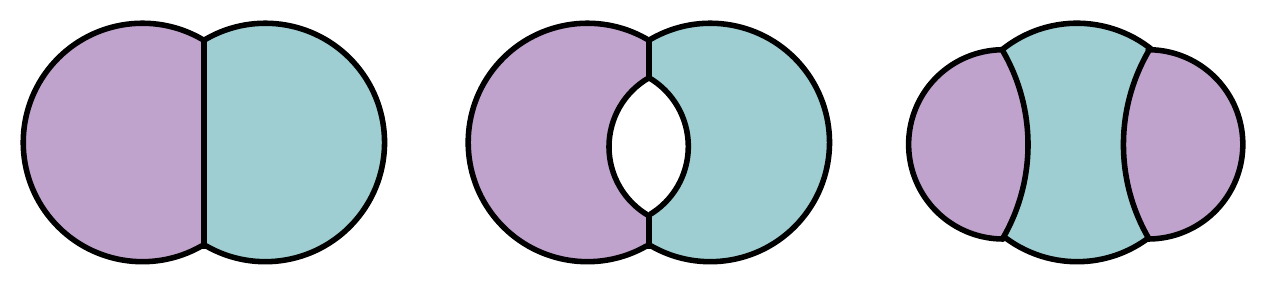}
\vspace{-5pt}
 \caption{\emph{Left.} The standard double bubble. \emph{Middle.} An empty pocket. \emph{Right.} A
   split bubble.
}
  \label{fig:2drules}
\end{figure}

In the same way that we are allowed to add nodes to minimal
networks to decrease length, we will allow \emph{empty pockets} and
\emph{bubble splitting} (Fig. \ref{fig:2drules}) if it helps us reduces length.
We will always ask that the bubble network is connected for physical
reasons.\footnote{If the network is not connected, the
  disconnected parts can ``float'' relative to each and will soon
  collide, forming a connected network.}
This leads to...

\vspace{10pt}
\begin{mybox2}
  \begin{statement}
    \emph{The Planar Minimal Bubble Problem.} \label{box:planar}
  \end{statement}
  Find the connected bubble network of smallest perimeter enclosing cells of area $A_1,
  A_2, \ldots, A_n$, allowing empty pockets and split bubbles.
\end{mybox2}
\vspace{5pt}

\noindent While we derived the $120^\circ$ rule directly from
minimizing length, the other salient property of bubble networks is
that walls are straight or arcs of circles.
You can see where this comes from using the physics of surface tension.

\vspace{10pt}
\begin{mybox}
  \begin{exercise}
    \emph{Young-Laplace I.} \label{ex:walls} 
  \end{exercise}
The molecules in a bubble wall are attracted to each other.
If you try to \emph{bend} the surface, it strains the molecular
bonds, which attempt to restore the unstretched state.
The amount of bending at a point can be quantified by finding
    a circle which fits snugly onto the curve (the green
    circle, below right).
\vspace{5pt}
\begin{center}
  \includegraphics[scale=0.18]{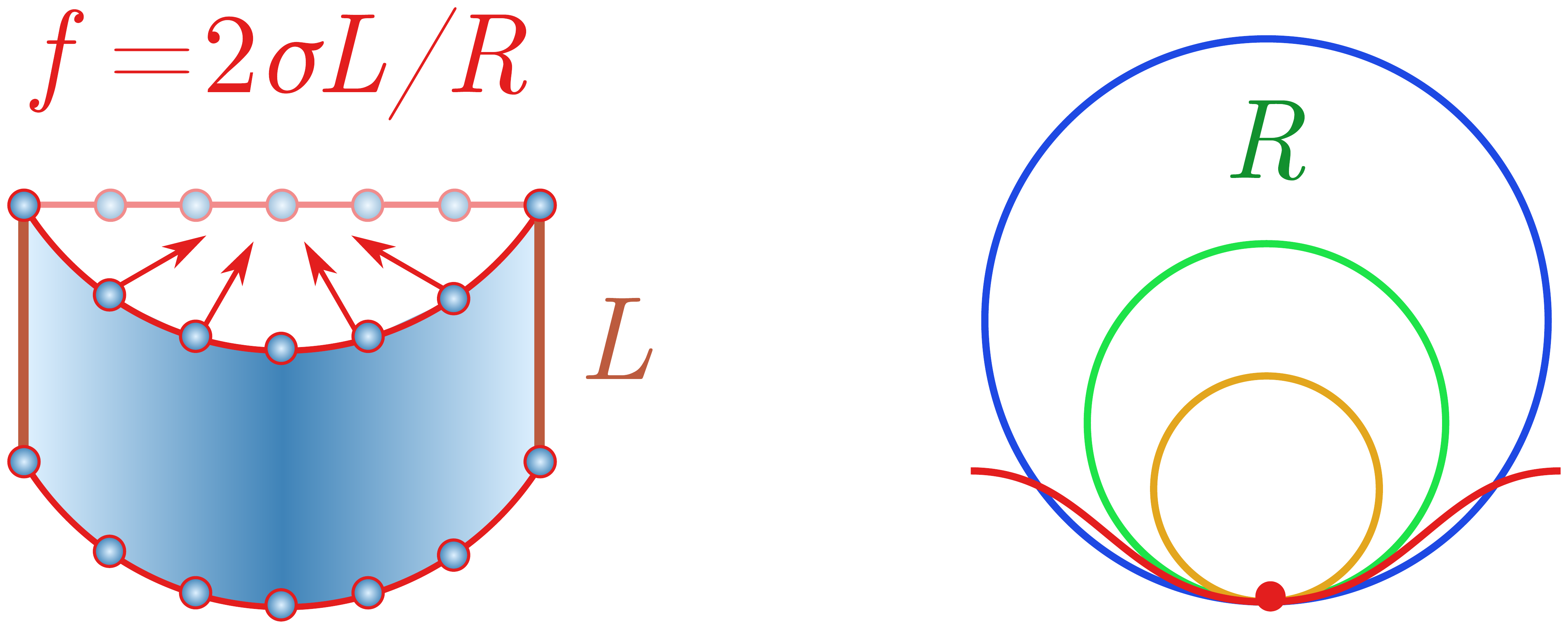}
\end{center}
    If this snug circle has radius $R$, we say the bend has
    \emph{radius of curvature} $R$.
For a bubble wall of height $L$, the restoring force per unit
length of curve is $f=2\sigma L/R$, where $\sigma$ is the \emph{surface
  tension}.
  \begin{enumerate}[label=(\alph*), itemsep=0pt]
  \item Show that if there are no other forces acting on the wall, it
    must be straight. 
  \item Now consider the effects of \emph{pressure} at a point on the
    wall.
    If the pressure on one side is $P_\text{out}$, and inside is
    $P_\text{in}$, argue the bend will have radius of curvature
    \begin{equation}
      R = \frac{2\sigma}{\Delta P},\label{eq:yll}
     \end{equation}
     where $\Delta P = P_\text{out} - P_\text{in}$.
    This is called the \emph{Young-Laplace law},
 after \textsc{Thomas
      Young} (1773--1829) and \textsc{Pierre-Simon Laplace} (1749--1827).
    Check it is consistent with (a).
  \item Within a cell, pressure differences equalize very quickly, so
    it is reasonable to assume pressure is constant on a face of the network.
   Deduce that bubble walls are either flat or arcs of
    circles.
\end{enumerate}
\vspace{0pt}
\end{mybox}
\vspace{5pt}

Although Exercise \ref{ex:walls} does involve surface tension, it says
nothing about minimizing surface area or solving the bubble
configuration problem.
It seems plausible that real bubbles do solve this problem, but for
the moment, let us view the bubble configurations as \emph{physical
  conjectures} about minimal-length solutions.
In other words, they are guesses made by Nature, awaiting the
rubber stamp of mathematical proof.

\begin{figure}[h]
  \centering
  \includegraphics[scale=0.6]{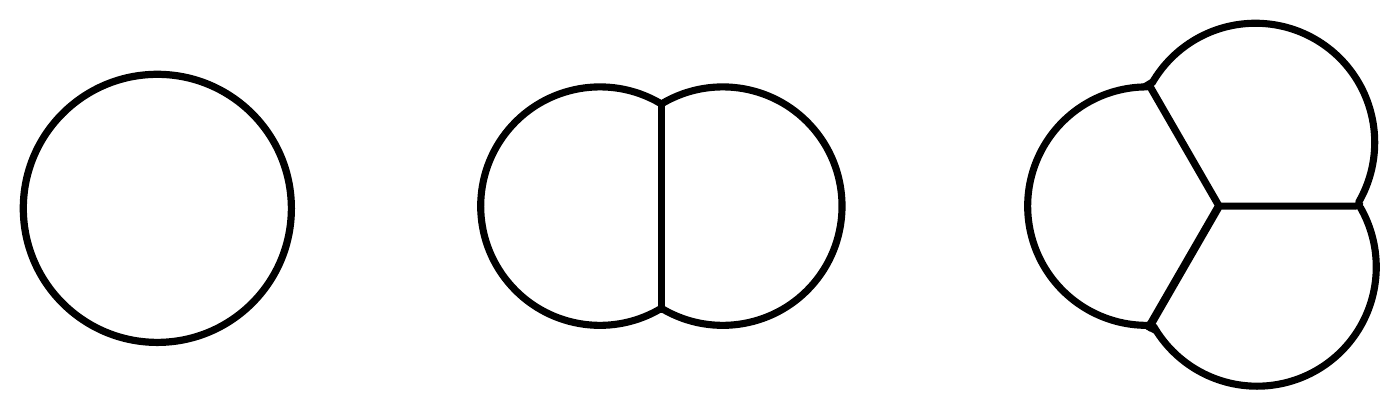}
\vspace{-5pt}
 \caption{\emph{Left.} A circle. \emph{Middle.} The standard double
   bubble. \emph{Right.} The standard triple bubble.
}
  \label{fig:2dconfigs}
\end{figure}

The simplest physical conjecture is for a single bubble of fixed area $A$.
The only smooth way to draw a single cell is a circle
(Fig. \ref{fig:2dconfigs} (left)), and when Nature is left to its own
devices, bubbles tend to assume this form.
You can also check (Exercise \ref{ex:connex}) that there is no way to split the single bubble, or
introduce air pockets, while maintaining a connected bubble network.

\vspace{10pt}
\begin{mybox}
  \begin{exercise}
    \emph{One bubble to rule them all.} \label{ex:connex}
  \end{exercise}
\begin{enumerate}[label=(\alph*), itemsep=0pt]
\item Show that, if we split a bubble into parts which contain a total
  area $A$, they cannot share any edges. Explain why the same goes for
  an empty air pocket and the region outside the bubble configuration.
\item Argue that splitting a single bubble, or adding empty pockets,
  violates (a).
\end{enumerate}
\vspace{0pt}
\end{mybox}
\vspace{5pt}

Mathematicians as far back as \textsc{Archimedes} (287--212 BC) have
suggested that the circle is the shape of smallest perimeter for a
fixed area $A$, a guess called the \emph{isoperimetric
  inequality}.\footnote{Saying the circle has the \emph{least perimeter} of
  all figures of area $A = \pi r^2$ is the same as saying it has \emph{most area}
  of all figures with the same perimeter $2\pi r$.
  ``Isoperimetric'' means ``same
  perimeter''.}
This guess wasn't verified until the 19th
century, 
but the modern proof is simple enough to present in outline.
The basic idea is to wobble a line and see how the length and enclosed
area change.
To start with, we consider wobbling the radius of a single circular
arc.

\vspace{10pt}
\begin{mybox}
  \begin{exercise}
    \emph{Stretched arcs.} \label{ex:onearc}
  \end{exercise}
Suppose an arc of length $L$ and radius $R$ is part of a curve
enclosing some area on the plane.
Consider extending the radius by a small amount $t$, where ``small''
means much smaller than $L$. 
  \begin{center}
    \includegraphics[scale=0.45]{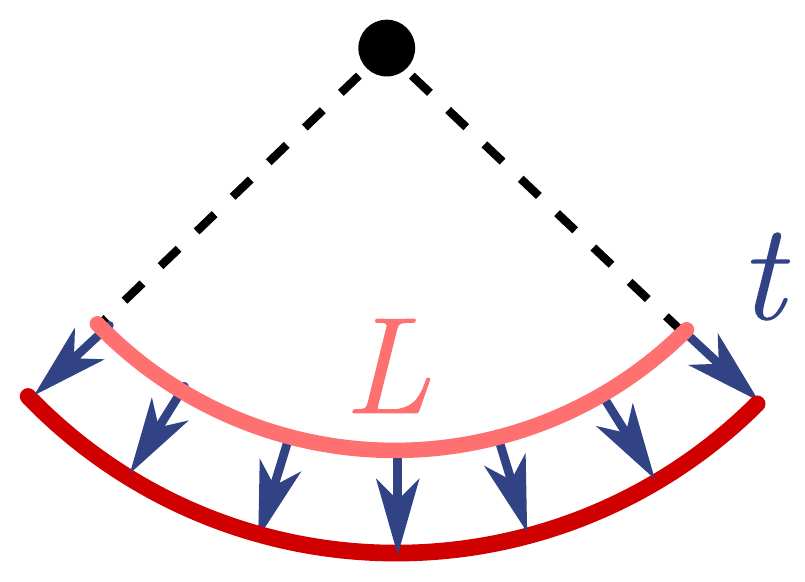}
  \end{center}
\begin{enumerate}[label=(\alph*), itemsep=0pt]
\item Show that the area enclosed changes by
  \begin{equation}
    \Delta A \approx Lt.\label{eq:deltaA}
  \end{equation}
\item Assuming that the angle subtended by the arc is the same,
  explain why the length of the arc changes by
  \begin{equation}
    \Delta L \approx \frac{L t}{R}.\label{eq:deltaL}
  \end{equation}
\item Check that (b) still makes sense for a straight line.
\end{enumerate}
For both $\Delta A$ and $\Delta L$, there are some additional
corrections, but these will appear as higher powers of $t$, starting
at $t^2$.
\vspace{0pt}
\end{mybox}
\vspace{5pt}

In general, we can take a curve on the plane and chop it up
into $k$ small pieces of length $L_1, L_2, \ldots, L_k$ and constant
radius of curvature $R_1, R_2, \ldots, R_k$, setting $R_i = \infty$
for any straight lines.\footnote{If we wanted to be rigorous, we would
  actually chop the line up into an \emph{infinite} number of pieces
  using calculus. 
}
Imagine we wobble the curve by \emph{independently} changing the radii
for each segment, adding $t_1, t_2, \ldots, t_k$.
Using (\ref{eq:deltaA}), the total change in area is
\begin{align}
  \Delta A  & = \Delta A_1 + \Delta A_2 + \cdots + \Delta A_k= L_1 t_1 + L_2 t_2 + \cdots + L_k t_k.   \label{eq:DeltaA}
\end{align}
From (\ref{eq:deltaL}), the total change in length is
\begin{align}
  \Delta L  & = \Delta L_1 + \Delta L_2 + \cdots + \Delta L_k = \frac{L_1 t_1}{R_1} + \frac{L_2 t_2}{R_2} + \cdots + \frac{L_k t_k}{R_k}.   \label{eq:DeltaL}
\end{align}
It's clear that if we deform the curve so as to preserve area, then
$\Delta A = 0$.


But here is the clever part: if the curve is a local minimum of
perimeter, then the perimeter looks like a \emph{quadratic} function of the
wobbling.\footnote{This is similar to the argument for equilateral triangles 
that the network
  length was an even function of wobble.}
But we are making $t$ small enough that we can ignore these quadratic
$t^2$ terms, and keep only the terms proportional to $t$.
Thus, in the approximation we have used to compute (\ref{eq:DeltaL}), a
perimeter-minimizing curve has $\Delta L = 0$.
You can show in the next exercise that, given the forms for $\Delta A$
and $\Delta L$, this is only possible if
\[
R_1 = R_2 = \cdots = R_k = R,
\]
i.e. the radius of curvature is constant.
Thus, the perimeter-minimizing curve has constant $R$.

\vspace{10pt}
\begin{mybox}
  \begin{exercise}
    \emph{Constant radius of curvature.} \label{ex:delta}
  \end{exercise}
If we vary a perimeter-minimizing curve, then $\Delta L = 0$.
If the wobbles also preserve area, then $\Delta A = 0$.
We will show that this implies all the radii $R_1, R_2, \ldots, R_k$
are the same.
\begin{enumerate}[label=(\alph*), itemsep=0pt]
\item Suppose that only $t_1$ and $t_2$ are nonzero in (\ref{eq:DeltaA}).
  Show that $\Delta A = 0$ implies
  \[
    t_1 = - \frac{L_2 t_2}{L_1}.
  \]
\item Now substitute this into (\ref{eq:DeltaA}), and from $\Delta L =
  0$, argue $R_1 = R_2$.
\item Extend this argument to show that $R_1 = R_2 = R_3 = \cdots = R_k$.
\end{enumerate}
\vspace{0pt}
\end{mybox}
\vspace{5pt}

\begin{figure}[h]
  \centering
  \includegraphics[scale=0.2]{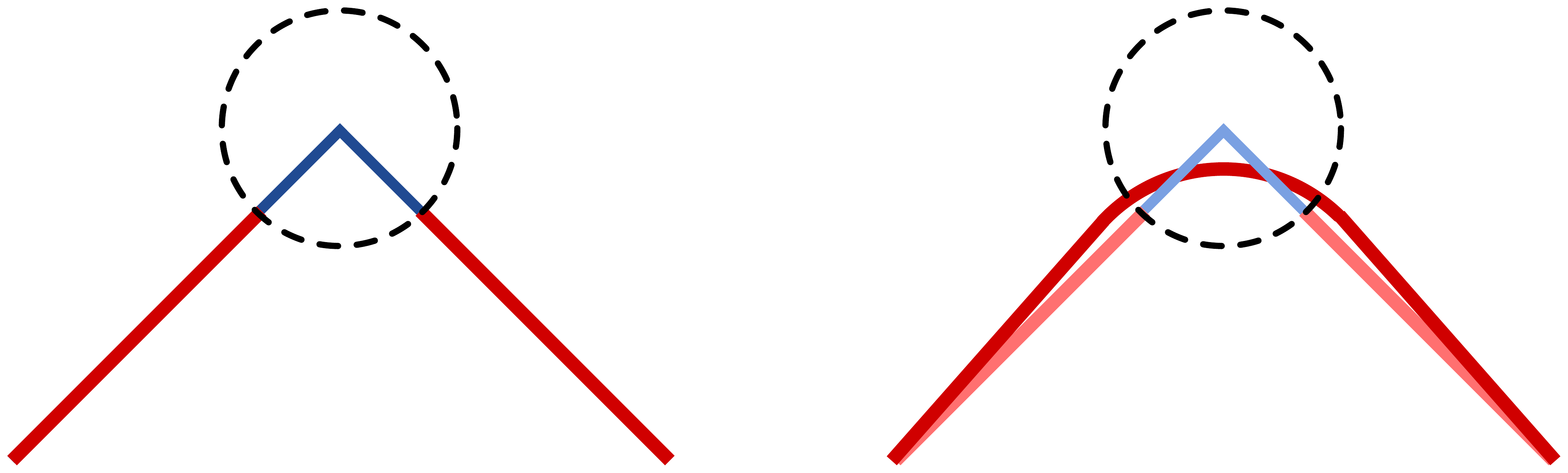}
\vspace{-5pt}
 \caption{Replacing a kink by a smooth edge which encloses the same area.
}
  \label{fig:kink}
\end{figure}

Does constant radius of curvature mean we have a circle?
Not necessarily.
You could join arcs of the same circle with a ``kink''.
But we can always approximate a kink as closely as we like by a smooth
edge which encloses the same area (Fig. \ref{fig:kink}).
This edge will have a \emph{different} radius of curvature, which
contradicts our argument!
The only smooth, closed curve we can draw, which has the same radius
of curvature $R$ at every point, is the circle of radius $R$ itself.
This more or less proves the isoperimetric
theorem.\footnote{Technically, we have only shown that \emph{if} there is a
  perimeter-minimizing shape of fixed area, it is a circle. But our
  approximation strategy can also be turned into a proof that the
  circle does minimize.}

\vspace{0pt}

So much for a single bubble.
The next simplest problems involve two and three bubbles of equal
area $A$.
The standard \emph{double bubble} (Fig. \ref{fig:2dconfigs} (middle))
and \emph{tripple bubble} (Fig. \ref{fig:2dconfigs} (right))
configurations are drawn below.
Since we can now introduce air pockets and splitting, as in
Fig. \ref{fig:2drules} for two bubbles, it is much harder to show
these simple arrangements are minimal.
The double bubble was only shown to be minimal in 1993 \cite{foisy1993}, and the triple
bubble in 2002 \cite{Wichiramala}.
As far as I know, no other finite planar bubble configurations are
known to be minimal.

\vspace{0pt}

\vspace{5pt}
\begin{figure}[h]
  \centering
  \includegraphics[scale=0.2]{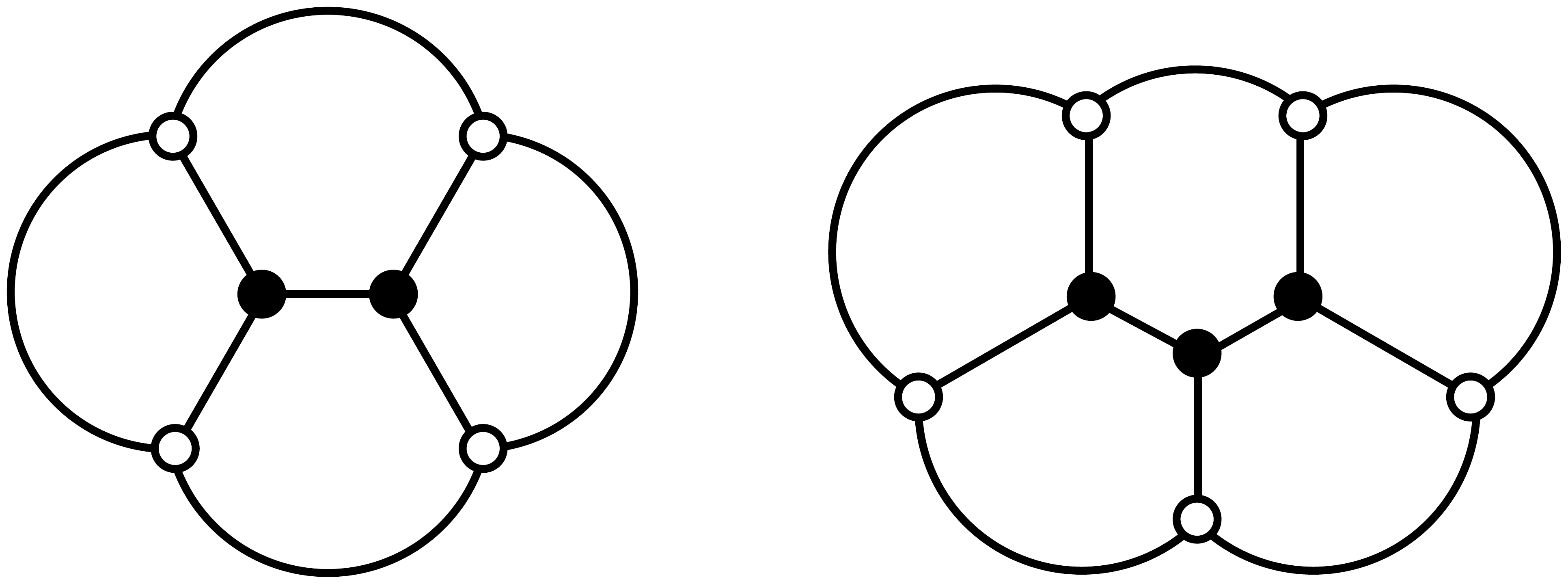}
\vspace{-5pt}
 \caption{Tinkertoys giving rise to bubbletoys.
}
  \label{fig:bubbletoy}
\end{figure}

Of course, we might say: forget mathematics, and let Nature be our
guide.
By carrying out simple experiments, we should be able to see which
configurations are predicted by physics.
Right?
Unfortunately, the same combinatorial explosion that plagued soap
bubble computers in \S \ref{sec:tinkertoys} afflicts planar bubble
configurations.
One argument is that every tinkertoy gives rise to a bubble
configuration, simply by adding arcs to the outside
as illustrated in Fig. \ref{fig:bubbletoy}.\footnote{More generally, we will have to bend the inner walls to make the sure the
pressure difference is balanced by tension. See Exercise \ref{ex:double} for
more details.
}
I call these \emph{bubbletoys}.
Incidentally, the two pictured bubbletoys are conjectured to be the
minimal planar configurations for four and five equal-area bubbles.
This suggests that solving the planar bubble problem is
\textsf{NP-hard}, and even finding a bubble configuration which
encloses the volumes $A_1, A_2, \ldots, A_n$ is
\textsf{NP-complete}, or possibly \textsf{NP-hard} as well.\footnote{
I haven't been able to find this statement in the literature, and
would be grateful if anyone could point me in the right direction. 
But it seems much harder than minimal networks! Unlike tinkertoys, where the problem could be easier when we stiched
together small tinkertoys, here, there are no external nodes so we
only have \emph{large} tinkertoys.
And finding the minimal configuration is \emph{much, much harder},
since we not only have an exponential number of tinkertoys, but an
infinite set of configurations that arise from splitting and empty
pockets.
No wonder we know almost nothing about bubbles!
}
Nature will take increasingly long times to converge on her
``conjectures'', solve the wrong problem, or both.
Either way, we cannot get physics to magically solve our
\textsf{NP-complete} problems for us!

\vspace{10pt}
\begin{mybox}
  \begin{exercise}
    \emph{Bubble radii and pressure cocycles.} \label{ex:double} \Mountain
  \end{exercise}
  We show the standard double and triple bubble for bubbles of
  different radii below.
  \vspace{-3pt}
  \begin{center}
    \includegraphics[scale=0.22]{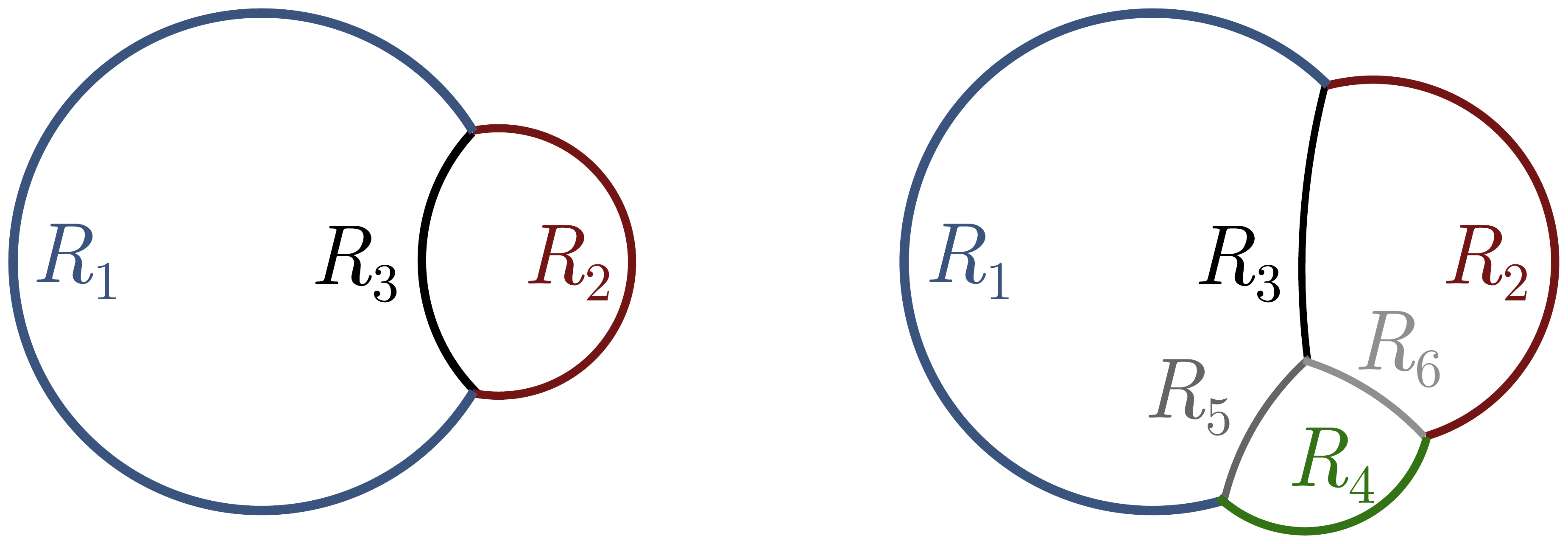}
  \end{center}
  This exercise uses physics to simply relate the bubble radii!
(There are also derivations from the $120^\circ$ rule, but they are much messier \cite{glassner}.)
  \begin{enumerate}[label=(\alph*), itemsep=0pt]
  \item Let's start with the double bubble. By considering pressure
    differences as across interfaces, explain why
    \begin{equation}
      \frac{1}{R_2} = \frac{1}{R_1} + \frac{1}{R_3}.\label{eq:cocy1}
    \end{equation}
    \emph{Hint.} Use Exercise \ref{ex:walls}(b).
 \item We can make this observation more general.
  Consider moving around a loop on a bubble network.
  Across each interface, there are pressure differences $\Delta P_1,
  \Delta P_2, \ldots, \Delta P_n$. 
  Show that, along the loop,
    \[
      \Delta P_1 + \Delta P_2 + \cdots \Delta P_n = 0.
    \]
    This is called the \emph{pressure cocycle condition} in the
    mathematics literature.
  \item Using the pressure cocycle condition for the triple bubble,
    calculate that in addition to (\ref{eq:cocy1}), we have
    \begin{align*}
      \frac{1}{R_4} = \frac{1}{R_1}+\frac{1}{R_5} = \frac{1}{R_2}+\frac{1}{R_6}.
    \end{align*}
    Check that the results of executing a loop around the inner
    junction are consistent with these relations.
  \end{enumerate}
\vspace{0pt}
\end{mybox}

\newpage

\section{Bubbles in three dimensions}
\label{sec:plateaus-laws}

So far, we've only considered two-dimensional networks, while the real
world has three dimensions.
Thankfully, removing the plexiglass changes less than you might expect!
Let's start by summarizing what we know about bubble networks.
The key result from \S \ref{sec:triangles} was the $120^\circ$ rule.
In \S \ref{sec:soap-bubbles}, 
we
learned that bubble walls are straight or arcs of circles, so that
they have constant radius of curvature (Exercise \ref{ex:walls}).
We also discovered from our treatment of the isoperimetric problem
that perimeter-minimizing wall do not have ``kinks''.
We can encode these insights as ``laws'' for bubble networks:

\vspace{10pt}
\begin{mybox2}
  \begin{statement}
    \emph{Bubble network laws I.} \label{box:bubble}
  \end{statement}
  \begin{enumerate}[itemsep=0pt]
 \item \emph{No kinks.} Edges are smooth, i.e. no vertices attached to
   one or two edges.
  \item \emph{Constant curvature.} Edges have constant radius of curvature.
  \item \emph{The $120^\circ$ rule.} Three edges meet at a junction, separated by $120^\circ$.
  \end{enumerate}
\vspace{-6pt}
\end{mybox2}
\vspace{5pt}

There is another way to motivate the $120^\circ$ rule that will prove
very useful in three dimensions.
The law forbidding kinks means that the fewest edges that can
meet at a junction is three.
Moreover, meeting at angles of $120^\circ$ is the most
\emph{symmetric} way for incoming edges to be separated.
Way back in \S \ref{sec:equil-triangl}, we saw that symmetry had an
important role to play in minimizing the length of the network on an
equilateral triangle, so perhaps it's unsurprising that the two are
connected here.
We call this the \emph{minsym principle}.
It lets us reformulate our network laws in a slightly different way:

\vspace{10pt}
\begin{mybox2}
  \begin{statement}
    \emph{Bubble network laws II.}
  \end{statement}
  \begin{enumerate}[itemsep=0pt]
 \item \emph{No kinks.} Edges are smooth, i.e. no vertices attached to
   one or two edges.
  \item \emph{Constant curvature.} Edges have constant radius of curvature.
  \item[$\bar{3}$.] \emph{Minsym.} At a junction, the minimal number of edges
    meet symmetrically.
  \end{enumerate}
\vspace{0pt}
\end{mybox2}

\noindent Generalizing to three dimensions is now ``easy''!

\subsection{Mean curvature}
\label{sec:mean}

Viewed through a dimensional lens, a planar bubble network is a set of
\emph{two-dimensional} cells separated by \emph{one-dimensional} bubble
walls. 
But when bubbles can roam around in three dimensions, the cells are
\emph{three-dimensional} volumes separated by \emph{two-dimensional} walls.
Although it seems like a whole differen kettle of fish,
three-dimensional bubbles are governed by almost exactly the same laws
as their planar counterparts.
The three-dimensional laws are called \emph{Plateau's laws}, after the
Belgian physicist \textsc{Joseph Plateau} (1801--1883) who guessed
them by assiduously observing bubbles \cite{plateau}.

``No kinks'' seems straightforward: bubble walls are smooth and cannot
suddenly terminate.\footnote{Unless there is something for them to end
  on, e.g. a bubble blower. We'll return to this problem below.}
But there are subtleties for the remaining two rules.
In a bubble network, edges have constant radius of curvature.
What is the analogous statement for surfaces?
It turns out in three and more dimensions, the notion of the curvature
of a surface is not unique, and different definitions are useful for
different applications.
For our purposes, the relevant notion is \emph{constant mean
  curvature}. This is a technical notion, and requires a bit more explanation.

\vspace{5pt}
\begin{figure}[h]
  \centering
  \includegraphics[scale=0.28]{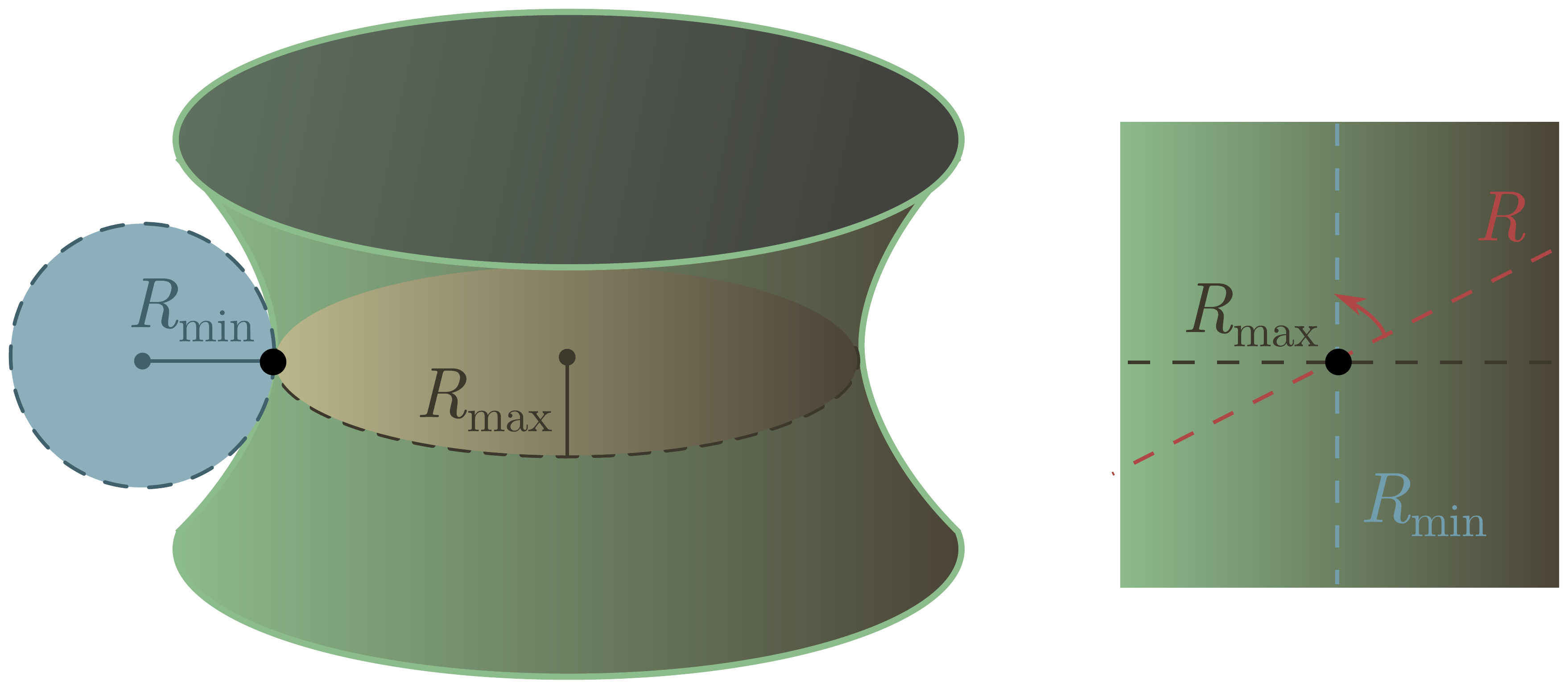}
\vspace{-5pt}
 \caption{\emph{Left.} A surface, with principal ``snug''
   circles. \emph{Right.} Straight slices through a point.}
  \label{fig:mean}
\end{figure}

Suppose we have a two-dimensional surface like the one in
Fig. \ref{fig:mean} (left).
If we take various straight slices through the black point
(shown in Fig. \ref{fig:mean} (right)), each will give rise to a
radius
of curvature, i.e. the radius of a circle which fits ``snugly'' onto
the curve at that point, and which is
perpendicular to the surface. 
As we rotate the red slice in Fig. \ref{fig:mean} (right), the
radius of curvature $R$ will vary, producing a maximum value
$R_\text{max}$ and minimum value $R_\text{min}$.
The reciprocals $1/R_\text{max}$ and $1/R_\text{min}$ are called the \emph{principal curvatures}.
Note that a radius curvature can be \emph{negative} if it is outside
the surface, as in Fig. \ref{fig:mean} (left).\footnote{Sometimes,
  outside and inside aren't well-defined, so you just make an
  aribtrary choice, and attach a minus sign to any circles which are
  outside. The sign of curvature depends on this choice.}

The \emph{mean curvature} $H$ is defined as the sum of
principal curvatures:
\begin{equation}
  \label{eq:11}
  H = \frac{1}{R_\text{max}} + \frac{1}{R_\text{min}}.
\end{equation}
A \emph{constant mean curvature (CMC)} surface is one where the mean
curvature $H$ is the same everywhere on the surface.
Notice that, as in Fig. \ref{fig:mean} (right), it is always
the case that the principle curvatures $R_\text{max}$ and
$R_\text{min}$ are measured along orthogonal slices.
We call this the \emph{orthogonal circle
  theorem}.\footnote{Unfortunately, it would take us too far afield to
  prove it here.}
To generalize the constant curvature rule from planar bubble
networks, we take bubble surfaces to be CMC.
You can explore some of the physics behind this in Exercise \ref{ex:yl2}.

\vspace{10pt}
\begin{mybox}
  \begin{exercise}
    \emph{Spheres are CMC.} \label{ex:sphere}
  \end{exercise}
Show that a sphere of radius $R$ has constant mean curvature
    $H = 2/R$.
\emph{Hint.} The slice normal to the sphere at any point is a great circle.
\end{mybox}
\vspace{5pt}

\begin{mybox}
  \begin{exercise}
    \emph{Young-Laplace II.} \label{ex:yl2} \Mountain
  \end{exercise}
In Exercise \ref{ex:walls}, we saw the Young-Laplace
    law (\ref{eq:yll}) for a bubble wall:
    \[
      \Delta P = \frac{2\sigma}{R},
    \]
    for $\Delta P = P_\text{out} - P_\text{in}$ and $R$ the radius of
    curvature of the wall.
  \begin{enumerate}[label=(\alph*), itemsep=0pt]
  \item Viewing a bubble wall as a surface, argue that $1/R_\text{max} = 0$.
  \item Using the orthogonal circle theorem, deduce that $H = 1/R$ is the
    mean curvature of the wall.
    Hence, the Young-Laplace law can be written
    \begin{equation}
    \Delta P = 2\sigma H.\label{eq:yll2}
   \end{equation}
    This turns out to be the correct form for an arbitrary surface!
  \item If we dip two identical circular bubble blowers in soap film (red below), the
    surface that results is typically like the one below left, rather
    than a cylinder:
    \begin{center}
      \includegraphics[scale=0.15]{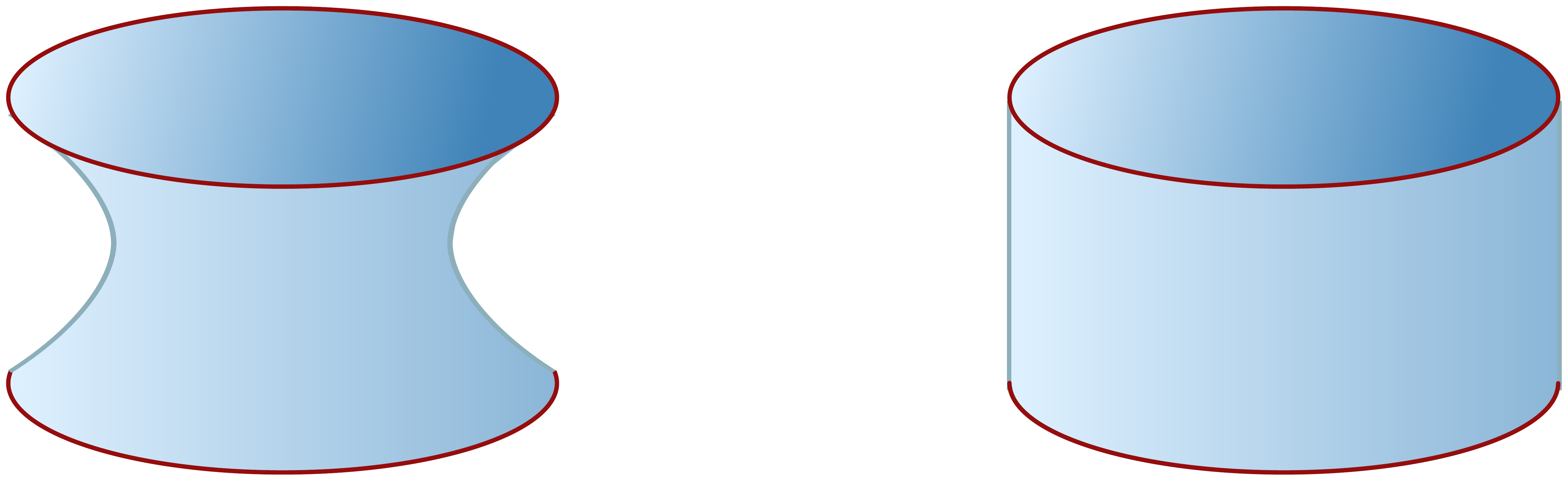}
    \end{center}
    Give a qualitative explanation, using (\ref{eq:yll2}) and mean
    curvature.
  \item Using Exercise \ref{ex:sphere}, what is the \emph{smallest} spherical bubble that can form in the
    atmosphere?
    Atmospheric pressure is $P = 10^5 \text{ N/m}^2$ and the
    surface tension of soapy water is $\sigma = 7 \times 10^{-2}
    \text{ N/m}$.
    Can a spherical bubble form in space?
  \end{enumerate}
\vspace{0pt}
\end{mybox}

\subsection{Plateau's laws}
\label{sec:laws}

\begin{figure}[h]
  \centering
  \includegraphics[scale=0.3]{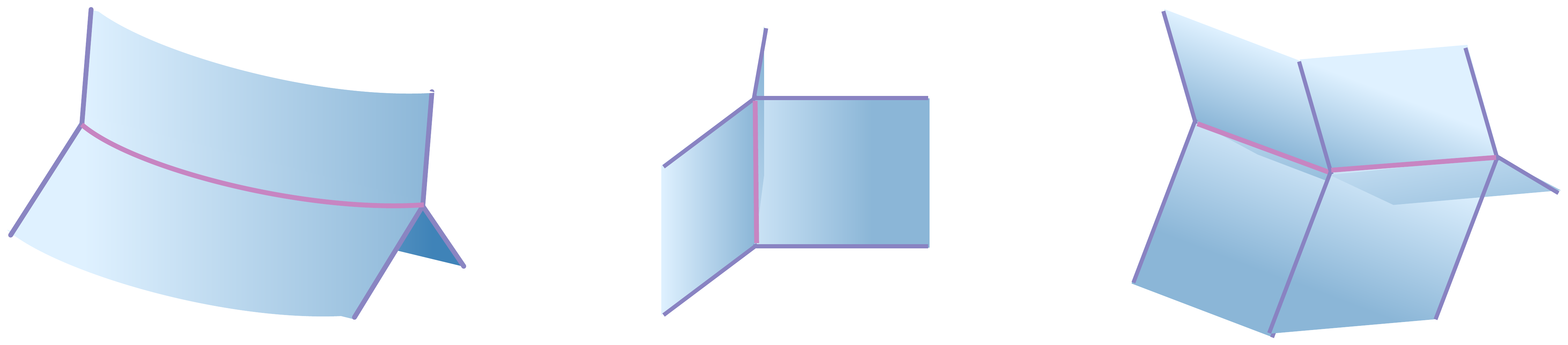}
\vspace{-5pt}
 \caption{\emph{Left.} Three faces meeting at a border. \emph{Middle.} Vertical bubble walls. \emph{Right.} A kink.}
  \label{fig:3faces}
\end{figure}

Finally, we have to generalize the ``minsym'' principle to three dimensions.
The ``no kink'' requirement means that we cannot have two walls meeting
at an angle, since that would introduce a kink, and if there is no
angle, they may as well be count as part of the same wall.
Thus, we must have at least three walls meet at any junction of walls.
According to the minsym principle, precisely three
faces should meet (minimum) separated by $120^\circ$ (symmetry), as in
Fig. \ref{fig:3faces} (left).
In fact, this is exactly what we need to get the $120^\circ$
rule in a planar bubble network, since the walls are secretly
two-dimensional and vertical oriented between the plexiglass plates
(Fig. \ref{fig:3faces} (middle)).

The edge along which three bubble walls meet is called a \emph{Plateau
  border}.
These borders themselves can intersect!
Minsym requires us to figure out the minimum number to avoid kinks,
and the most symmetric arrangement thereof.
Clearly, we need at least three, since otherwise we can arrange a
junction of three walls with a kink, as in Fig. \ref{fig:3faces}
(right).

\vspace{5pt}

\begin{figure}[h]
  \centering
  \includegraphics[scale=0.5]{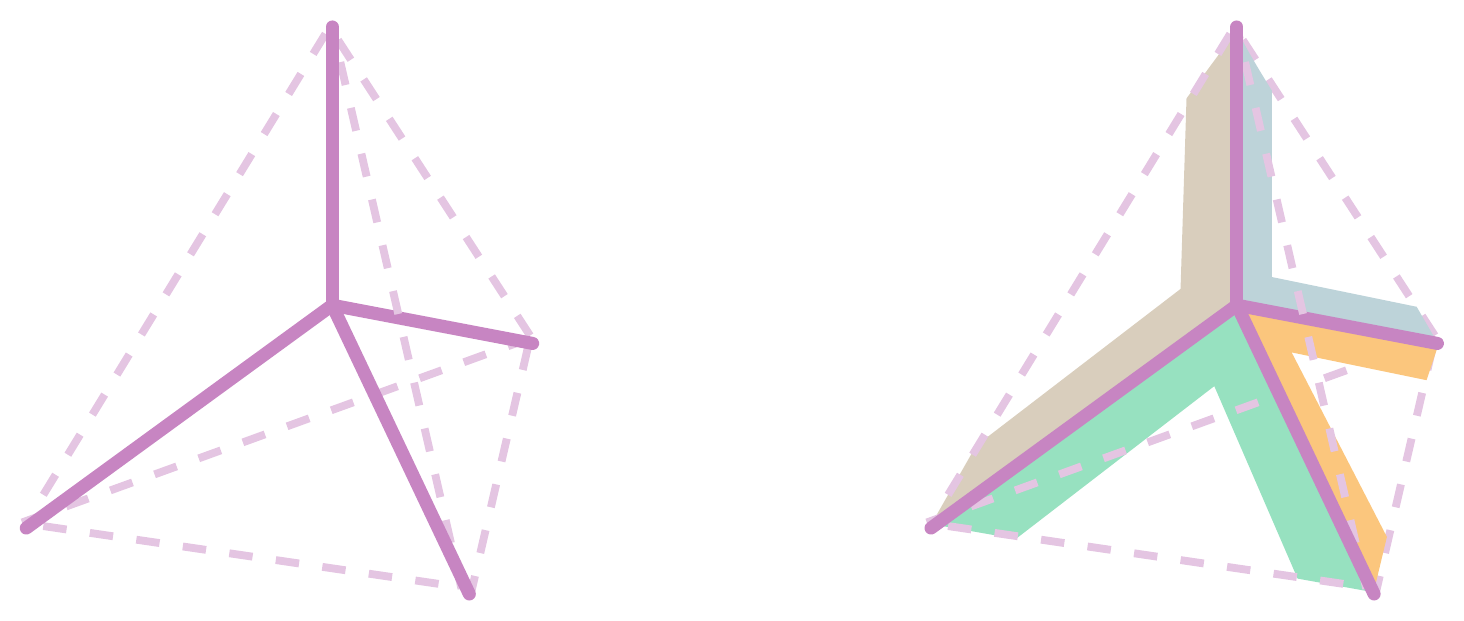}
\vspace{-5pt}
 \caption{
\emph{Left.} Four Plateau borders meeting tetrahedrally.
\emph{Right.} Smoothly connected walls.}
  \label{fig:tetra}
\end{figure}

Can we have exactly three?
It's not hard to see that the three sets of three faces cannot be
connected smoothly, simply because we have an odd number of faces!
You can check the details in Exercise \ref{ex:plat}.
This exercise also shows that it \emph{is} possible to connect the
faces smoothly for four sets of Plateau borders.
Thus, the minsym principle suggests that \emph{precisely} four Plateau
borders should meet in the most symmetric arrangement.
Symmetry is maximized by shooting out the Plateau borders \emph{tetrahedrally},
i.e. from the center towards the corners of a regular tetrahedron (Fig. \ref{fig:tetra}).
If you like vectors, you can play around with the geometry in Exercise \ref{ex:simplex}.

\vspace{5pt}

Having defined constant mean curvature, and worked out the
implications of the minsym principle in three dimensions, we are
finally in a position to state the laws Plateau discovered \cite{plateau}:

\vspace{10pt}
\begin{mybox2}
  \begin{statement}
    \emph{Plateau's laws.}
  \end{statement}
  \begin{enumerate}[itemsep=0pt]
 \item \emph{No kinks.} The faces in a soap film are smooth. 
  \item \emph{Constant curvature.} Any face has constant mean curvature. 
  \item \emph{Minsym I.} Three faces always meet at a Plateau border, separated
    by $120^\circ$.
  \item \emph{Minsym II.} Plateau borders always meet tetrahedrally at a vertex.
  \end{enumerate}
\vspace{-6pt}
\end{mybox2}
\vspace{5pt}

\noindent These are empirical observations about bubbles.
While the constant curvature condition follows from the Young-Laplace
law (Exercise \ref{ex:yl2}), and Minsym I from the $120^\circ$ rule,
it is not at all obvious that a tetrahedral arrangement of Plateau
borders minimizes area.
Minimizing subject to what constraints?
(Feel free to have a guess now.)
Is every configuration satisfying Plateau's laws a local minimum,
subject to these constraints?
And does every such locally minimal solution satisfy Plateau's laws?
(These are harder to figure out without a doctorate in math.)
Read on to find out!

\vspace{10pt}
\begin{mybox}
  \begin{exercise}
    \emph{Plateau borders.} \label{ex:plat}
  \end{exercise}
Let's check we need four Plateau borders in order to
smoothly connect walls.
Consider some number of Plateau borders meeting at a vertex.
    Each border has three associated bubble walls.
  \begin{enumerate}[label=(\alph*), itemsep=0pt]
  \item 
    Explain why ``no kinks'' requires each wall to connect smoothly to
    another. 
  \item Argue that this is impossible for an odd number of
    borders meeting at a node.
  \item Show explicitly it is possible for four Plateau borders to smoothly
    connect.
   \emph{Hint.} Add the two remaining walls in Fig. \ref{fig:tetra} (right).
  \end{enumerate}
\vspace{0pt}
\end{mybox}

\vspace{10pt}

\begin{mybox}
  \begin{exercise}
    \emph{Simplices.} \label{ex:simplex} \VarIceMountain
  \end{exercise}
  The equilateral triangle and the tetrahedron are part of a family of
  symmetric shapes called \emph{simplices}.
  We can describe them using vector analysis.
  \begin{center}
    \includegraphics[scale=0.5]{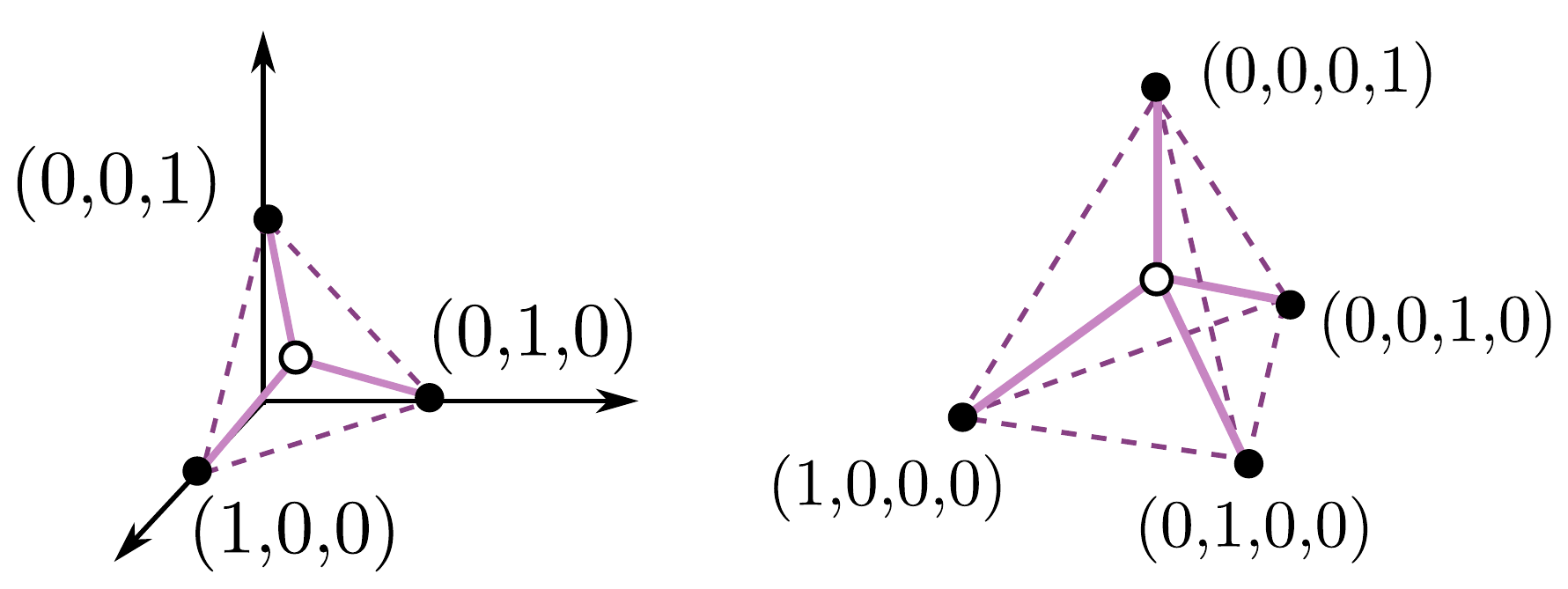}
  \end{center}
  \begin{enumerate}[label=(\alph*), itemsep=0pt]
  \item We can embed the vertices of an equilateral triangle in three dimensions as
    \[
      \Delta_3 = \{(1, 0, 0), (0, 1, 0), (0, 0, 1)\}.
    \]
    Why is this a maximally symmetric arrangement of three points?
  \item The \emph{center} of the triangle is just the average of the
    vertices.
    Show that the vectors from center to vertices have length
    $\sqrt{2}$ and are given by
    \[
      V_3 = \left\{\tfrac{1}{3}(-2, 1, 1), \tfrac{1}{3}(1,-2, 1), \tfrac{1}{3}(1, 1, -2)\right\}.
    \]
  \item 
    Using the formula
    \[
      \theta = \cos^{-1}\left(\frac{\mathbf{u}\cdot \mathbf{v}}{|\mathbf{u}||\mathbf{v}|}\right) ,
    \]
    check that the vectors in $V_3$ make angle $120^\circ =
    \cos^{-1} (-1/2)$ with each other.
  \item We can embed the tetrahedron in \emph{four} dimensions as
    \[
      \Delta_4 = \{(1, 0, 0,0), (0, 1, 0,0), (0, 0, 1,0), (0,0,0,1)\}.
    \]
    Show that the vectors from center to vertex have length
    $\sqrt{3}$, and make angles
    \[
      \theta = \cos^{-1}\left(-\frac{1}{3}\right) \approx 109.5^\circ.
    \]
  \end{enumerate}
The tetrahedron is just a higher-dimensional version of an equilateral triangle!
    We can continue in this fashion, defining the
    \emph{$n$-simplex} $\Delta_n$ as a maximally symmetric arrangement
    of $n$ points in $n$ dimensions:
    \[
      \Delta_n = \{(1, 0, \ldots, 0), (0, 1, \ldots, 0), \ldots, (0, \ldots, 0, 1)\}.
    \]
  \begin{enumerate}[itemsep=0pt]
  \item[(e)] Check that the distance from the center of $\Delta_n$ to each vertex is $\sqrt{n}$, and
    that any two such vectors make an angle
    \[
    \theta = \cos^{-1}\left(-\frac{1}{n}\right).
  \]
 As $n$ gets large, confirm these vectors are almost orthogonal.
\item[(f)] Extrapolate the minsym principle to higher-dimensional
  foams.
  In other words, if the universe has $n$ dimensions, make a guess at
  Plateau's laws.
 \end{enumerate}
\vspace{0pt}
\end{mybox}
\vspace{5pt}

\subsection{Bubbles and wireframes}
\label{sec:spheres-bubbletoys}

As you might have guessed, Plateau's laws are related to the
\emph{three-dimensional} version of the planar bubble configuration problem
outlined in Box \ref{box:planar}.
Instead of enclosing areas $A_1, A_2, \ldots, A_n$, we want
to enclose volumes $V_1, V_2, \ldots, V_n$ with a bubble film of
minimal surface area.
As before, we allow for empty pockets and split bubbles.
We state the optimization problem as follows:

\vspace{10pt}
\begin{mybox2}
  \begin{statement}
    \emph{The Minimal Bubble Problem.} \label{box:config}
  \end{statement}
  Find the connected bubble film of smallest area enclosing volumes $V_1,
  V_2, \ldots, V_n$, allowing air pockets and split bubbles.
\end{mybox2}
\vspace{5pt}

\begin{figure}[h]
  \centering
  \includegraphics[scale=0.25]{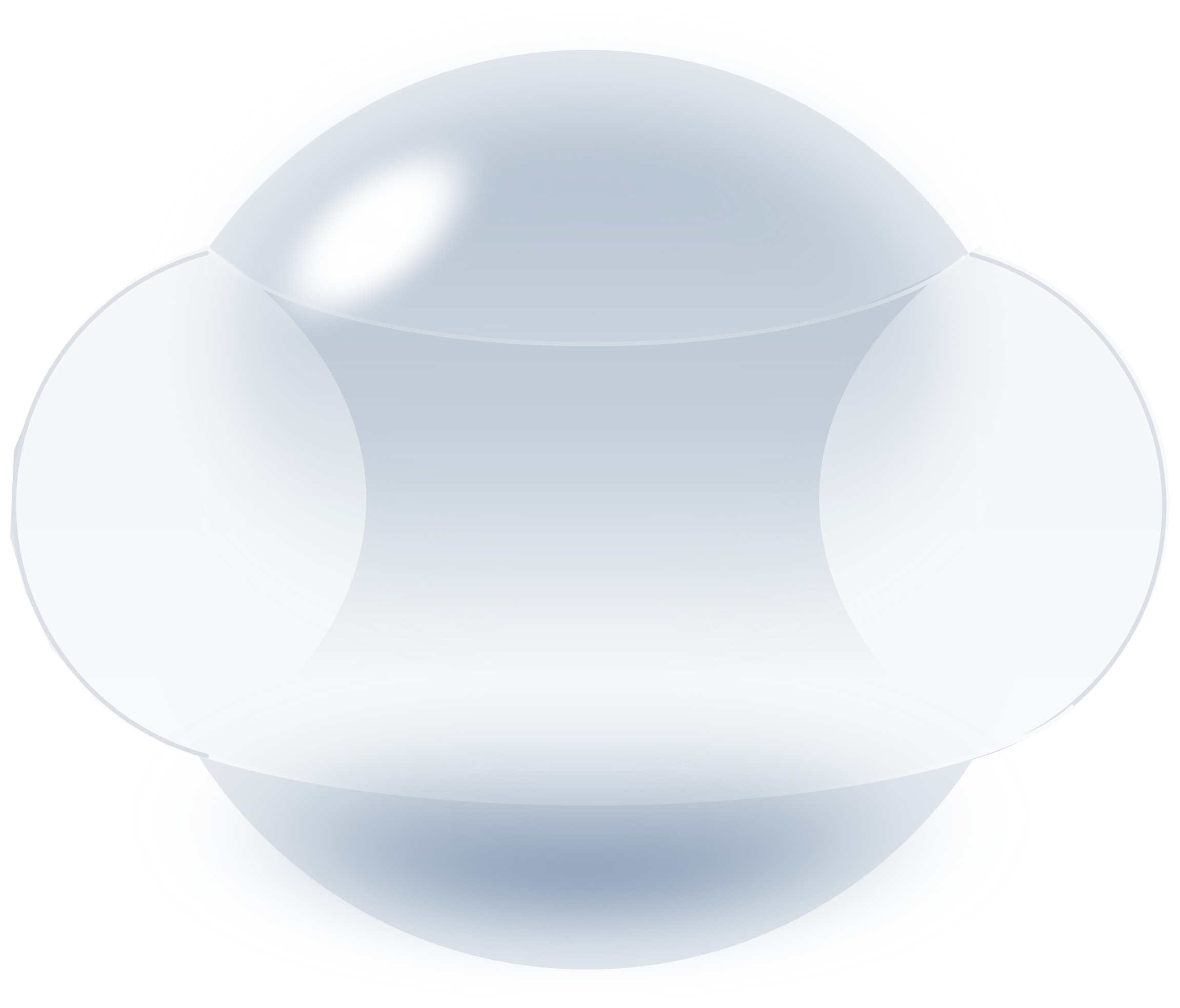}
\vspace{-5pt}
 \caption{An unstable double bubble, consisting of an apple core wrapped
   in a donut.}
  \label{fig:belt}
\end{figure}

\noindent Surfaces with bubbles of fixed volume, and which locally
minimize area, also satisfy Plateau's laws, as mathematican
\textsc{Jean Taylor} proved in her 1976 tour-de-force
\cite{taylor}.\footnote{I want to point out that the $120^\circ$ rule was
  more or less proved as soon the problem was
  stated. It took over 100 years for the teatrahedral rule to go from
  empirical observation to mathematical proof. It's much harder!}
The converse is not true, since we can find bubbles
satisfying Plateau's laws that are not stable (Fig. \ref{fig:belt}).

The planar bubble configuration problem (Box \ref{box:planar})
is a special case of the three-dimensional bubble configuration problem, where we put the foam between plates.
This implies that local minima satisfy the bubble network laws (Box
\ref{box:bubble}), since these are simply Plateau's laws in the case
where bubble walls are vertical, and because they are vertical, there are no
vertices at which Plateau borders intersect.
And since planar bubbles are hard, three-dimensional bubbles are hard!
Physically speaking, we expect that foams will take longer and longer
to converge to a minimum, or answer the wrong question, if we force them to compute for us.

\begin{figure}[h]
  \centering
  \includegraphics[scale=0.15]{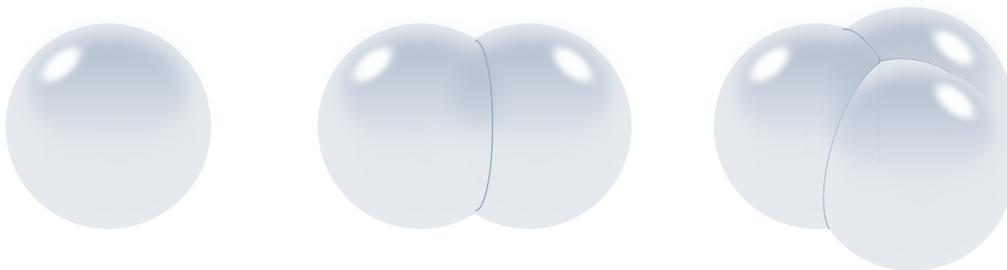}
\vspace{-5pt}
 \caption{\emph{Left.} A single spherical bubble. \emph{Middle.} The
   standard double bubble. \emph{Right.} The standard triple bubble.}
  \label{fig:bubbles}
\end{figure}

Even when Nature does make conjectures, they can be bewilderingly hard to
prove.
The simplest example is a single bubble of volume $V$.
Experience suggests that a lone bubble is always spherical, as in Fig. \ref{fig:bubbles} (left).
The corresponding conjecture is that the area-minimizing surface of
volume $V$ is a sphere.
This is the three-dimensional version of the isoperimetric theorem for circles in \S \ref{sec:circles-bubbletoys}.

\vspace{5pt}

The proof is remarkably similar.
We split the surface into many small patches of area $A_1, A_2,
\ldots, A_k$, with mean curvature $H_1, H_2, \ldots, H_k$.
If these areas are ``pushed out'' a distance $t_1,
t_2, \ldots, t_k$ normal to the surface, the volume and
area change as\footnote{It's not hard to see that volume changes this
  way, but area is the tricky one, and I won't prove it here.}
\begin{align*}
  \Delta V & = A_1 t_1 + A_2 t_2 + \cdots + A_k t_k \\
  \Delta A & = A_1 t_1 H_1 + A_2 t_2 H_2 + \cdots + A_k t_k H_k.
\end{align*}
If the wobbling preserves volume, then $\Delta V = 0$, and if area is
locally minimized, then $\Delta A = 0$ as before.
We can then repeat our argument word for word to conclude that $H_1 =
H_2 = \cdots = H_k$. Mean curvature is the same everywhere, and we
have a CMC surface!\footnote{This proof actually generalizes to higher
dimensions, where the mean curvature is $H = 1/R_1 + 1/R_2
+ \cdots + 1/R_n$ for $n$ mutually orthogonal principal curvatures.}

In the plane, there was exactly one way to have a smooth curve with
constant radius of curvature.
In three dimensions, there are all sorts of exotic CMC surfaces.
But it turns out that the sphere is the \emph{only} CMC surface that enclose
a finite volume,
as proved by
\textsc{Aleksandr Aleksandrov}
(1912--1999) in 1958 \cite{Aleksandrov}.\footnote{Well, almost. If the
  surface is allowed to intersect itself, there is an odd three-lobed
  donut called the \emph{Wente torus}, but this is something of an
  embarassment so we ignore it.}
The same argument we gave in Exercise \ref{ex:connex} shows that empty
pockets and splitting bubbles will not help.
Thus, we have proved the isoperimetric theorem in three dimensions: the sphere
is the surface of smallest area enclosing a given volume $V$.

The next simplest problem is two bubbles of equal volume $V$.
Again, Nature seems to prefer the ``standard double bubble'', with two
spheres fused at a single Plateau border
(Fig. \ref{fig:bubbles} (middle)) over its non-standard competitors.
One of these competitors is the ``donut'' configuration, where a single
bubble is squeezed into the shape of an apple core by a donut-shaped
ring around the outside (Fig. \ref{fig:belt}).
This bubble satisfies Plateau's laws, but turns out to be unstable, and 
jiggling the donut will cause it to collapse into the 
standard double bubble \cite{Sullivan1999}. 
It wasn't until 2002 that the standard double bubble was shown to
be minimal \cite{hutchings2004proof}.
Similarly, we often observe the standard triple bubble for three cells
of volume $V$ (Fig. \ref{fig:bubbles} (right)).
No one knows if this is truly minimal, so it remains the \emph{triple
  bubble conjecture}.

\vspace{5pt}

If there is a three-dimensional 
bubble configuration problem, it stands to reason
there should be a three-dimensional minimal network problem. 
In the minimal network problem, Box \ref{ex:min-net}), we had to find
a network of shortest length connecting some set of fixed nodes. 
A node is a \emph{zero-dimensional} object---it has no extent at all!
If we raise the number of dimensions of the configurable object, going
from a one-dimensional graph to a two-dimensional surface, 
perhaps we should raise the dimensions of the fixed object, going from
fixed points to \emph{fixed curves}.
The suggests the following task: 

\vspace{10pt}
\begin{mybox2}
  \begin{statement}
    \emph{The Wireframe Problem.} \label{box:plateau-prob}
  \end{statement}
  Given some fixed curves $C_1, C_2, \ldots, C_n$ in three-dimensional
  space, find a soap film of minimal area that connects them.
\vspace{5pt}
\end{mybox2}
\vspace{5pt}

\begin{figure}[h]
  \centering
  \includegraphics[scale=0.2]{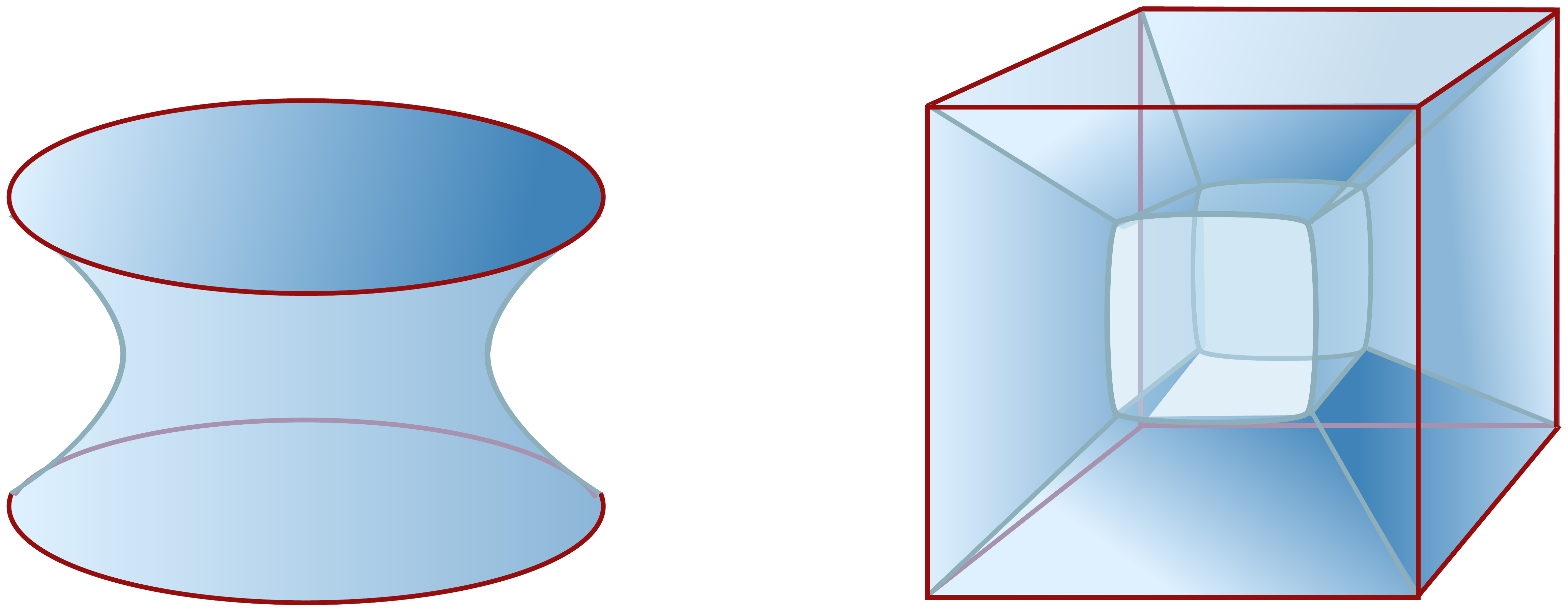}
\vspace{-5pt}
 \caption{\emph{Left.} The catenoid, a surface with mean curvature
   zero, which forms between identical wireframe rings. \emph{Right.} The
   tesseract formed from a wireframe cube.}
  \label{fig:wires}
\end{figure}

These fixed curves are called \emph{wireframes}, since physically
speaking, we can implement them with twisted pieces of wire.
Dunking wire into soapy water gives soap bubbles something to hold
onto, and as with plexiglass and screws, we have an analogue computer
to solve our problem for us. 
We give two very beautiful examples in Fig. \ref{fig:wires}: the
\emph{catenoid}, a surface forming between two rings, and the
\emph{tesseract} formed
when we dip a wireframe cube.
The cube creates a second, slightly puffed out\footnote{To ensure
  borders meet tetrahedrally.} inner cube, and then connects
corresponding corners with Plateau borders.

\vspace{5pt}

Since this generalizes the minimal network problem (the screws are a
particularly boring type of wireframe), the wireframe problem is \textsf{NP-hard}.
We can dip some arbitrarily complicated piece of wire into the soap,
but when we pull it out, it may take a very long time---longer than
the age of the universe in some cases---for the soap bubbles to
converge on a stable solution.
Or it will solve a different problem altogether.
This is yet another physical prediction!

But it's not obvious there is a solution at all.
If we dunk some random wireframe into the water, the bubble film that
connects them must satisfy Plateau's laws, except along the wire
itself, in the same way that minimal networks satisfy the $120^\circ$
rule at a hub but not at a fixed node.
But do Plateau's laws always allow a solution?
Perhaps we can defeat Nature by giving it some wacky curve it cannot
connect with soap film.
The intuitive physical argument is that we can simply dip our
wireframe in and see what comes out. But as we've just argued, for a
complicated enough problem, it may take a very, very, very long time to converge.
And I find an argument less convincing if I am guaranteed to die before it
successfully terminates!

The question of the \emph{existence} of a solution to the wireframe
task is called \emph{Plateau's problem}.
In the 1930s, mathematicians \textsc{Jesse Douglas} (1897--1965) and
\textsc{Tibor Radó} (1895--1965) independently showed these solutions
always exist \cite{douglas, rado}.
Even if it takes longer than the lifetime of universe, Nature will
eventually get there. 

\newpage
\subsection{Space-filling foams}
\label{sec:infinite-foam}



In this final section, we'll consider the \emph{three-dimensional
  honeycomb problem}, that is, how to optimally partition space into
equal-volume cells.
To minimize surface area per cell, the partition must satisfy Plateau's
laws.\footnote{Clearly the surface area is infinite, so how can it be
  minimal? Like I say, our goal is to minimize\emph{ average surface
  area per cell}. This obeys Plateau's laws because the laws are either
  \emph{local}, applying in the vicinty of a point (minsym I), line
  (minsym II), or face (no kinks).
  The CMC rule, on the other hand, can be derived at the level of a \emph{cell}.
}
The $120^\circ$ rule had dramatic consequences for large bubble
networks in the plane.
We will see that, in three dimensions, the tetrahedral law has
similar (if less dramatic) implications for the structure of three-dimensional foams.
To explore these, we first need to extend Euler's formula
(\ref{eq:euler}) to include multiple bubbles.
In a finite configuration of soap bubbles, let $N$ denote the number of vertices
(where Plateau borders join), $E$ the number of Plateau borders, $F$ the
number of bubble faces, and $C$ the number of
enclosed bubble cells.
As before, we will count the region outside the bubble configuration
as a cell as well.

Recall from \S \ref{sec:foams} that Euler's formula applies to
a polyhedron like the cube, possessing two cells: the inside and the outside.
We can divide up the internal cell 
by adding inner walls. 
If there are $C$ cells altogether, there are $C - 1$ internal cells,
and $C - 2$ ``extra'' internal cells compared to a regular polyhedron.
For each extra cell, we can remove an internal face so that two
neighbouring cells become one.
This leaves something we can flatten into a planar graph, with $F' = F - (C - 2)$
faces, and hence by Euler's formula
\[
N - E + F' = 2.
\]
Rearranging gives Euler's ``foamula'':
\begin{equation}
  \label{eq:foamula}
  N - E + F = 2 + (C-2) = C.
\end{equation}
Equation (\ref{eq:foamula}) is true for any polyhedron with
mutiple internal cells, whether or not it satisfies Plateau's laws.


The $120^\circ$ rule, tetrahedral rule, and foamula together show that
bubble faces tend to have \emph{less} than six sides.
More precisely, if $F_\text{avg}$ is the average number of faces per
bubble cell, $E_\text{avg}$ the average number of edges around the
boundary of a cell, and $e_\text{avg}$ the average number of edges per
face, you can show in Exercise \ref{ex:subhex} that
\begin{equation}
  \label{eq:foamula2}
F_\text{avg} = \frac{1}{3}E_\text{avg} + 2 = \frac{12}{6- e_\text{avg}}.
\end{equation}
Since $F_\text{avg}$ is positive, it follows that $e_\text{avg} < 6$,
so faces tend to be sub-hexagonal. 

\vspace{10pt}
\begin{mybox}
  \begin{exercise}
    \emph{Sub-hexagonal faces.} \label{ex:subhex}\Mountain
  \end{exercise}
    \begin{enumerate}[label=(\alph*), itemsep=0pt]
    \item From Plateau's fourth law and the handshake lemma (\ref{eq:handshake}), argue that $E = 2N$.
    \item Let $F_\text{avg}$ denote the average number of faces per
      cell and $E_\text{avg}$ the average number of edges per cell.
      Show that
      \[
        F_\text{avg} = \frac{2F}{C}, \quad E_\text{avg} =
        \frac{3E}{C}.
        \]
   \emph{Hint.} You may assume that, like in a bubble network, a face in a bubble foam always has
   different cells on either side.
 \item From (\ref{eq:foamula}), deduce the relation between average number of edges and faces:
   \[
     3F_\text{avg} - E_\text{avg} = 6.
\]
     This is analogous to the result $3F - E = 6$ for bubble networks.
   \item Let $e_\text{avg}$ be the average number of edges per face.
	Derive the relation
        \[
          F_\text{avg} = \frac{12}{6- e_\text{avg}}.
        \]
        \emph{Hint.} Write $E_\text{avg}$ in terms of $F_\text{avg}$
        and $e_\text{avg}$. Don't forget to handshake!
    \end{enumerate}
\end{mybox}
\vspace{5pt}

Since we are discussing \emph{averages}, they continue to make sense even if
the foam is infinite!
The simplest infinite foams are those in which each bubble is the
same, so the bubbles form a \emph{space-filling tessellation}.
This is the three-dimensional analogue of the plane tessellations we
saw in Fig. \ref{fig:tess}.
In our quest for infinite foams, we will first rule out simple
tessellations which simply extrude these plane tessellations into
three-dimensional prisms.

\vspace{10pt}
\begin{mybox}
  \begin{exercise}
    \emph{Prisms.} \label{ex:prism}
  \end{exercise}
    \begin{enumerate}[label=(\alph*), itemsep=0pt]
    \item One way to tessellate space is to take a regular tessellation of the
plane, then extend it in the perpendicular direction to form a layer
of prisms (as below).
We can then stack these layers on top of each other to tessellate space.
\begin{center}
  \includegraphics[scale=0.3]{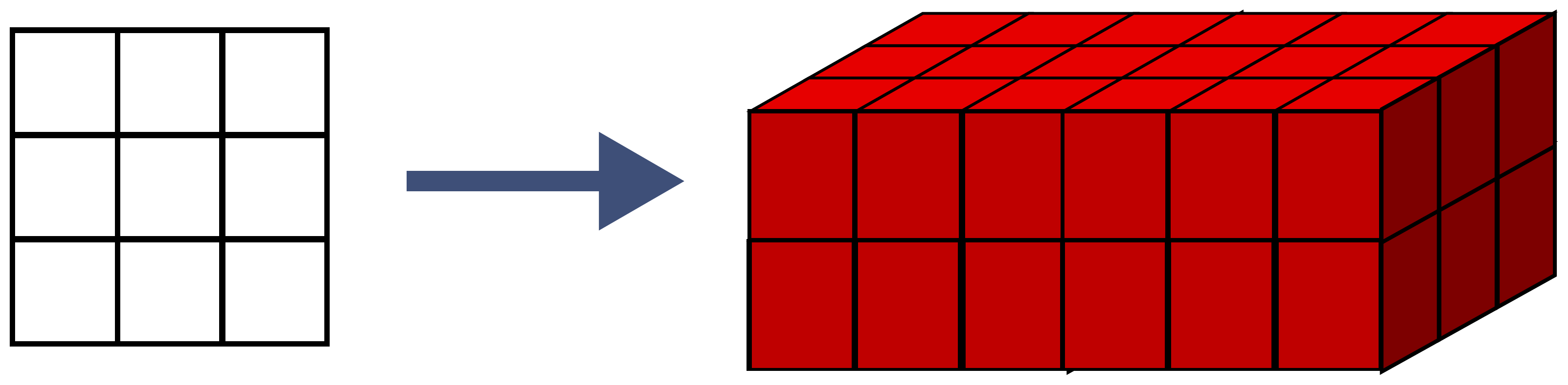}
\end{center}
Explain why no prism-based tessellation satisfies Plateau's laws.
\item  The \emph{gyrobifastigium} is made from two triangular prisms
  joined with a twist at their bases. All faces of the solid are
  regular polygons.
  Like prisms, you can arrange gyrobifastigia into layers, and stack
  layers
to fill space:
\vspace{-5pt}
\begin{center}
  \includegraphics[scale=0.25]{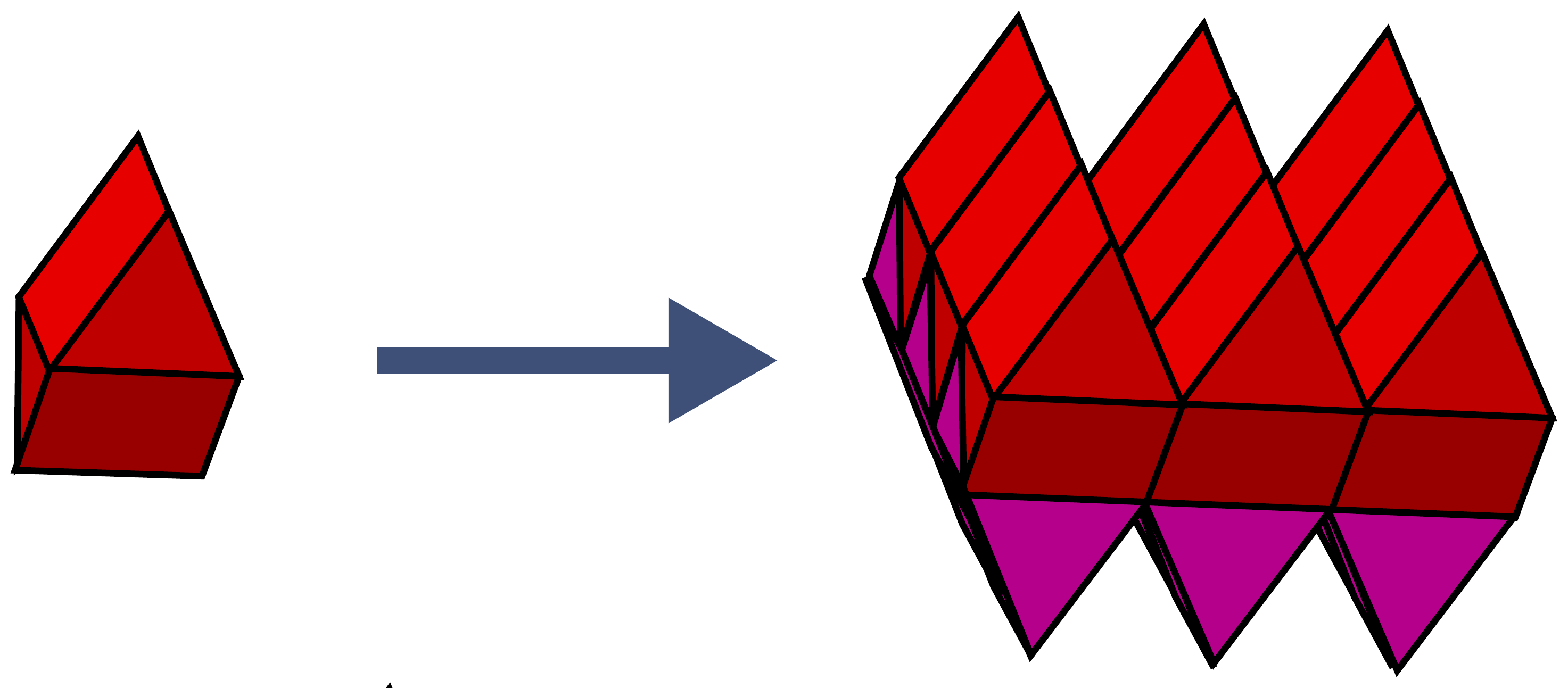}
\end{center}
\vspace{-5pt}
  Does this tessellation satisfy Plateau's laws?
\end{enumerate}
\vspace{0pt}
\end{mybox}
\vspace{5pt}

\begin{figure}[h]
  \centering
  \includegraphics[scale=0.4]{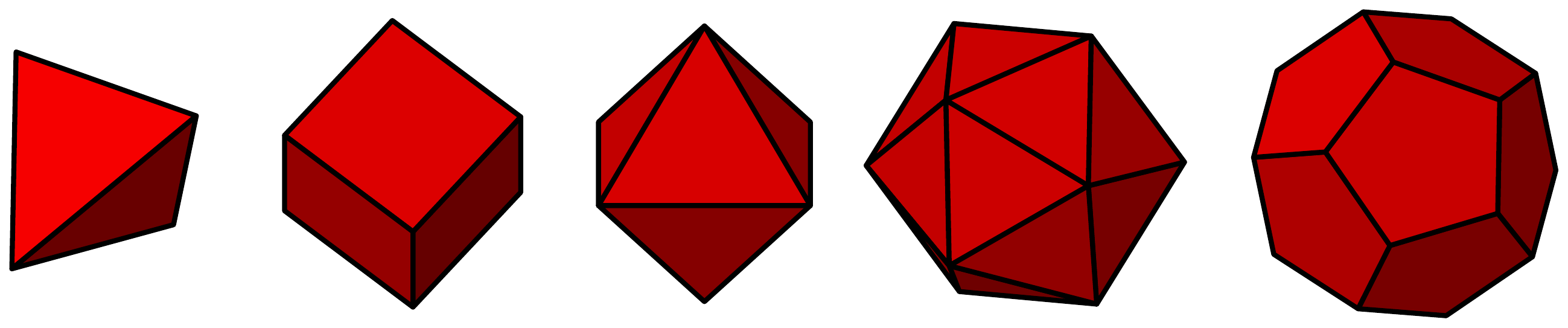}
\vspace{-5pt}
 \caption{
From left to right: tetrahedron, cube, octahedron, icosahedron, dodecahedon.}
  \label{fig:platonic}
\end{figure}

Our next step is to consider the analogue of
regular polygons, the \emph{Platonic solids} (Fig. \ref{fig:platonic}).
These are polyhedra whose faces are identical regular polygons.
Could any of these describe an infinite soap foam?
Of these solids, only the cube can tessellate space by itself, but
since this is a prism-based tessellation (it is an extruded rectangular tiling), Exercise \ref{ex:prism} rules it out.
We need to work harder!

\begin{figure}[h]
  \centering
  \includegraphics[scale=0.5]{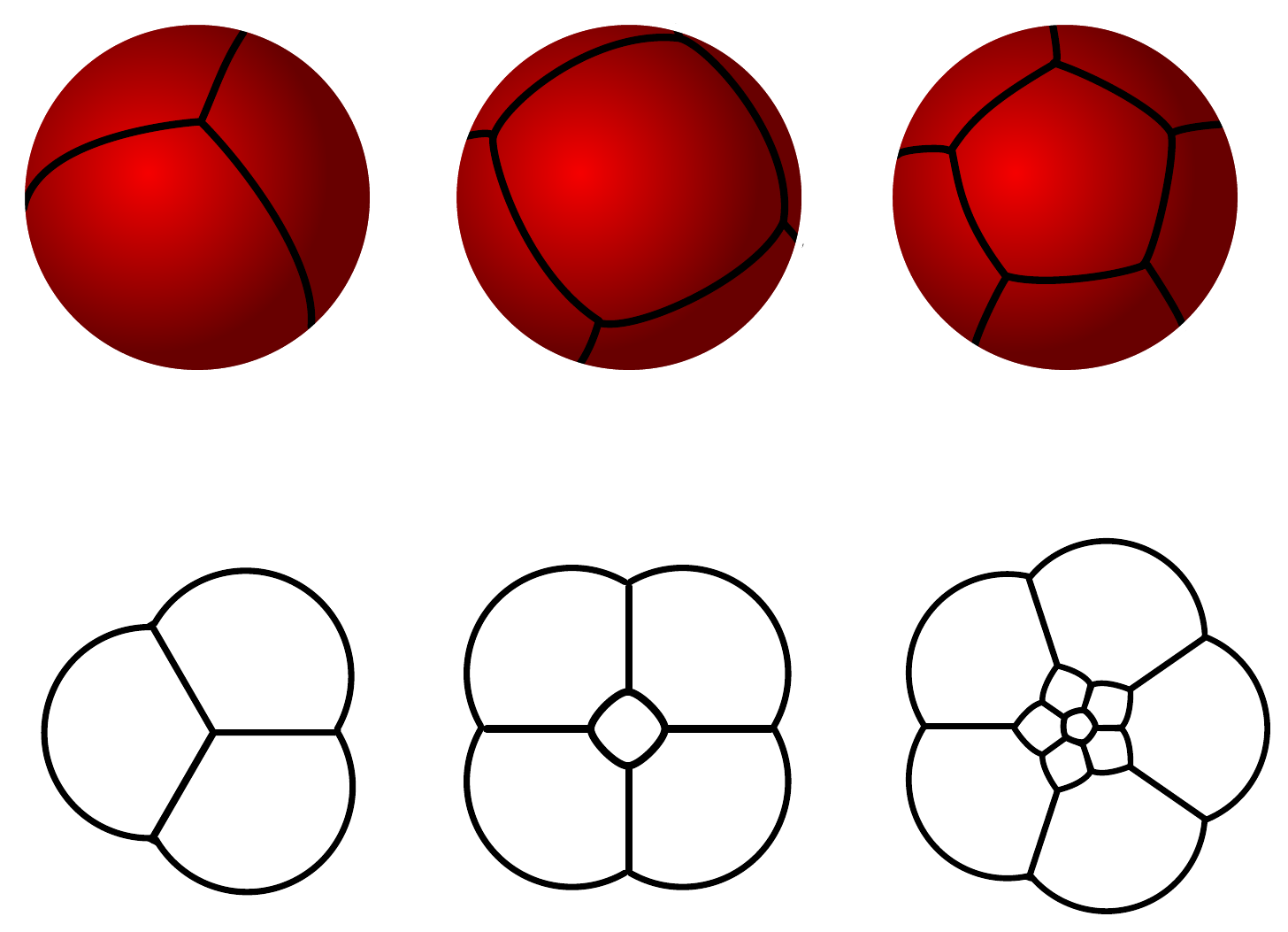}
\vspace{-5pt}
 \caption{
The tetrahedron, cube, and dodecahedron as tessellations of a sphere
(above) and bubble networks (below).}
  \label{fig:sphere-bubble}
\end{figure}

Before we move on, it would be remiss not to mention that the
Platonic solids can also be interpreted 
as \emph{regular tessellations of
the sphere}.\footnote{This is the positively curved counterpart to the
hyperbolic tiling we saw in Exercise \ref{ex:hyp-hon}.}
(You can collect all such tessellations in
Exercise \ref{ex:platonic}.)
In fact, the solids with three edges meeting at a
node---the tetrahedron, cube, dodecahedron displayed in
Fig. \ref{fig:sphere-bubble}---obey the $120^\circ$ rule.
If we ``flatten'' them in the same way we did the
cube\footnote{Technically, we have drawn a very special flattening
  called the \emph{stereographic projection}. This is what the
  vertices would look like to an observer positioned on top of the
  sphere, or if a lantern at the same point cast the shadows of each edge
  onto the plane.}
(Fig. \ref{fig:cube}), we get the bubble networks shown on the bottom
row of Fig. \ref{fig:sphere-bubble}.
It is remarkable that the tetrahedral pattern, which so beautifully exhibits the
$120^\circ$ rule, also shows up as a triple bubble configuration
(Fig. \ref{fig:2dconfigs}), in Plateau's laws, and as a spherical tessellation!

\vspace{5pt}

\vspace{5pt}

\vspace{10pt}
\begin{mybox}
  \begin{exercise}
    \emph{Platonic solids.} \VarMountain \label{ex:platonic}
  \end{exercise}
In this exercise, we'll classify the regular tessellations of the
sphere.
We'll use Euler's formula, $N - E + F = 2$.
\begin{enumerate}[label=(\alph*), itemsep=0pt]
\item Suppose each face of the tessellation has $a$ edges, and each
  node joins up with $b$ edges. Show that
  \[
    2E = Na = Fb.
  \]
\item Using Euler's formula, deduce that
  \begin{equation}
    E = \frac{2 ab}{2(a+b)-ab}.\label{eq:platonic}
  \end{equation}
\item Since $E$ is a whole number, so is the RHS of (\ref{eq:platonic}).
  We will find all possible solutions.
  To begin with, argue that we can interchange the roles of $a$ and
  $b$, so a tessellation with $a$ edges per face and $b$ edges per
  node also gives a tessellation with $b$ edges per face and $a$ edges
  per node.
  These tessellations are said to be \emph{dual} to each other.
\item If $a = 2$, what are the possible values of $b$? Draw the
  corresponding patterns on the sphere and their duals.
\item The denominator of (\ref{eq:platonic}) must be positive. Show
  that this implies
  \[
    a < \frac{2b}{b-2},
  \]
  and hence there are no solutions for $b \geq 6$ and $a \geq 3$.
\item The numerator in (\ref{eq:platonic}) is even, for $a, b$ whole
  numbers.
  Argue that, in order for $E$ to be a whole number, at most one of
  $a$ and $b$ can be odd.
\item Finally, using part (f), conclude 
  that only five combinations of $a$ and $b$ are allowed for $3 \leq
  a, b \leq 5$.
  Check that each of these gives a Platonic solid.
\end{enumerate}
\end{mybox}
\vspace{5pt}

Like Exercise \ref{ex:hyp-hon}, the best partition depends on how much
honey we want to store in each cell. 
But for the equal area cells represented by Fig. \ref{fig:platonic},
are the regular tessellations the best way to split up equal cells on
the surface of the sphere? 
Or does some irregular tiling do better? 
The ``spherical honeycomb conjecture'' is that regular tessellations
are best.
Although known to be true for a dodecahedron and tetrahedron
\cite{hales2002honeycomb, engelstein2009}, it remains a conjecture for
the cube.\footnote{And in case you're wondering, our hexagonality
  argument from \S \ref{sec:hexagon} does not apply simply because the
  sphere has finite surface area. There's not enough space for a network to get large!}

\begin{figure}[h]
  \centering
  \includegraphics[scale=0.43]{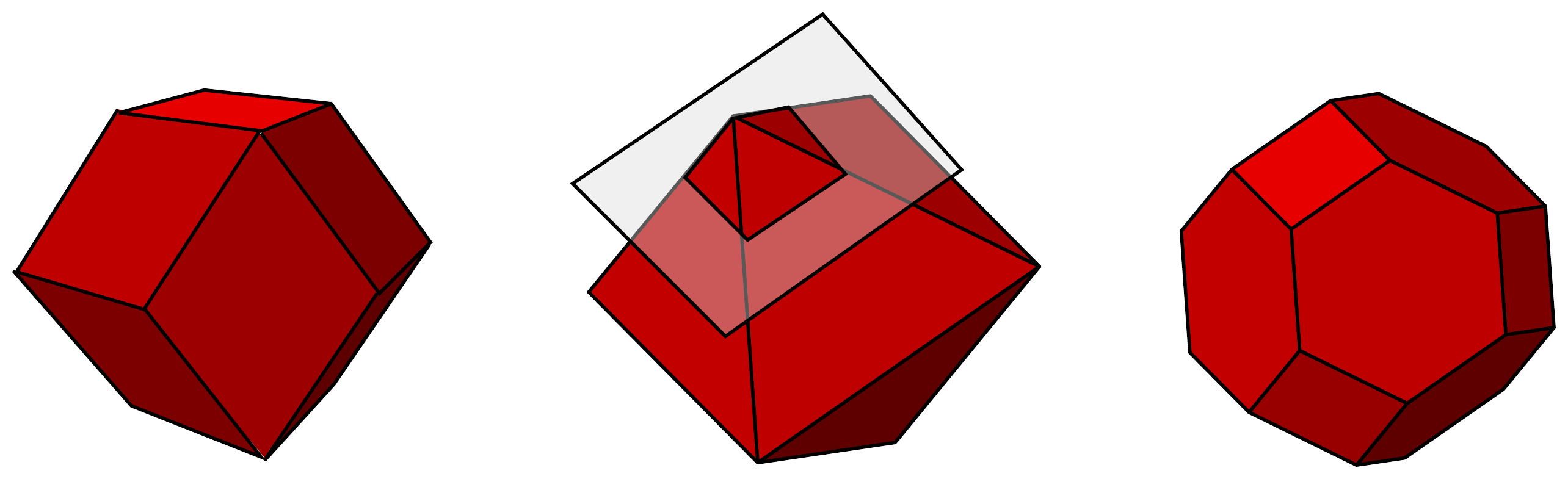}
\vspace{-5pt}
 \caption{
\emph{Left.} The rhombic dodecahedron, with 12 rhombic
faces. \emph{Middle.} Snipping off the corner of an
octahedron. \emph{Right.} Doing
this for all six corners gives the truncated
octahedron.}
  \label{fig:semi}
\end{figure}

Enough about spheres.
Let's return to the problem of space-filling foams, where, if you
recall, we had concluded there was no way to tessellate space
with Platonic solids so as to satisfy Plateau's laws.
Platonic solids are maximally symmetric, in the sense that every
vertex looks alike, and every face looks alike.
But there are many more possibilities when we relax these contraints!
The next simplest shapes are the ``semi-regular'' polyhedra,
comprise by the 13 \emph{Archimedean solids}, whose vertices all look
alike but faces differ, and
the 13 \emph{Catalan solids}, whose faces all look alike but vertices differ.\footnote{In
  fact, these are dual to each other in the sense of Exercise \ref{ex:platonic}.}
Only one from each class can tessellate space:
\begin{itemize}[itemsep=0pt]
\item the \emph{rhombic dodecahedron}, a Catalan solid with twelve
  rhombic faces;
\item the \emph{truncated octahedron}, an Archimedean solid we get by
  snipping off an octahedron's corners.
\end{itemize}  
These are shown in Fig. \ref{fig:semi}, including the ``snipping'' of
a single octahedral corner.

\vspace{5pt}

\begin{figure}[h]
  \centering
  \includegraphics[scale=0.63]{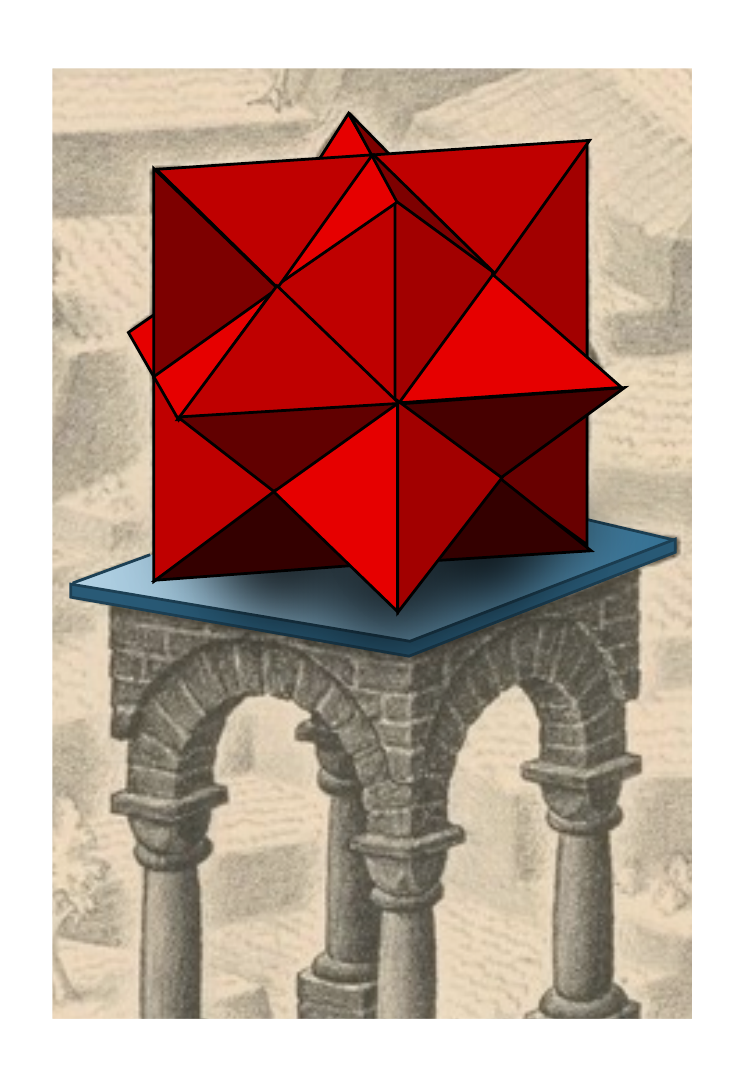}
\vspace{-20pt}
 \caption{
Stellated rhombic dodecahedron. Adapted from \emph{Waterfall} (1961),
M. C. Escher.}
  \label{fig:escher}
\end{figure}

There is one more possibility left in our who's who of space-filling solids.
Take each face of the rhombic dodecahedron and extrude it to
form a pyramid, with each rhombic face replaced by four
triangles.
The result is the \emph{stellated rhombic dodecahedron}, with
``stellated'' meaning ``star-like''.
It is also called \emph{Escher's solid}, since it features in Escher's
marvellous lithograph \emph{Waterfall} (Fig. \ref{fig:escher}).
Remarkably, the extrusions interlock in such a way that Escher's solid
continues to tessellate space, which I like to call a ``testellation''.
Admittedly,
  I've included Escher's solid mainly for the sake of this pun!
We can now use (\ref{eq:foamula2}) to eliminate all but one
candidate on our shortlist.

\vspace{5pt}

\vspace{-5pt}
\begin{mybox}
  \begin{exercise}
    \emph{The Kelvin structure.}
  \end{exercise}
  We have three remaining candidates for an infinite foam with regular
  cells:
  \begin{itemize}[itemsep=0pt]
  \item the rhombic dodecahedron, with two rhombic faces;
  \item Escher's solid, with 48 triangular faces; and
  \item the truncated octahedron, with eight hexagonal faces and six squares.
  \end{itemize}
  Show that only the truncated octahedron satisfies
  (\ref{eq:foamula2}).
\vspace{0pt}
\end{mybox}
\vspace{5pt}

\begin{figure}[h]
  \centering
  \includegraphics[scale=0.26]{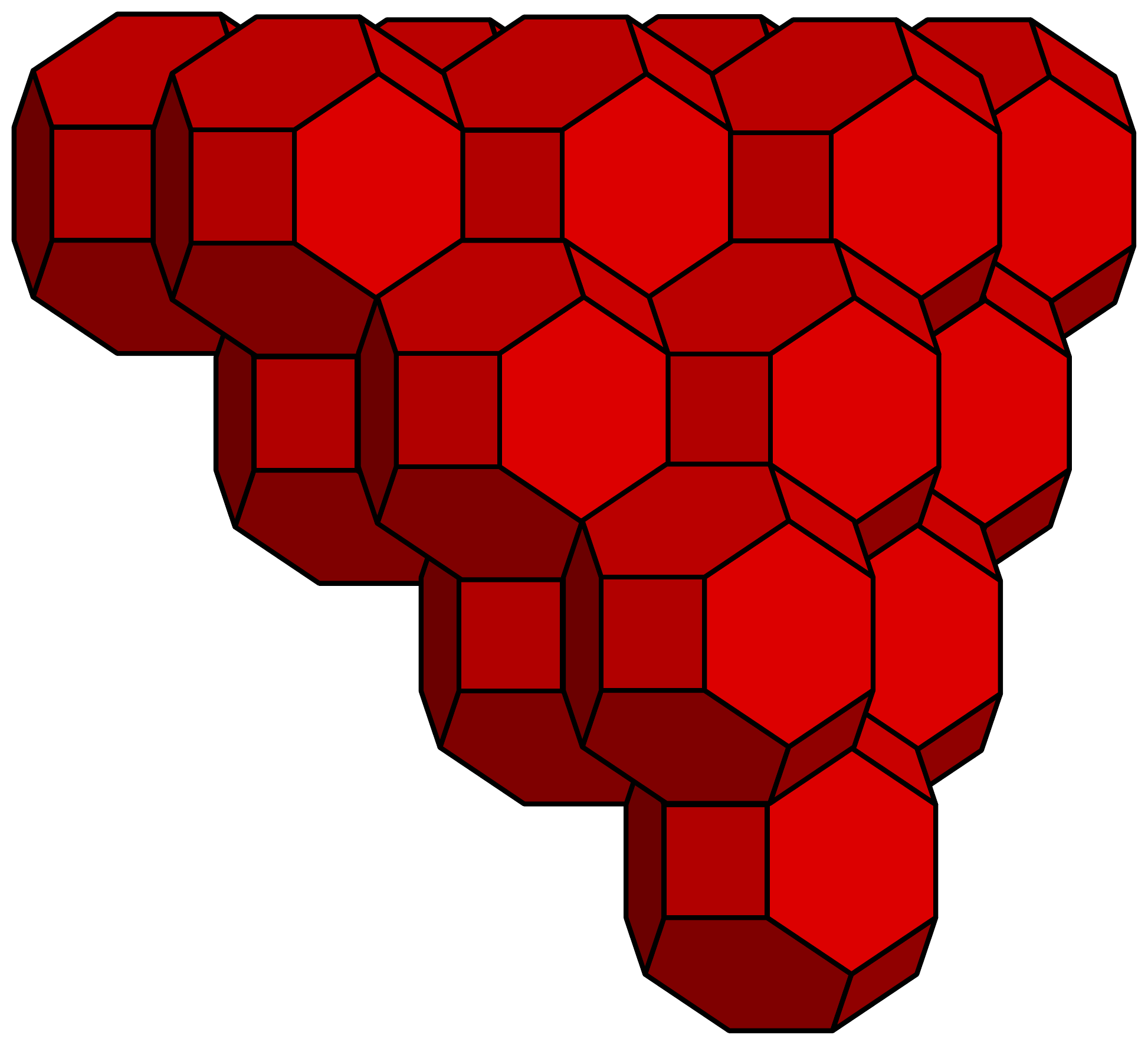}
\vspace{-0pt}
 \caption{The Kelvin structure, made by tiling truncated octahedra.}
  \label{fig:kelvin}
\end{figure}

This truncated octahedron tessellation (Fig. \ref{fig:kelvin}) is called the \emph{Kelvin
  structure} in honor of physicist \textsc{William Thomson, 1st Baron
  Kelvin} (1824--1907), who conjectured it was
the most efficient way to separate equal volume cells.\footnote{Note
  that we need to bend the edges a little to ensure they meet at
  $\theta \approx 109.5^\circ$, in accord with Plateau's laws. 
}
Kelvin's conjecture, often called the \emph{Kelvin problem}, is the
three-dimensional version of the honeycomb
conjecture.\footnote{Appropriate to four-dimensional bees.}
To prove it, we must show there are no \emph{irregular}, equal-volume
tessellations of space more efficient than the Kelvin structure.

\vspace{5pt}

\begin{figure}[h]
  \centering
  \includegraphics[scale=0.18]{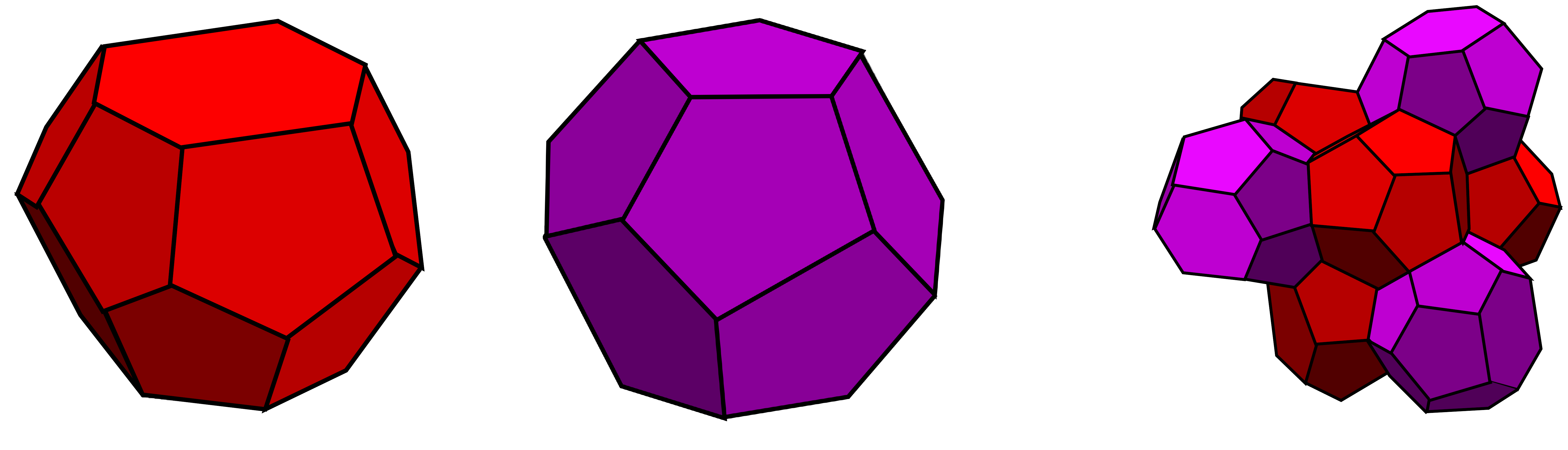}
\vspace{-0pt}
 \caption{\emph{Left.} The $14$-hedron $A_2$. \emph{Middle.} The
   irregular dodecahedron $A_0$. \emph{Right.} A chunk of the Weaire-Phelan structure.}
  \label{fig:weaire}
\end{figure}

Like the hexagonal lattice, the Kelvin structure is the only regular tessellation
which is locally minimal.
But the space of possibilities is much richer in three dimensions than
in two.
In 1993, \textsc{Denis Weaire} and \textsc{Robert Phelan} 
discovered \cite{weaire} they could improve on the Kelvin structure by weaving together
two funny-shaped cells of equal volume:
  \begin{itemize}[itemsep=0pt]
  \item an irregular dodecahedron $A_{0}$, with twelve pentagonal faces; and
  \item a $14$-hedron $A_{2}$ with two hexagonal and twelve pentagonal faces.
  \end{itemize}
The arrangement is called the \emph{Weaire-Phelan structure},
shown in Fig. \ref{fig:weaire}.

\vspace{5pt}

While we have (cautiously) extolled the virtues of soap bubble
computers for solving minimization problems, Weaire and Phelan took
the amusing approach of \emph{simulating} foams! 
They were using the software of the physical universe,
but not the hardware.
Experimentally speaking, the problem is that real soap films are
finnicky, and it is challenging to arrange equal-volume
bubbles \cite{weaire2}.
Even when you ask it to solve the correct problem, 
it often returns the Kelvin structure instead!
Nature is obstreperous, just as \textsf{NP} Hardness predicts (\S
\ref{sec:comp-with-bubbl}).

\begin{figure}[h]
  \centering
  \includegraphics[scale=0.42]{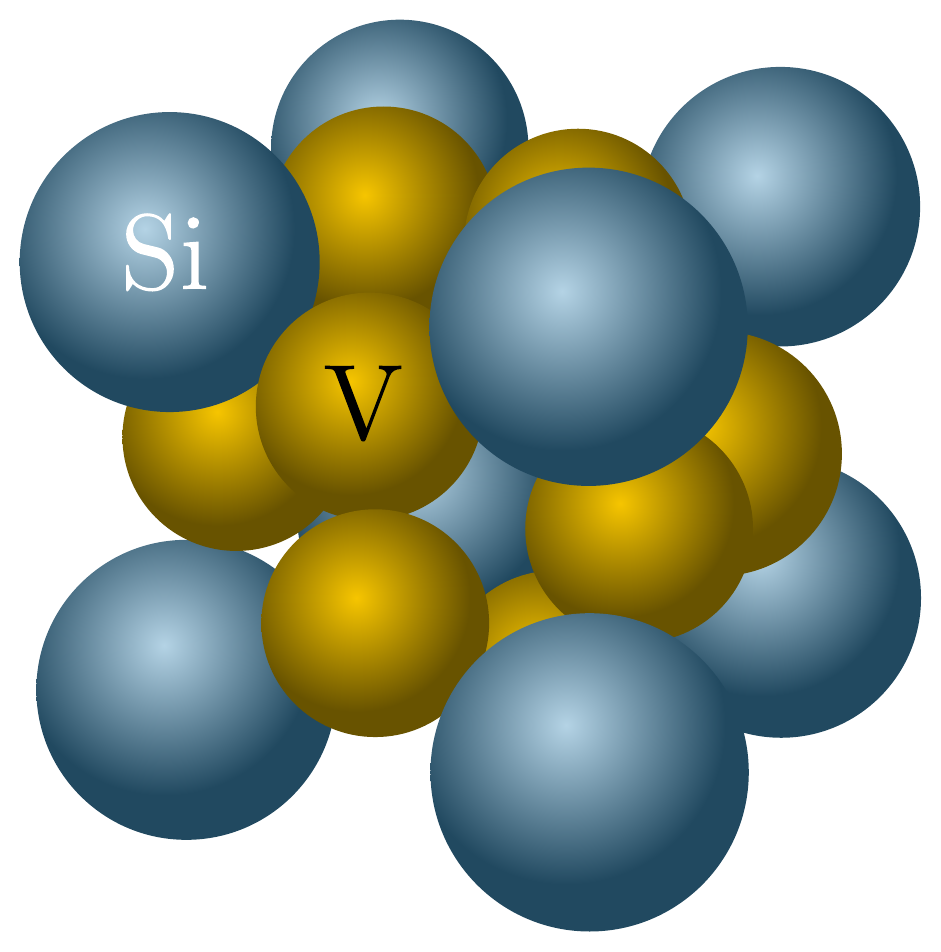} 
\hspace{55pt}
\includegraphics[scale=0.16]{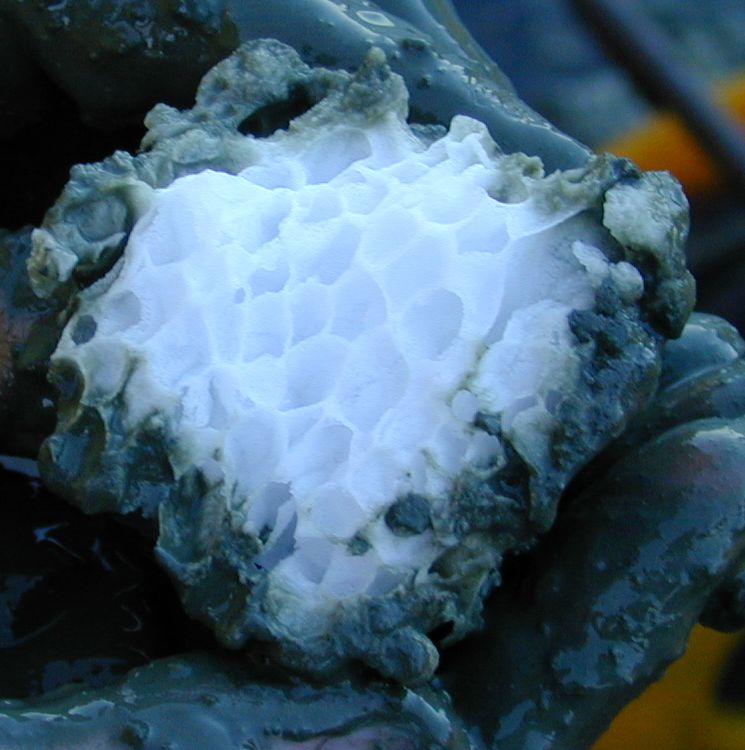}
\vspace{-0pt}
 \caption{\emph{Left.} The A15 phase in superconducting vanadium
   silicide. \emph{Right.} Deep sea methane clathrate hydrate, from \href{https://commons.wikimedia.org/wiki/File:Gashydrat_mit_Struktur.jpg}{Wikipedia}.
}
  \label{fig:a15}
\end{figure}

But stubborn though it is, Nature is also wise. It knew about
Weaire and Phelan's oddity long before the tool-wielding monkeys it
evolved to observe itself!
In 1931, chemists noticed that layers of tungsten\footnote{This is the
  same metal ancient light bulb filaments are made from.} formed by
electrolysis had an unusual chemical structure; a couple
of years later, the same structure was observed in chromium silicide
$\mathrm{Cr}_3\mathrm{Si}$, and in 1953, in the
superconducting\footnote{This means vanadium silicide exhibits \emph{no electrical
  resistance}, at least when cooled below $-256\, {}^\circ$C.}
compound vanadium silicide $\mathrm{V}_3\mathrm{Si}$.
This got the physicists excited! 
Since then, many more superconducting compounds with the same
underlying atomic arrangement have been discovered.
Chemists \textsc{F. C. Frank} and \textsc{J. S. Kasper} began investigating the
mathematical properties of these silicide arrangements, and some close cousins,
together called \emph{tetrahedrally closed-packed (TCP) structures} \cite{Frank1, Frank2}.
The 
original TCP structure is called the
\emph{A15 phase}, shown in Fig. \ref{fig:a15} (left).
Here is the punchlne: this is precisely what you get if
you put an atom at the center of each polyhedron in Weaire and Phelan
space-filling pattern!

\vspace{5pt}

\begin{figure}[h]
  \centering
  \includegraphics[scale=0.18]{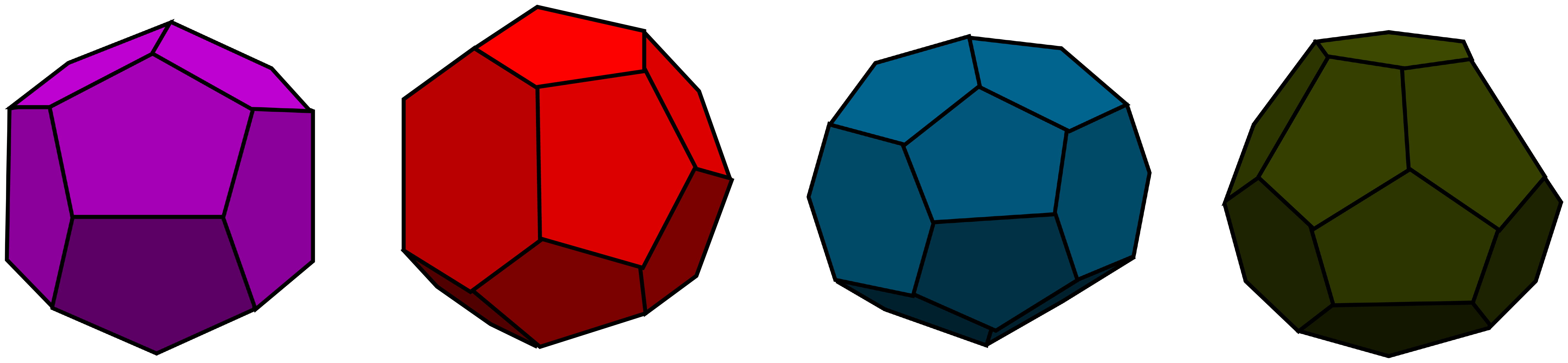}
\vspace{-0pt}
 \caption{The polyhedra $A_{0}, A_{2}, A_{3}$ and $A_{4}$
   appearing in TCP structures.}
  \label{fig:tcp}
\end{figure}

Above, we introduced the $12$- and $14$-sided polyhedra $A_{0}$ and $A_{2}$.
Frank and Kasper built the TCP structures out of \emph{four}
polyhedra, pictured below (Fig. \ref{fig:tcp}).
Each has twelve pentagonal faces, and $0$, $2$, $3$ or $4$
hexagonal faces, with $A_i$ referring to the solid with $i$ hexagons.
No one has classified all the combinations possible with
these TCP polyhedra, though it seems there may be an \emph{infinite}
number!
Weaire-Phelan is the current TCP record-holder, but whether it is
globally optimal in the vasts of TCP space is an open problem.
See \cite{Sullivan1999} for further discussion.
While exploring this full space of possibilities is well beyond us
(and indeed, professional mathematics), the ``foamula''
(\ref{eq:foamula}) gives some simple constraints, derived in Exercise
\ref{ex:tcp}.

\vspace{5pt}

\vspace{0pt}
\begin{mybox}
  \begin{exercise}
    \emph{TCP structures.} \label{ex:tcp} 
  \end{exercise}
  \begin{enumerate}[label=(\alph*), itemsep=0pt]
  \item Suppose we can build a tessellation out of the TCP polyhedra $A_{0},
A_2, A_3, A_4$ in the ratio
\[
  a_0 : a_2 : a_3 : a_4.
\]
Using (\ref{eq:foamula2}), what are the constraints on the possible
ratios?
\item The Weiare-Phelan structure (A15 phase) interleaves $A_0$
  and $A_2$ polyhedra. What is their ratio?
\item The \emph{Z phase} has $A_0$, $A_2$ and $A_3$ in the ratio
  $a_0:2:2$.
  What is $a_0$?
\item The \emph{C15 phase} uses $A_0$ and $A_{4}$ polyhedra.
  What is the ratio?
\item Finally, show that we can write
  \emph{any} ratio for a TCP structure as a combination of A15, C15
  and Z ratios.
\end{enumerate}
\vspace{-5pt}
  Although different structures can have the same ratio, this is a
  useful way to understand the space of possibilities.
\vspace{0pt}
\end{mybox}
\vspace{5pt}

There is a second route to Weaire-Phelan through chemistry.
Instead of placing atoms at the centre of the alternating polyhedra, we can place them at
\emph{vertices} where Plateau borders join.
The 
Weaire-Phelan structure is then called the
\emph{Type I clathrate structure}, and the class of compounds they
occur in the \emph{clathrate hydrates}.\footnote{The Type I clathrate is built from
  $A_0$ and $A_2$ cages, as we expect. There is also a Type II
  clathrate structure, built
  from $A_0$ and $A_4$ cages, which can assemble into a C15 phase.}
Roughly speaking, this means ``water cage'',\footnote{``Clathrate'' is from the Latin \emph{clathratus}, meaning ``with
bars'', while ``hydrate'' is from the Ancient Greek \emph{hydor}
($\ddot{\breve{\upsilon}}\delta\omega\varrho$) for water.
Mixing Greek and Latin like this is considered very poor form in some circles.
}
since clathrate compounds are tiny, elaborate cages made from
ice.
Regular ice doesn't form cages, since the hydrogen bonds
are too
strong, collapsing the cage into the usual crystalline arrangement.
But if you trap a few \emph{gas} atoms inside---such as methane,
carbon dioxide, or neon---it weakens the bonds enough for the cage to
persist!
It's a jail that only exists when it has a prisoner.
A chunk of deep sea methane hydrate is pictured in Fig. \ref{fig:a15} (right).

Clathrates are found in all sorts of exotic locales, from the deep ocean
floor to the outer solar system. Since these ice cages can trap
natural gases like methane, they provide a
vast but non-renewable energy source \cite{clathrate}.
Ironically, 
clathrates also offer a possible means of capturing carbon dioxide and
therefore mitigating climate change.
So, our journey, which started on the train, 
has led via a 
graph of associated minimization problems 
to superconductors,
trans-Neptunian objects and climate change.
Though 
long by some measures,
I suspect we have followed the
shortest path connecting these fixed vertices in concept space.

\begin{center}
  \includegraphics[scale=0.13]{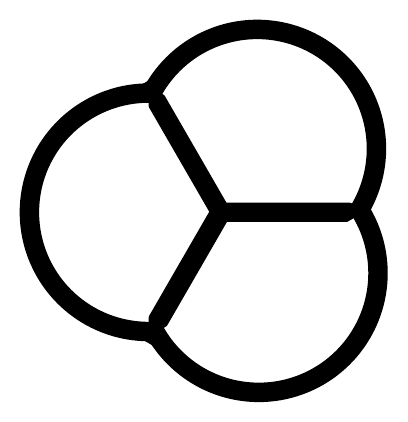} 
\vspace{-5pt}
\end{center}

  \begin{quote}
    \begin{small}
      ``Have you guessed the riddle yet?'' the Hatter said, turning to
      Alice again.  ``No, I give it up,'' Alice replied: ``what’s the
      answer?''  ``I haven’t the slightest idea,'' said the Hatter.
      ``Nor I,'' said the March Hare.  Alice sighed wearily. ``I think
      you might do something better with the time,'' she said, ``than
      waste it in asking riddles that have no answers.''
    \end{small}
  \end{quote}
\vspace{5pt}
\newpage

\bibliographystyle{acm}
\bibliography{steiner}

\end{document}